\documentclass[longauth]{aa} 

\usepackage{graphicx}
\usepackage[varg]{txfonts}
\usepackage{multirow}
\graphicspath{{./Figures/}}
\usepackage{twoopt}
\usepackage{natbib}
\bibpunct{(}{)}{;}{a}{}{,} 
\usepackage[breaklinks=true]{hyperref} 
\makeatletter
\newcommandtwoopt{\citeads}[3][][]{\href{https://ui.adsabs.harvard.edu/\#abs/#3}%
{\def\hyper@linkstart##1##2{}%
\let\hyper@linkend\@empty\citealp[#1][#2]{#3}}}
\newcommandtwoopt{\citepads}[3][][]{\href{https://ui.adsabs.harvard.edu/\#abs/#3}%
{\def\hyper@linkstart##1##2{}%
\let\hyper@linkend\@empty\citep[#1][#2]{#3}}}
\newcommandtwoopt{\citetads}[3][][]{\href{https://ui.adsabs.harvard.edu/\#abs/#3}%
{\def\hyper@linkstart##1##2{}%
\let\hyper@linkend\@empty\citet[#1][#2]{#3}}}
\newcommandtwoopt{\citeyearads}[3][][]%
{\href{https://ui.adsabs.harvard.edu/\#abs/#3}
{\def\hyper@linkstart##1##2{}%
\let\hyper@linkend\@empty\citeyear[#1][#2]{#3}}}
\makeatother
\hypersetup{colorlinks = true, allcolors={cyan}} 
\newcommand{\W}{$\mathrm{W3(H_2O)}$}
\newcommand{\WOH}{W3(OH)}
\newcommand{\mc}{$\mathrm{CH_3CN}$}
\newcommand{\mck}[1]{$\mathrm{CH_3CN}\,(12_#1-11_#1)$}
\newcommand{\mckr}[2]{$\mathrm{CH_3CN}\,(12_K-11_K)\,K = #1-#2$}

\newcommand{\mcisokr}[2]{$\mathrm{C{H_3}^{13}CN}\,(12_K-11_K)\,K = #1-#2$}
\newcommand{\mf}{$\mathrm{HCOOCH_3}$}
\newcommand{\water}{$\mathrm{H_2O}$}

\newcommand{\eg}    {e.g.,}

\newcommand{\jpb}   {Jy~beam$^{-1}$}
\newcommand{\jpp}   {Jy~pixel$^{-1}$}
\newcommand{\kms}   {km~s$^{-1}$}
\newcommand{\lo}    {$L_{\sun}$}
\newcommand{\mo}    {$M_{\sun}$}
\newcommand{\hii}   {\ion{H}{ii}}
\newcommand{\tq}{Toomre~$Q$}

\begin{document} 

   \title{Core fragmentation and Toomre stability analysis of W3(H$_2$O)} 

   \subtitle{A case study of the IRAM NOEMA large program CORE\thanks{Based on observations from an IRAM large program.
   IRAM is supported by INSU/CNRS (France), MPG (Germany), and IGN (Spain).}
   }

   \author{A.~Ahmadi\inst{1,2}
          \and
          H.~Beuther\inst{1}
          \and
          J.~C.~Mottram\inst{1}
          \and
          F.~Bosco\inst{1,2}
          \and
         H.~Linz\inst{1}
         \and
         Th.~Henning\inst{1}
         \and
         J.~M.~Winters\inst{3}
         \and
         R.~Kuiper\inst{4}
         \and
         R.~Pudritz \inst{5}
         \and
         \'A.~S\'anchez-Monge \inst{6}
         \and
         E.~Keto \inst{7}
         \and
         M.~Beltran \inst{8}        
         \and
          S.~Bontemps \inst{9} 
         \and
          R.~Cesaroni \inst{8}
         \and
         T.~Csengeri \inst{10} 
         \and
         S.~Feng \inst{11}
         \and
         R.~Galvan-Madrid \inst{12}
         \and
         K.~G.~Johnston \inst{13} 
         \and
         P.~Klaassen\inst{14}
         \and
         S.~Leurini \inst{15}
         \and
         S.~N.~Longmore \inst{16}
         \and
         S.~Lumsden \inst{13}
         \and
         L.~T.~Maud \inst{17}
         \and
         K.~M.~Menten \inst{10}
        \and
        L.~Moscadelli \inst{8}
        \and
        F.~Motte \inst{18}
        \and
        A.~Palau \inst{12}
        \and
        T.~Peters \inst{19}
        \and
        S.~E.~Ragan \inst{20}
        \and
        P.~Schilke \inst{6}
        \and
        J.~S.~Urquhart \inst{21}
        \and
        F.~Wyrowski \inst{10}
        \and
        H.~Zinnecker \inst{22, 23}
}
   \institute{Max Planck Institute for Astronomy, K\"onigstuhl 17,
              69117 Heidelberg, Germany, e-mail: \href{mailto:ahmadi@mpia.de}{\nolinkurl{ahmadi@mpia.de}}
  \and
Fellow of the International Max Planck Research School for Astronomy and Cosmic Physics at the University of Heidelberg (IMPRS-HD)
  \and
    IRAM, 300 rue de la Piscine, Domaine Universitaire de Grenoble, 38406 St.-Martin-d'H\`eres, France
   \and
  Institute of Astronomy and Astrophysics, University of T\"ubingen, Auf der Morgenstelle 10, 72076, T\"ubingen, Germany
    \and
    Department of Physics and Astronomy, McMaster University, 1280 Main St. W, Hamilton, ON L8S 4M1, Canada
    \and
    I. Physikalisches Institut, Universit\"at zu K\"oln, Z\"ulpicher Str. 77, D-50937, Köln, Germany
    \and
    Harvard-Smithsonian Center for Astrophysics, 160 Garden St, Cambridge, MA 02420, USA
    \and
    INAF, Osservatorio Astrofisico di Arcetri, Largo E. Fermi 5, I-50125 Firenze, Italy
    \and
     Laboratoire d'Astrophyisique de Bordeaux - UMR 5804, CNRS - Universit\'e Bordeaux 1, BP 89, 33270 Floirac, France
     \and
    Max-Planck-Institut f\"ur Radioastronomie, Auf dem H\"ugel 69, 53121 Bonn, Germany   
    \and
    Max-Planck-Institut f\"ur Extraterrestrische Physik, Gissenbachstrasse 1, 85748 Garching, Germany
    \and
    Instituto de Radioastronom\'ia y Astrof\'isica, Universidad Nacional Auton\'oma de M\'exico, PO Box 3-72, 58090 Morelia, Michoacan, Mexico   
     \and
    School of Physics \& Astronomy, E.C. Stoner Building, The University of Leeds, Leeds LS2 9JT, UK   
    \and
    UK Astronomy Technology Centre, Royal Observatory Edinburgh, Blackford Hill, Edinburgh EH9 3HJ, UK
     \and
    INAF - Osservatorio Astronomico di Cagliari, via della Scienza 5, 09047, Selargius (CA), Italy   
     \and
    Astrophysics Research Institute, Liverpool John Moores University, 146 Brownlow Hill, Liverpool L3 5RF, UK   
    \and
    Leiden Observatory, Leiden University, PO Box 9513, 2300 RA Leiden, The Netherlands
     \and
     Universit\'{e} Grenoble Alpes, CNRS, Institut de Plan\'{e}tologie et d'Astrophysique de Grenoble, F-38000 Grenoble, France
    \and
    Max-Planck-Institut f\"ur Astrophysik, Karl-Schwarzschild-Str. 1, D-85748 Garching, Germany
    \and
    School of Physics and Astronomy, Cardiff University, Cardiff CF24 3AA, UK
    \and
    Centre for Astrophysics and Planetary Science, University of Kent, Canterbury, CT2 7NH, UK
     \and
     SOFIA Science Center, Deutsches SOFIA Institut, NASA Ames Research Center, Moffett Field, CA, 94035, USA
     \and
     Universidad Autonoma de Chile, Av. Pedro de Valdivia 425, Santiago, Chile
 }

  \authorrunning{A. Ahmadi et al.}
  \titlerunning{Core fragmentation and Toomre stability analysis of \W}
   
  \date{Received; Accepted}

  \abstract
   {The fragmentation mode of high-mass molecular clumps and the properties of the central rotating structures surrounding the most luminous objects have yet to be comprehensively characterised.}
   {We study the fragmentation and kinematics of the high-mass star-forming region \W, as part of the IRAM NOEMA large program CORE.}
   {Using the IRAM NOrthern Extended Millimeter Array (NOEMA) and the IRAM 30-m telescope, the CORE survey has obtained high-resolution observations of 20 well-known highly luminous star-forming regions in the 1.37~mm wavelength regime in both line and dust continuum emission.}
   {We present the spectral line setup of the CORE survey and a case study for \W. At $\sim$$0\farcs35$ (700~AU at 2.0~kpc) resolution, the \W\ clump fragments into two cores (West and East), separated by $\sim$2300~AU. Velocity shifts of a few \kms\ are observed in the dense-gas tracer, \mc, across both cores, consistent with rotation and perpendicular to the directions of two bipolar outflows, one emanating from each core. The kinematics of the rotating structure about \W\,W shows signs of differential rotation of material, possibly in a disk-like object. The observed rotational signature around \W\,E may be due to a disk-like object, an unresolved binary (or multiple) system, or a combination of both. We fit the emission of \mckr{4}{6} and derive a gas temperature map with a median temperature of $\sim$165~K across \W. We create a \tq\ map to study the stability of the rotating structures against gravitational instability. The rotating structures appear to be Toomre unstable close to their outer boundaries, with a possibility of further fragmentation in the differentially-rotating core, \W\,W. Rapid cooling in the Toomre-unstable regions supports the fragmentation scenario.} 
   {Combining millimeter dust continuum and spectral line data toward the famous high-mass star-forming region \W, we identify core fragmentation on large scales, and indications for possible disk fragmentation on smaller spatial scales.}
   

\maketitle

\section{Introduction}

Fundamental questions pertaining to the fragmentation of high-mass clumps and the accretion processes that result in the birth of the most massive stars ($M\gtrsim 8$~\mo) still remain unanswered. This is in part due to the clustered nature of star-formation and the typically large distances involved. For a long time, the existence of high-mass stars had been puzzling as it was thought that the expected intense radiation pressure would prevent the accretion of enough material onto the protostar (\eg\ \citeads{1974A&A....37..149K}; \citeads{1987ApJ...319..850W}). More recently, two- and three-dimensional (magneto)hydrodynamical simulations of collapsing cores have validated the need for accretion disks in the formation of very massive stars, analogous to low-mass star formation (\eg\ \citeads{2002ApJ...569..846Y}; \citeads{2009Sci...323..754K}; \citeads{2010ApJ...711.1017P}; \citeads{2010ApJ...722.1556K}, \citeyear{2011ApJ...732...20K}; \citeads{2013ApJ...772...61K}; \citeads{2016ApJ...823...28K}). 
Furthermore, different fragmentation processes can contribute to the final stellar mass distribution within a single region, including fragmentation from clouds down to core scales (\eg\ \citeads{2010A&A...524A..18B}; \citeads{2013ApJ...762..120P}, \citeyear{2015MNRAS.453.3785P}; \citeads{2018arXiv180501191B}; see review by \citeads{2017arXiv170600118M}), and disk fragmentation at smaller spatial scales (\eg\ \citeads{2003ApJ...595..913M}; see review by \citeads{2016ARA&A..54..271K}).  

In the disk-mediated accretion scenario, the non-isotropic treatment of the radiation field reduces the effect of radiation pressure in the radial direction, such that radiation can escape through the poles along the disk rotation axis, while the disk is shielded due to the high densities. Observationally, the existence of such disks is expected due to ubiquitous observations of collimated outflows (\eg\ \citeads{2002A&A...387..931B}; \citeads{2009A&A...504..127F}; \citeads{2011A&A...530A..12L}; \citeads{2014prpl.conf..451F}; \citeads{2015MNRAS.453..645M}), which has also been predicted by theoretical models (\eg\ \citeads{2007prpl.conf..277P}). Although some accretion disks in differential Keplerian-like rotation about B-type (proto)stars have been found in recent years (\eg\ \citeads{2012ApJ...752L..29C}; \citeads{2013A&A...552L..10S}; \citeads{2014A&A...571A..52B}; see reviews by \citeads{2007prpl.conf..197C}, and \citeads{2016A&ARv..24....6B}), the existence of such rotating structures around the most massive, O-type protostars is still elusive, with only a few cases reported so far (\citeads{2015ApJ...813L..19J}; \citeads{2016MNRAS.462.4386I}; \citeads{2017A&A...602A..59C}). 

As higher resolution observations are becoming more accessible, thus allowing structures to be resolved on scales $<$1000~AU, it is important to determine whether disks around intermediate to high-mass stars (OB-type) are ubiquitous and if so, to characterise their properties. What is the typical extent of these disks? Are they in differential rotation about a centrally-dominating protostar, similar to their low-mass counterparts and if so, over what range of radii? Is there any scale where a core stops fragmenting?\footnote{Here, a core is defined as a gravitationally-bound region that forms a single or multiple stars, following \citeads{2000prpl.conf...97W}.} At what scales do we see the fragmentation of disks? Are close binary/multiple systems an outcome of disk fragmentation as suggested by, for example, \citetads{2018MNRAS.473.3615M}? If so, stability analyses of these high-mass rotating cores and disks are needed to shed light on fragmentation at disk scales. These questions can only be answered with a statistical approach for a large sample of high-mass star-forming regions.

We have undertaken a large program at IRAM, called CORE \citepads{2018arXiv180501191B}, making use of the IRAM NOrthern Extended Millimeter Array (NOEMA, formerly Plateau de Bure Interferometer) at 1.37~mm in both line and continuum emission to study the early phases of star formation for a sample of 20 highly luminous ($L>10^4$~\lo) star-forming regions at high angular resolution ($\sim$0.4\arcsec), to analyse their fragmentation and characterise the properties of possible rotating structures. Additionally, observations with the IRAM 30-m telescope are included to complement the interferometric data, allowing us to understand the role of the environment by studying high-mass star formation at scales larger than those covered by the interferometer. Observations in the 1.3~mm wavelength regime of the CORE project began in June 2014 and finished in January 2017, consisting of a total of more than 400 hours of observations with NOEMA. The sample selection criteria and initial results from the observed level of fragmentation in the full sample are presented in \citetads{2018arXiv180501191B}, and details of the 30-m observations and the merging of single-dish with the interferometric observations can be found in Mottram et al. (in prep.). In this work, we describe our spectral setup and present a case study of one of the most promising star-forming cloud in our sample, \W. 

\W, also known as the ``Turner-Welch object'', resides in the W3 high-mass star-forming region and was initially identified through observations of the dense-gas tracer HCN at 88.6~GHz \citepads{1984ApJ...287L..81T}. It is located $\sim$0.05~pc (5\arcsec) east of the well-known ultra-compact \hii\ region (UCHII) \WOH. The name, \W, stems from the existence of water masers in the vicinity of the source \citepads{1981ApJ...245..857D}, allowing for an accurate distance measurement of 2.0~kpc for this region (\citeads{2006ApJ...645..337H}; \citeads[cf.][]{2006Sci...311...54X}). The relative proper motions of these masers are further explained by an outflow model oriented in the east-west direction \citepads{2006ApJ...645..337H}. A continuum source elongated in the east-west direction and spanning the same extent as the water maser outflow has been observed in sub-arcsecond VLA observations in the radio regime with a spectral index of --0.6, providing evidence for synchrotron emission (\citeads{1995ApJ...443..238R}; \citeads{1999ApJ...513..775W}). This source of synchrotron emission has been characterised by a jet-like model due to its morphology, and the point symmetry of its wiggly bent structure about the center hints at the possibility of jet precession. Moreover, \citetads{2004A&A...418.1045S} attribute this radio emission to a circumstellar jet or wind ionised by the embedded (proto)star at this position. Additional radio continuum sources have been detected in the vicinity of the synchrotron jet, the closest of which is to the west of the elongated structure and has a spectral index of 0.9 (\citeads{1999ApJ...513..775W}; \citeads{2006ApJ...639..975C}), consistent with a circumstellar wind being ionised by another embedded protostellar source. In fact, the high angular resolution ($\sim$0\farcs7) observations of \citetads{1999ApJ...514L..43W} in the 1.36~mm band allowed for the detection of three continuum peaks in thermal dust emission, one of which peaks on the position of the water maser outflow and synchrotron jet, and another on the position of the radio continuum source with positive spectral index, confirming the existence of a second source at this position. The detection of two bipolar molecular (CO) outflows further supports the protobinary scenario, suggesting that \W\ may be harbouring (at least) two rotating structures \citepads{2011ApJ...740L..19Z}. The two cores within \W\ have individual luminosities on the order of $2\times10^4$ \lo, suggesting two 15~\mo\ stars of spectral type B0\footnote{The luminosity and spectral type calculations are described in detail in Section~\ref{ss: mass_estimates}.}.

In this paper, we aim to study the fragmentation properties of \W\ and the kinematics of the rotating structures within it. We use this source as a test-bed for what will be expanded in a forthcoming paper which will focus on the kinematic properties of a larger sample within our survey. The structure of the paper is as follows. Section~\ref{s: obs_reduction} presents our spectral line setup within the CORE survey with the details of our observations and data reduction for \W.  The observational results are described in Section~\ref{s: obs_results}. The kinematics, temperature, and stability analysis of \W\ is presented in Section~\ref{s: analysis_discussion}. The main findings are summarized in Section~\ref{s: conclusion}. 

\begin{table}
\caption{Observations of \W\ and \WOH.}
\centering
\label{t:obs_list}
\begin{tabular}{lccc}
\hline\hline
Observation Date & Array & Time On-source & Bandpass \\
& & (h) & Calibrator \\
\hline
2014-Oct-31  & D & 3.9  & 3C454.3 \\
2015-Mar-18  & A & 2.6  & 3C84\\
2015-Apr-3 & B & 0.9 & 3C84 \\
2015-Apr-6 & B & 1.3 & 3C84\\
2016-Mar-11 & A & 2.2 & 3C84\\
\hline
\end{tabular}
\tablefoot{The phase and flux calibrators were 0059+581 and MWC349, respectively, for all observations.}
\end{table}

\begin{table}
\caption{Correlator units and frequency ranges observed with NOEMA.}
\centering
\label{t:correlator_table}
\begin{tabular}{lccc}
\hline\hline
Correlator & Spectral Unit & Pol. & Frequency Range \\
& & & (MHz) \\
\hline 
Narrow-band & L01 & H & 220\,690.6--220\,769.7\\
 & L02 & H & 220\,630.6--220\,709.7\\
 & L03 & H & 220\,570.6--220\,649.7\\
 & L04 & H & 220\,130.6--220\,209.7\\
 & L05 & H & 218\,860.6--218\,939.7\\
 & L06 & H & 218\,415.6--218\,494.7\\
 & L07 & H & 218\,280.6--218\,359.7\\
 & L08 & H & 218\,180.6--218\,259.7\\
\hline
WideX & L09 & H & \multirow{2}{*}{218\,878.6--220\,859.5}\\
 & L10 & V & \\
 & L11 & H & \multirow{2}{*}{217\,078.6--219\,059.4}\\
 & L12 & V & \\
\hline
\end{tabular}
\tablefoot{H and V correspond to horizontal and vertical polarisations.}
\end{table}

\section{Observations and data reduction \label{s: obs_reduction}}

   \begin{figure*}
   \centering
   \includegraphics[width=\hsize]{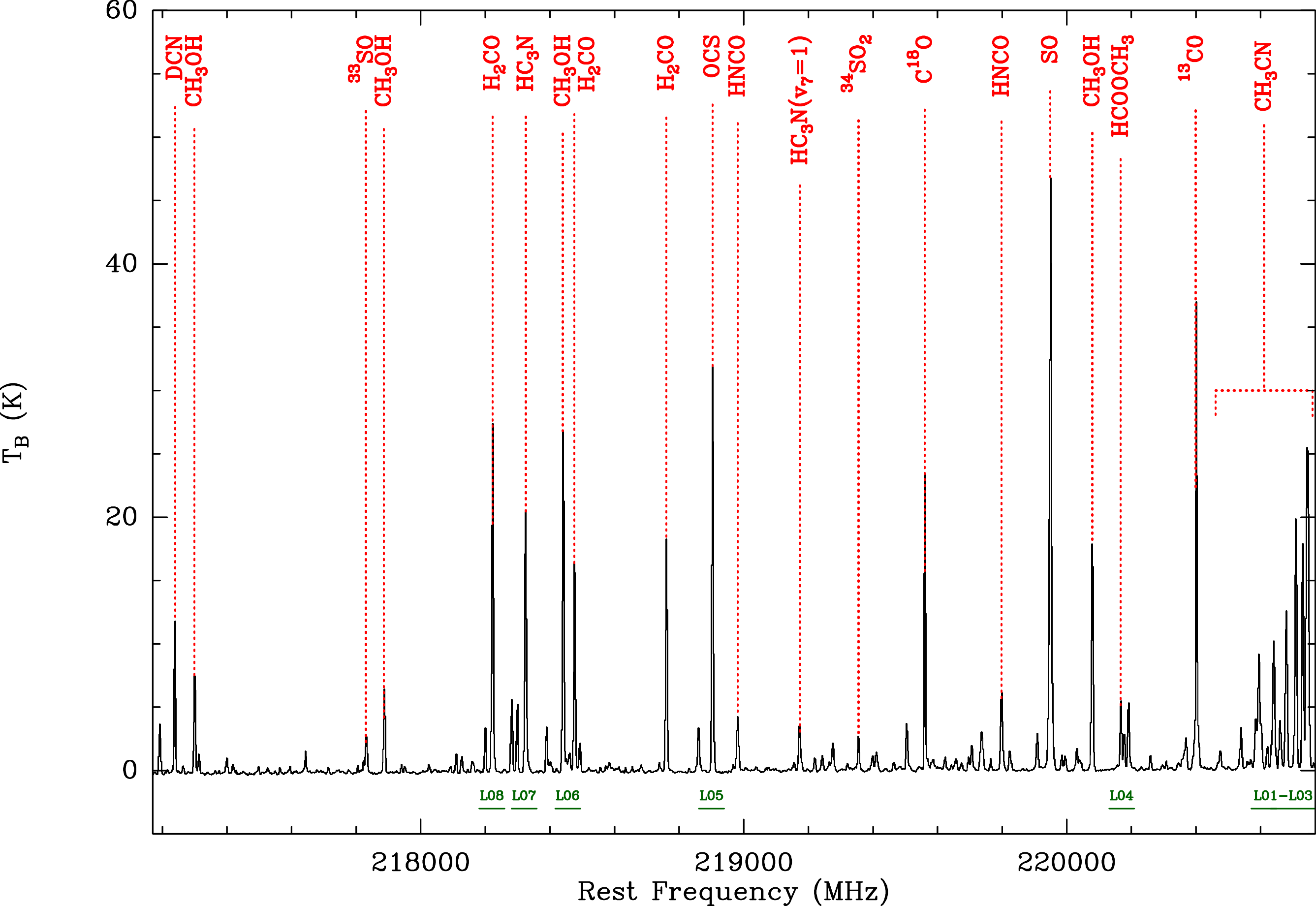}
      \caption{Full WideX spectrum of \W\ averaged over a $4\arcsec \times 4\arcsec$  region encompassing two cores, \W~E and \W~W, showing the chemical-richness of the source. The coverage of the narrow-band correlator units are shown as horizontal green lines and labeled accordingly. The units of the spectrum has been converted from \jpb\ to K by multiplying the flux by 188 $\mathrm{K/Jy}$ under the Rayleigh-Jeans approximation.}
         \label{f: broad_bands_together}
   \end{figure*}
%

   \begin{figure}
   \centering
   \includegraphics[width=\hsize]{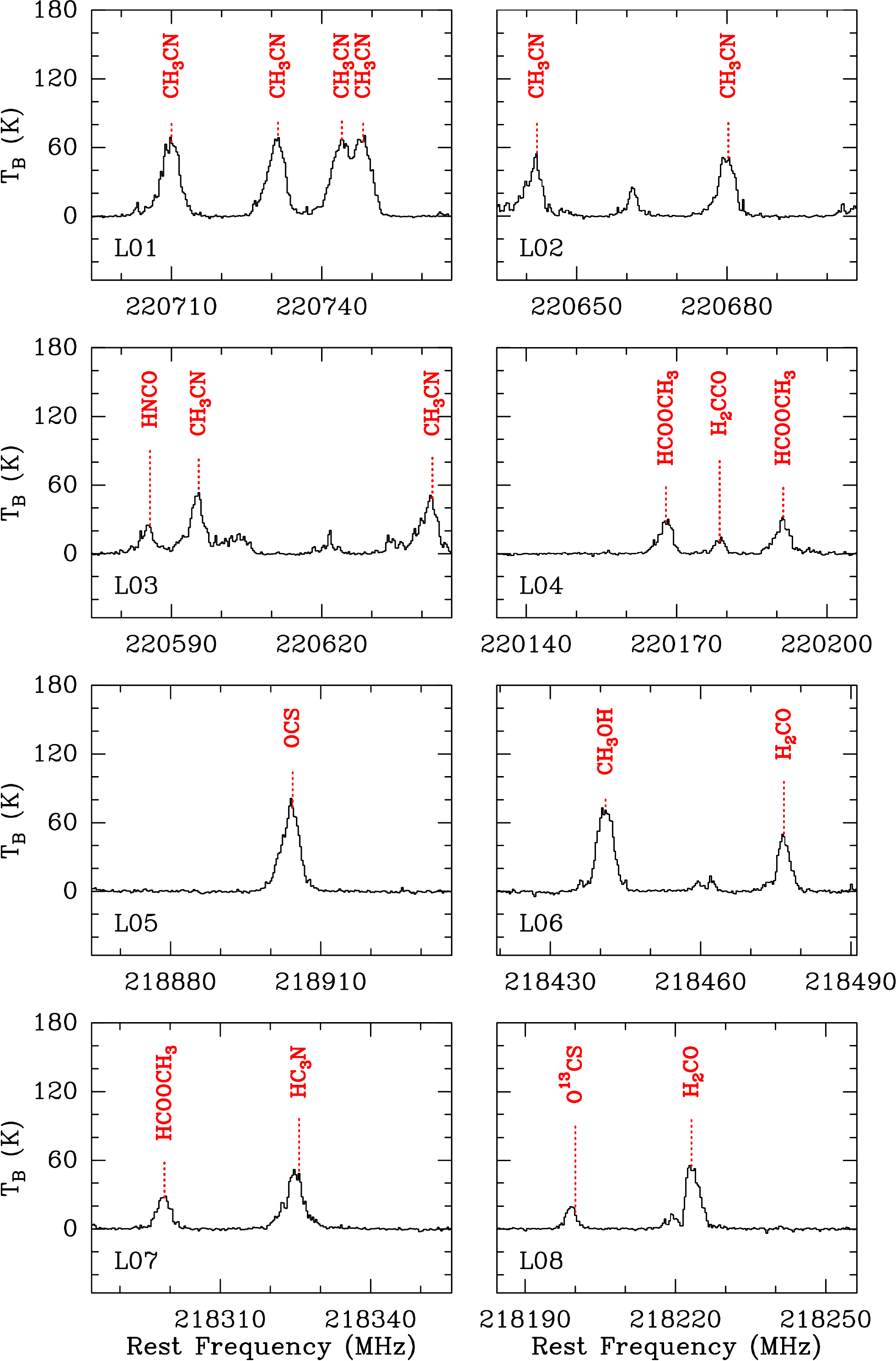}
      \caption{Spectra of the frequency range covered by the narrow-band correlator for the pixel at the phase center toward \W. The name of the correlator unit is listed in the bottom left corner of each panel, with some of the detected lines marked.}
         \label{f: narrow_bands}
   \end{figure}

\subsection{NOEMA observations}
Observations of \W\ at 1.37~mm were made between October 2014 and March 2016 in the A-, B-, and D-array configurations of NOEMA in track-sharing mode with W3\,IRS4. The compact D-array observations were made with six antennas while seven antennas were used for the more extended A- and B-array observations. Baselines in the range of $19-760$~m were covered, therefore the NOEMA observations are not sensitive to structures larger than 12\arcsec\ (0.1~pc) at 220~GHz. On-source observations were taken in roughly 20-minute increments distributed over an observing run and interleaved with observations of various calibration sources. The phase center for the observations of \W\ is $\alpha$(J2000) =  02$^{\rm h}$ 27$^{\rm m}$ 03$\fs87$, $\delta$(J2000) = 61$\degr$ 52$'$ 24$\farcs$5. A summary of the observations can be found in Table~\ref{t:obs_list}.

The full CORE sample of 20 regions has been observed with both a narrow- and a wide-band correlator, simultaneously. The wide-band correlator, WideX, has four units, each with 1.8~GHz bandwidth, covering two overlapping ranges in frequency in both horizontal and vertical polarisations (H and V) with a fixed spectral resolution of 1.95~MHz ($\sim$2.7~\kms\ at 219~GHz). The full coverage of the WideX correlator is shown in Fig.~\ref{f: broad_bands_together} with bright lines marked. The narrow-band correlator has 8 units, each with 80~MHz bandwidth and a spectral resolution of 0.312~MHz ($\sim$0.43 \kms), placed in the 1.37~mm wavelength regime. The frequency coverage of the correlator bands are listed in Table~\ref{t:correlator_table}. The narrow-band correlator can only process the signal from six antennas; therefore, in cases for which the sources were observed with more than six antennas, the correlator automatically accepts the signal from the antennas that yield the best \textit{uv}-coverage. Important lines covered by the narrow-band receiver are listed in Table~\ref{t:narrow_lines_info} and presented in Fig.~\ref{f: narrow_bands} for the pixel at the phase center toward \W.

Data reduction and imaging were performed with the \textsc{clic} and \textsc{mapping} programs of the \textsc{gildas}\footnote{\url{http://www.iram.fr/IRAMFR/GILDAS}} software package developed by IRAM and Observatoire de Grenoble. The continuum was extracted by identifying line-free channels in the range 217\,078.6$-$220\,859.5~MHz covered by all four spectral units of the WideX correlator. As we are interested in achieving the highest possible angular resolution, we CLEANed the cubes using the CLARK algorithm \citepads{1980A&A....89..377C} with a uniform weighting (robust parameter of 0.1)\footnote{This corresponds to the \textsc{casa} robust weighting with a robustness parameter of --2.} yielding a synthesized beam size of $0\farcs43\times0\farcs32$, PA=86\degr, and an rms noise of 3.2~m\jpb\ for the continuum emission using the combined set of observations in the A-, B-, and D-array configurations (hereafter ABD). We also imaged the data from the A- and B-array configurations together (hereafter AB), as well as the A-array only, for which the synthesized beam sizes and rms noise values are summarized in Table~\ref{t: config_specs}.

Continuum subtraction for the lines was performed in the \textit{uv}-plane, by subtracting the emission in the line-free channels in the spectral unit in which the line was observed. Due to line contamination in spectral unit L03, we used the continuum from spectral unit L02 to remove the continuum from spectral unit L03, under the assumption of there being no significant spectral slope between the two adjacent spectral windows. For the WideX continuum subtraction, we subtracted the continuum obtained from line-free channels in all four spectral units. All narrow-band spectra have been resampled to a spectral resolution of 0.5~\kms\ and when imaged with the CLARK algorithm and uniform weighting have a negligibly smaller synthesized beam than the continuum images. The average rms noise of the line images in the ABD configuration is 11.2~m\jpb\ \kms. The synthesized beam size and the average rms noise of the line data for all imaged combinations of array configurations are listed in Table~\ref{t: config_specs}. 
 
\subsection{30-m observations}
Observations of \W\ with the 30-m telescope were obtained on 13 March 2015 centered on the same position as the phase center of the interferometric observations. We used the Eight MIxer Receiver (EMIR) covering the range $213-236$~GHz, reaching a spectral resolution of 0.3~\kms.   In this work, we have merged the NOEMA observations of $^{13}$CO with the single-dish observations using the \textsc{mapping} software and CLEANed the merged cube with the Steer-Dewdney-Ito (SDI) method \citepads{1984A&A...137..159S} in order to recover more of the extended features to study molecular outflows. Further details of the 30-m observations and data reduction as well as the merging process can be found in Mottram et al. (in prep). The resulting merged image has an angular resolution of  $1\farcs14\times0\farcs92$, PA=49\degr, and an rms noise of 8.4~m\jpb\ \kms. We also make use of our single-dish $^{13}$CO data which have been reduced and converted to brightness temperatures for a detailed outflow analysis presented in Section \ref{ss: outflow}. 

\begin{table}
\caption{Bright lines covered in the narrow-band correlator setup.}
\centering
\label{t:narrow_lines_info}
\begin{tabular}{lccr}
\hline\hline
Molecule & Transition & Rest Frequency & $E_u/k$\\
& & (MHz) & (K) \\
\hline
$\mathrm{O^{13}CS}$ & 18--17 & 218\,199.00 & 99.5\\
$\mathrm{H_2CO}$ & 3$_{0,3}$--2$_{0,2}$ & 218\,222.19 & 21.0\\
$\mathrm{HCOOCH_3}$ & 17$_{3,14}$--16$_{3,13}$A & 218\,297.89 & 99.7\\
$\mathrm{HC_3N}$ & 24--23 & 218\,324.72 & 131.0\\
$\mathrm{CH_3OH}$ & 4--3 & 218\,440.05 & 45.5\\
$\mathrm{H_2CO}$ & 3$_{2,2}$--2$_{2,1}$ & 218\,475.63 & 68.1\\
OCS & 18--17 & 218\,903.36 & 99.81\\
$\mathrm{HCOOCH_3}$ & 17$_{4,13}$--16$_{4,12}$E & 220\,166.89 & 103.1\\
$\mathrm{H_2CCO}$ & 11$_{1,11}$--10$_{1,10}$ & 220\,177.57 & 76.5\\
$\mathrm{HCOOCH_3}$ & 17$_{4,13}$--16$_{4,12}$A &  220\,190.29 & 103.2\\
HNCO & 10$_{1,9}$--9$_{1,8}$ & 220\,584.75 & 101.5\\
$\mathrm{CH_3CN}$ & 12$_6$--11$_6$ & 220\,594.42 & 325.9\\
$\mathrm{CH_3^{13}CN}$ & 12$_3$--11$_3$ & 220\,599.98 & 133.1\\
$\mathrm{CH_3^{13}CN}$ & 12$_2$--11$_2$ & 220\,621.14 & 97.4\\
$\mathrm{CH_3^{13}CN}$ & 12$_1$--11$_1$ & 220\,633.83 & 76.0\\
$\mathrm{CH_3^{13}CN}$ & 12$_0$--11$_0$ & 220\,638.07 & 68.8\\
$\mathrm{CH_3CN}$ & 12$_5$--11$_5$ & 220\,641.08 & 247.4\\
$\mathrm{CH_3CN}$ & 12$_4$--11$_4$ & 220\,679.29 & 183.2\\
$\mathrm{CH_3CN}$ & 12$_3$--11$_3$ & 220\,709.02 & 133.2\\
$\mathrm{CH_3CN}$ & 12$_2$--11$_2$ & 220\,730.26 & 97.4\\
$\mathrm{CH_3CN}$ & 12$_1$--11$_1$ & 220\,743.01 & 76.0\\
$\mathrm{CH_3CN}$ & 12$_0$--11$_0$ & 220\,747.26 & 68.9\\
\hline
\end{tabular}
\tablefoot{Rest frequencies and upper energy levels have been obtained from the Cologne Database for Molecular Spectroscopy (CDMS) (\citeads{2001A&A...370L..49M}; \citeyearads{2005JMoSt.742..215M}), with the exception of those for $\mathrm{HCOOCH_3}$ transitions which were acquired from the Jet Propulsion Laboratory \citepads{1998JQSRT..60..883P}.}
\end{table}

\begin{table*}[!ht]
\caption{Details of CLEANed images.}
\centering
\label{t: config_specs}
\begin{tabular}{lccccc}
\hline\hline
& \multicolumn{2}{c}{Continuum} & & \multicolumn{2}{c}{Line (Narrow-band)}\\ 
\cline{2-3} \cline{5-6} & Synthesized & rms Noise & & Synthesized & Average rms Noise\\
Configuration & Beam & (m\jpb) & & Beam & (m\jpb\ \kms) \\
\hline
ABD & $0\farcs43\times0\farcs32$, PA=86\degr & 3.2 & & $0\farcs42\times0\farcs31$, PA=87\degr &  11.2\\
AB & $0\farcs41\times0\farcs30$, PA=86\degr & 2.6 & & $0\farcs38\times0\farcs28$, PA=87\degr& 8.6\\
A & $0\farcs39\times0\farcs28$, PA=88\degr & 2.5 & & $0\farcs36\times0\farcs26$, PA=88\degr& 8.0  \\
\hline
\end{tabular}
\end{table*}

\section{Observational Results \label{s: obs_results}} 
In the following, we present our detailed analysis for \W, and when applicable, we also showcase our observational results for \WOH. Our analysis mainly uses the continuum and \mc\ spectral line emission. Maps of the other lines are shown in Appendix~\ref{a: moment_maps_ABD}.

   \begin{figure*}
   \centering
   \includegraphics[width=0.9\hsize]{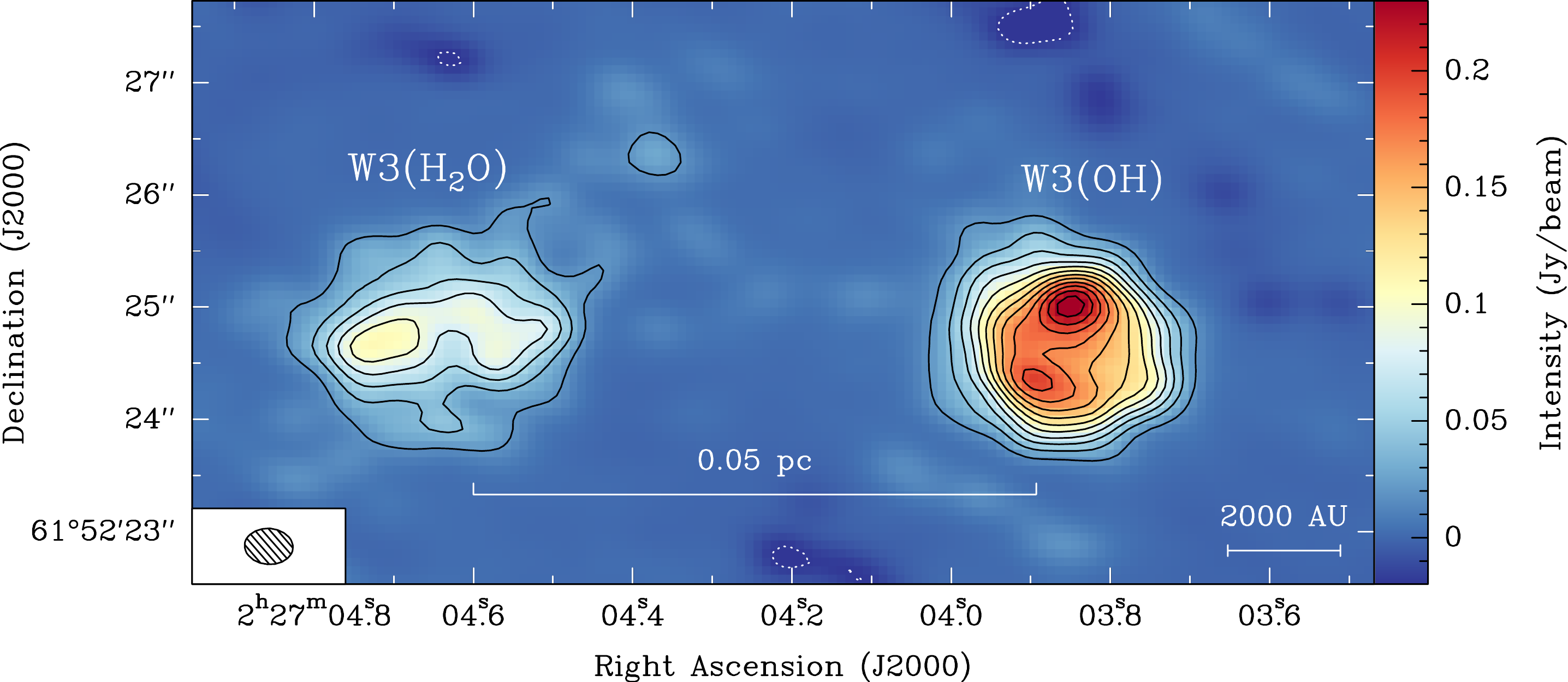}
      \caption{NOEMA 1.37~mm (219~GHz) continuum image toward \W\ and \WOH\ in the ABD configuration. The solid contours start at $6\sigma$ and increase in steps of $6\sigma$ ($1\sigma$ = 3.2~m\jpb). The dotted contours show the same negative levels. A scale-bar and the synthesized beam ($0\farcs43\times0\farcs32$, PA=86\degr) are shown in the bottom.}
         \label{f: w3_cont_ABD}  
   \end{figure*}

   \begin{figure*}
   \centering
   \includegraphics[width=\hsize]{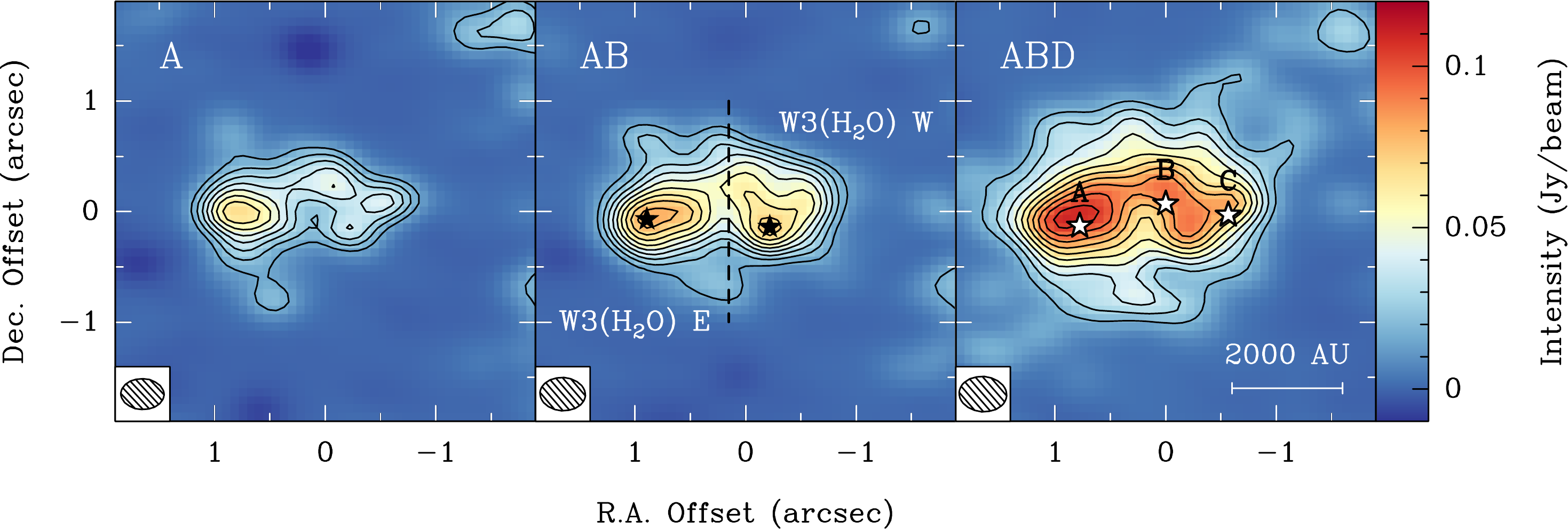}
      \caption{1.37~mm continuum image toward \W\ observed with the A (\emph{left}), AB (\emph{center}), and ABD (\emph{right}) configurations of NOEMA. The solid contours start at $6\sigma$ and increase in steps of $3\sigma$ (see Table~\ref{t: config_specs}). Synthesized beams are shown in the bottom left corners of each panel, along with a scale bar in the bottom right of the right-hand panel. The black stars in the middle panel correspond to the positions of the continuum peaks, marking the locations of the two individual cores, \W\ W and \W\ E, with the dashed line as the approximate separation boundary. The white stars in the right panel correspond to the positions of the continuum peaks A, B, and C from \citetads{1999ApJ...514L..43W}. The offset zero position is the phase center of the observations: $\alpha$(J2000) =  02$^{\rm h}$ 27$^{\rm m}$ 03$\fs87$, $\delta$(J2000) = 61$\degr$ 52$'$ 24$\farcs$5.}
         \label{f: w3h2o_cont_allconfig}
   \end{figure*}

\subsection{Continuum emission}
Figure~\ref{f: w3_cont_ABD} shows the 1.37~mm (219~GHz) continuum emission map of \W\ and \WOH\ in the ABD configuration. At this wavelength, the continuum emission in our field of view is dominated in \WOH\ by free-free emission, while the emission in \W\ is due to dust \citepads{1999ApJ...514L..43W}. In the following, we focus on the fragmentation and kinematics of the younger region, \W. 

Figure~\ref{f: w3h2o_cont_allconfig} shows a comparison of the continuum emission maps of \W\ obtained by imaging the ABD, AB, and A-array only observations. The integrated flux within the $6\sigma$ contours is 1220~mJy, 656~mJy, and 364~mJy for ABD, AB, and A-array images, respectively. The fragmentation of \W\ into two cores, separated by $\sim$2300~AU, is best seen in the AB image at 700~AU scales. The two cores are labeled \W\ E and \W\ W, and their peak continuum positions are depicted by stars in Fig.~\ref{f: w3h2o_cont_allconfig}, marking the positions of embedded (proto)stars. The peak position of \W~E is $\alpha$(J2000) =  02$^{\rm h}$ 27$^{\rm m}$ 04$\fs73$, $\delta$(J2000) = 61$\degr$ 52$'$ 24$\farcs$66, and that of \W~W is $\alpha$(J2000) =  02$^{\rm h}$ 27$^{\rm m}$ 04$\fs57$, $\delta$(J2000) = 61$\degr$ 52$'$ 24$\farcs$59. The approximate separation boundary between \W~E and W is marked with a vertical dashed line. The integrated flux within 6$\sigma$ contours and the separation boundary are 377~mJy and 279~mJy for \W~E and \W~W, respectively. Furthermore, there exists an additional emission peak to the northwest of \W\ at an offset of $-1\farcs6$, 1\farcs6 ($\alpha$(J2000) =  02$^{\rm h}$ 27$^{\rm m}$ 04$\fs37$, $\delta$(J2000) = 61$\degr$ 52$'$ 26$\farcs$35) which is best seen in the ABD image as it has the best sensitivity. This is most likely a site for the formation of lower mass stars. 

Radiative transfer models by \citetads{2006ApJ...639..975C} for \W\ in the 1.4~mm wavelength regime show that the averaged dust optical depth is less than 0.09, therefore, we assume the thermal dust emission to be optically thin in our observations. The positions of continuum peaks A and C from \citetads{1999ApJ...514L..43W} coincide well with the continuum peaks \W~E and \W~W in our observations to within a synthesized beam (see Fig.~\ref{f: w3h2o_cont_allconfig}). \citeauthor{1999ApJ...514L..43W} had attributed the third peak in their observations in between the other two core to an interplay of high column density and low temperatures in the central region and concluded \W\ to be harbouring two cores at the positions of continuum peaks A and C. Our doubly-peaked continuum image in AB, with a better spatial resolution than that of \citetads{1999ApJ...514L..43W} by a factor of 3.3, supports this interpretation. Furthermore, peak B coincides well with the northeast extension of \W~W in our observations, best seen in the middle panel of Fig.~\ref{f: w3h2o_cont_allconfig}. 

   \begin{figure*}
   \centering
   \includegraphics[width=0.9\hsize]{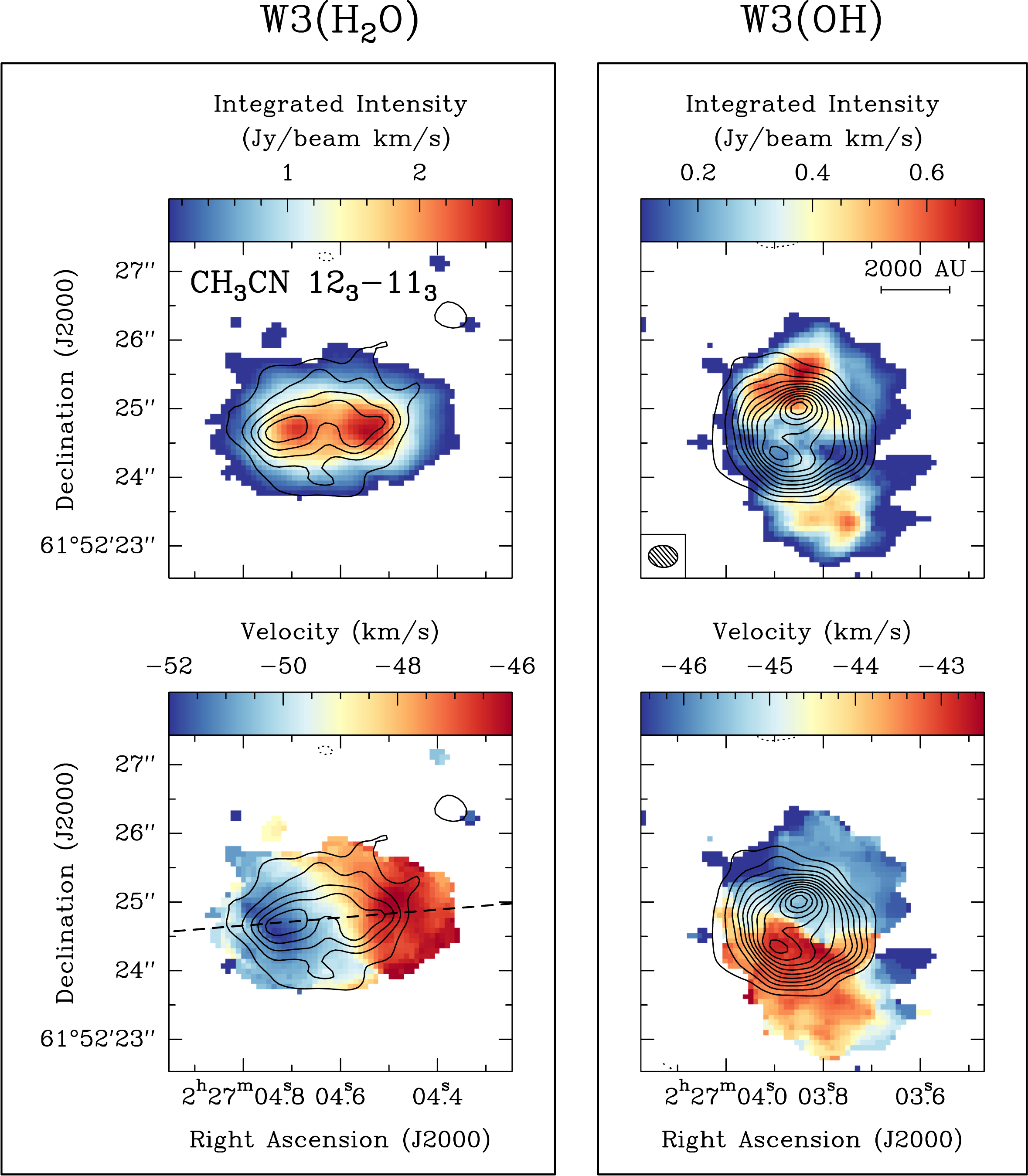}
      \caption{\emph{Top}: Integrated intensity (zeroth moment) map of \mck{3} for \W\ (\emph{left}) and \WOH\ (\emph{right}) in the ABD configuration. \emph{Bottom:} Intensity-weighted peak velocity (first moment) map of \mck{3} for \W\ (\emph{left}) and \WOH\ (\emph{right}) in the ABD configuration. The dashed line corresponds to the cut made for the PV plot of \W\ presented in Fig.~\ref{f: w3h2o_PV_ABD}. The solid contours correspond to the 1.37~mm continuum, starting at $6\sigma$ and increasing in steps of $6\sigma$ ($1\sigma$ = 3.2~m\jpb). The dotted contours correspond to the same negative levels. A scale-bar and the synthesized beam ($0\farcs43\times0\farcs32$, PA=86\degr) are shown in the top right panel.}
         \label{f: w3_mom1_ABD}
   \end{figure*}

\subsection{Line emission}

\W\ is one of the most chemically-rich sources in our sample (see Fig.~\ref{f: broad_bands_together}) with detections of sulfur-bearing species such as $\mathrm{^{33}SO}$ and $\mathrm{^{34}SO_2}$, complex species such as $\mathrm{HCOOCH_3}$, and vibrationally excited lines of $\mathrm{HC_3N}$, among many others. Fig.~\ref{f: w3_mom1_ABD} shows integrated intensity (zeroth moment) and intensity-weighted peak velocity (first moment) maps of \mck{3} for \W\ and \WOH. The zeroth moment map confirms the fragmentation of \W\ into two cores. The moment maps of most lines covered by the narrowband receiver (see Table~\ref{t:narrow_lines_info} and Fig.~\ref{f: narrow_bands}) are presented in Appendix~\ref{a: moment_maps_ABD}. All moment maps have been created inside regions where the signal-to-noise is greater than 5$\sigma$. The integrated intensity maps of most tracers for \W\ also show two peaks coincident with the locations of \W~E and \W~W. While the continuum emission is stronger for \W~E, some dense gas tracers (\eg\ \mc, HC$_3$N) show stronger line emission towards \W~W. 

The bottom panels in Fig.~\ref{f: w3_mom1_ABD} show the intensity-weighted peak velocity (first moment) map of the region in \mck{3}. We chose to do our kinematic analyses on this transition as it is the strongest unblended line in the methyl cyanide (\mc) $K$-ladder. There is a clear velocity gradient in the east-west direction across \W, and in the NW-SE direction across \WOH. The systemic velocities of both clumps are determined by averaging the spectra of \mck{3} over a $4\arcsec \times 4\arcsec$ area centered on each source and fitting a Gaussian line to the resulting averaged spectrum. In this way, \W\ and \WOH\ have average velocities of --49.1 and --45.0~\kms, respectively. 

The velocity gradient across \W\ is detected in most of the high spectral resolution lines in our survey (see Fig.~\ref{af: all_narrow_moment1} in Appendix~\ref{a: moment_maps_ABD}) and spans $\sim$6000~AU in size, corresponding to an amplitude of 170~\kms~$\mathrm{pc}^{-1}$. The velocity gradient resolved in \WOH\ has an amplitude of $\sim$100~\kms~$\mathrm{pc}^{-1}$ and is roughly perpendicular to the motion of the ionised gas in the east-west direction as traced by the H92$\alpha$ line in observations of \citetads{1995ApJ...444..765K}, and is perpendicular to the direction of the ``champagne flow'' observed to the northeast at radio frequencies (\citeads{1995ApJ...444..765K}; \citeads{1999ApJ...513..775W}). Hence, in \WOH\ we seem to be witnessing the large-scale motion of the remnant molecular gas. The interpretation of line emission for \WOH\ is complex, because most of the continuum emission at 1.37~mm is due to free-free emission which affects the appearance of molecular lines. In practice, free-free emission at 1.3~mm reduces the molecular line emission which is reflected by the reduced integrated line emission towards the peaks of \WOH. As this is beyond the scope of this paper, we refrain from further analysis of \WOH.

   \begin{figure*}
   \centering
   \includegraphics[width=0.95\hsize]{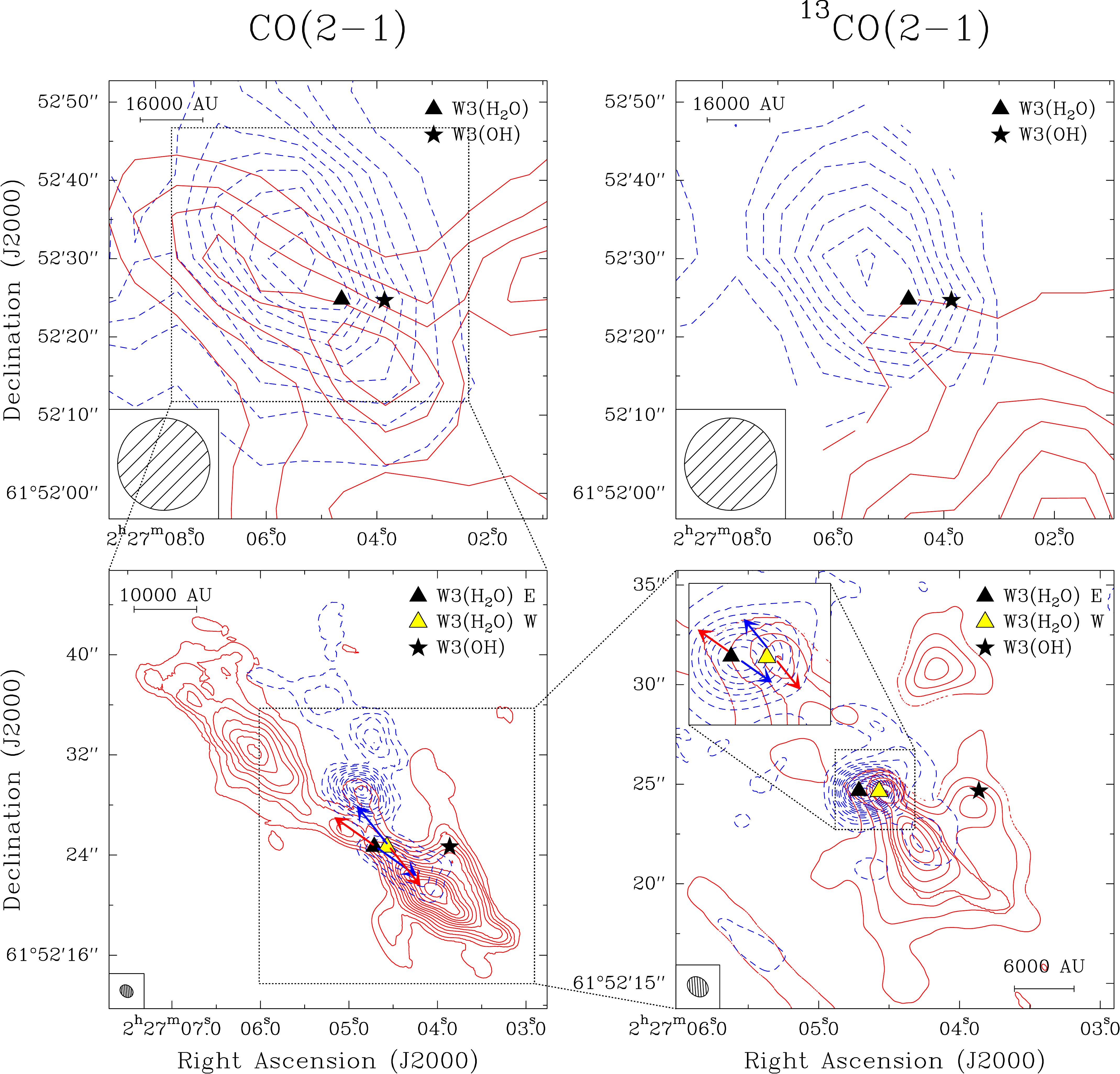}
      \caption{\emph{Top left:} Intensity map of CO\,(2--1) emission from IRAM 30-m integrated over the velocity range of $-40$ to $-20$ \kms\ for the redshifted and $-80$ to $-55$ \kms\ for the blueshifted gas. Contours start at $5\sigma$ and increase in steps of $5\sigma$ ($1\sigma$ = 1.9~K \kms). 
      \emph{Top right:} Intensity map of $^{13}$CO\,(2--1) emission from IRAM 30-m integrated over the velocity range of $-40$ to $-20$ \kms\ for the redshifted and $-80$ to $-55$ \kms\ for the blueshifted gas. Contours start at $3\sigma$ and increase in steps of $2\sigma$ ($1\sigma$ = 0.42~K \ \kms). 
      \emph{Bottom left:} Intensity map of CO\,(2--1) emission obtained by \citetads{2011ApJ...740L..19Z} with the SMA integrated over the velocity range of $-45$ to $-25$ for the redshifted and $-75$ to $-55$ for the blueshifted gas. Contours start at $5\sigma$ and increase in steps of $5\sigma$ ($1\sigma$ = 1.1~m\jpb\ \kms). The blue and red arrows highlight the positions and directions of the CO outflows emanating from each source. 
      \emph{Bottom right:} Intensity map of $^{13}$CO\,(2--1) emission from IRAM 30-m merged with NOEMA observations integrated over the velocity range of $-41$ to $-30$ for the redshifted and $-70$ to $-53$ for the blueshifted gas. Contours start at $3\sigma$ and increase in steps of $3\sigma$ ($1\sigma$ = 0.03~m\jpb\ \kms\ for redshifted and 0.17~m\jpb\ \kms\ for blueshifted sides). The zoom panel in the top left corner highlights the launching positions of each outflow, with the red and blue arrows showing the directions of the two bipolar outflows from \citetads{2011ApJ...740L..19Z}. The rms noise used in drawing the contours of the integrated intensity maps have been determined by first creating the maps without any constraints on the minimum emission level (threshold of 0) and calculating the noise in an emission-free part of the resulting map.}
         \label{f: w3h2o_outflows}
   \end{figure*}

\subsection{Outflow structure \label{ss: outflow}} 
In Fig.~\ref{f: w3h2o_outflows} we show integrated intensity (zeroth moment) maps of outflow-tracing molecules ($^{12}$CO and $^{13}$CO) for the redshifted and blueshifted gas. The minimum intensity below which a pixel is not considered in the creation of the moment maps is based on $5\sigma$ rms noise level in emission-free channels. The single-dish CO\,(2--1) map (see top left panel of Fig.~\ref{f: w3h2o_outflows}) shows the existence of a bipolar outflow in the overall vicinity of \W\ but also encompassing \WOH, in approximately northeast-southwest direction. The $^{13}$CO\,(2--1) single-dish map (see top right panel of Fig.~\ref{f: w3h2o_outflows}) shows less extended emission than the $^{12}$CO as the $^{13}$C isotopologue is a factor of $\sim$76 less abundant \citepads{1982A&A...109..344H}, and the map highlights the general northeast-southwest direction of the outflow even better.

The integrated intensity map of $^{12}$CO from SMA interferometric data of \citetads{2011ApJ...740L..19Z} allows for the detection of two bipolar outflows (see bottom left panel of Fig.~\ref{f: w3h2o_outflows}). The outflow emanating from \W~E has its blueshifted side to the southwest and its redshifted lobe to the northeast, while, the second outflow emanating from \W~W has its blueshifted side extending to the northeast with its redshifted side to the southwest \citepads{2011ApJ...740L..19Z}. The difference between the position angles of the two outflows (in the plane of the sky) is 25\degr. Furthermore, the resulting zeroth moment map of $^{13}$CO emission from our combined NOEMA and 30-m single-dish observations, presented in the bottom right panel of Fig.~\ref{f: w3h2o_outflows}, confirms the findings of \citetads{2011ApJ...740L..19Z} with regards to the directions and origin of the redshifted outflow lobe from \W~W and the origins of the blueshifted outflow lobe from \W~E. However, we miss much of the emission that is detected in the $^{12}$CO SMA  interferometric data, mainly due to the lower abundance of the $^{13}$C isotopologue, and thus its lower sensitivity to the outflowing gas. The same coloured arrows obtained from \citeauthor{2011ApJ...740L..19Z} are redrawn in a zoom panel inside the bottom-right panel of this figure, highlighting that the two outflows are in fact emanating from different positions. 

\begin{figure}
   \centering
   \includegraphics[width=\hsize]{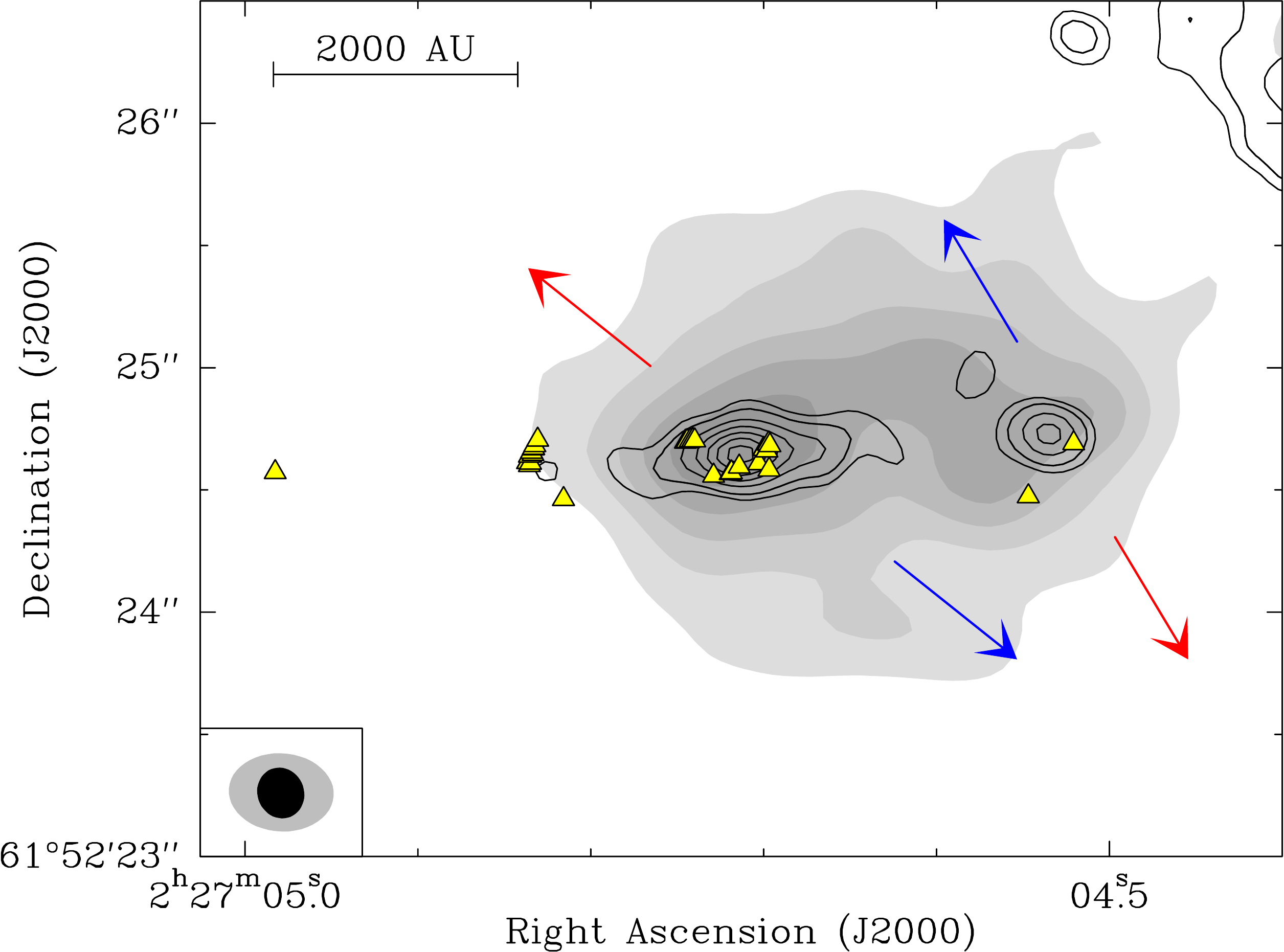}
      \caption{NOEMA 1.37~mm (219~GHz) continuum image toward \W\ in grey with levels starting at $6\sigma$ and increasing in steps of $6\sigma$. The black contours correspond to the cm emission from \citetads{1999ApJ...513..775W}. The positions of \water\ masers obtained from \citetads{2006ApJ...645..337H} are plotted as yellow triangles. The blue and red arrows show the directions of bipolar molecular outflows from \citetads{2011ApJ...740L..19Z} (see Fig.~\ref{f: w3h2o_outflows}). The synthesized beam size of the cm emission ($0\farcs21\times0\farcs19$, PA=68\degr) is shown in black in the bottom left corner. The synthesized beam size of our mm continuum image is shown in grey in the bottom left corner.}
         \label{f: w3_cm_masers}
   \end{figure}

In Fig.~\ref{f: w3_cm_masers}, we show how the cm emission aligns with the mm continuum emission. The directions of bipolar molecular outflows emanating from the cores are shown by red and blue arrows, and the positions of water masers shown by yellow triangles. The elongated radio source centered on \W~E has a spectral index of --0.6 and is a source of synchrotron emission, best described by a jet-like model due to its morphology (\citeads{1995ApJ...443..238R}; \citeads{1999ApJ...513..775W}). The radio source centered on \W~W, however, has a rising spectral index, possibly due to a circumstellar wind being ionised by an embedded (proto)stellar source at this position. Furthermore, the discrepancy between the direction of the synchrotron jet-like object and the molecular outflow could be due to the existence of multiple objects in this core, unresolved by our observations. Interestingly, it has been shown that synchrotron radiation can be produced not only via jets, but also through the acceleration of relativistic electrons in the interaction of disk material with a stellar wind \citepads{2004A&A...418.1045S}, providing an alternative explanation for the maser and cm emission.

\section{Analysis and Discussion \label{s: analysis_discussion}} 

\subsection{Dense gas kinematics \label{ss: kinematics}} 

The kinematics of \W\ can be further analyzed by looking at position-velocity (PV) diagrams for various transitions of dense gas and potentially disk-tracing molecules such as \mc\ and \mf\ \citepads[\eg]{2005ApJ...628..800B}. In the following, we divide our focus between the large-scale kinematics of the entire \W\ region where we put forward arguments for the observed velocity gradients in \mc\ and \mf\ being due to rotation instead of infall (Section~\ref{ss: large_scale_kinematics}), and the small-scale kinematics of the two separate cores within \W\ (Section~\ref{ss: small_scale_kinematics}). Although the alignment of the elongated radio emission with the water masers can be described by an outflow model oriented in the east-west direction, the detection of two molecular outflows emanating from the positions of the two continuum peaks, roughly perpendicular to the observed velocity gradient in \mc\ at large scale and perpendicular to the observed velocity gradients on smaller scales (see Section~\ref{ss: small_scale_kinematics}), make it unlikely that the motions in \mc\ would be due to expansion or outflow. Furthermore, \mc\ and \mf\ line profiles do not show the broad components typically seen in emission from expanding gas, and such species are nevertheless too complex to exist in an ionised jet. We therefore conclude that the observed velocity gradient is most likely to be due to rotation, which we will assume for the remainder of this paper.

   \begin{figure}
   \centering
   \includegraphics[width=\hsize]{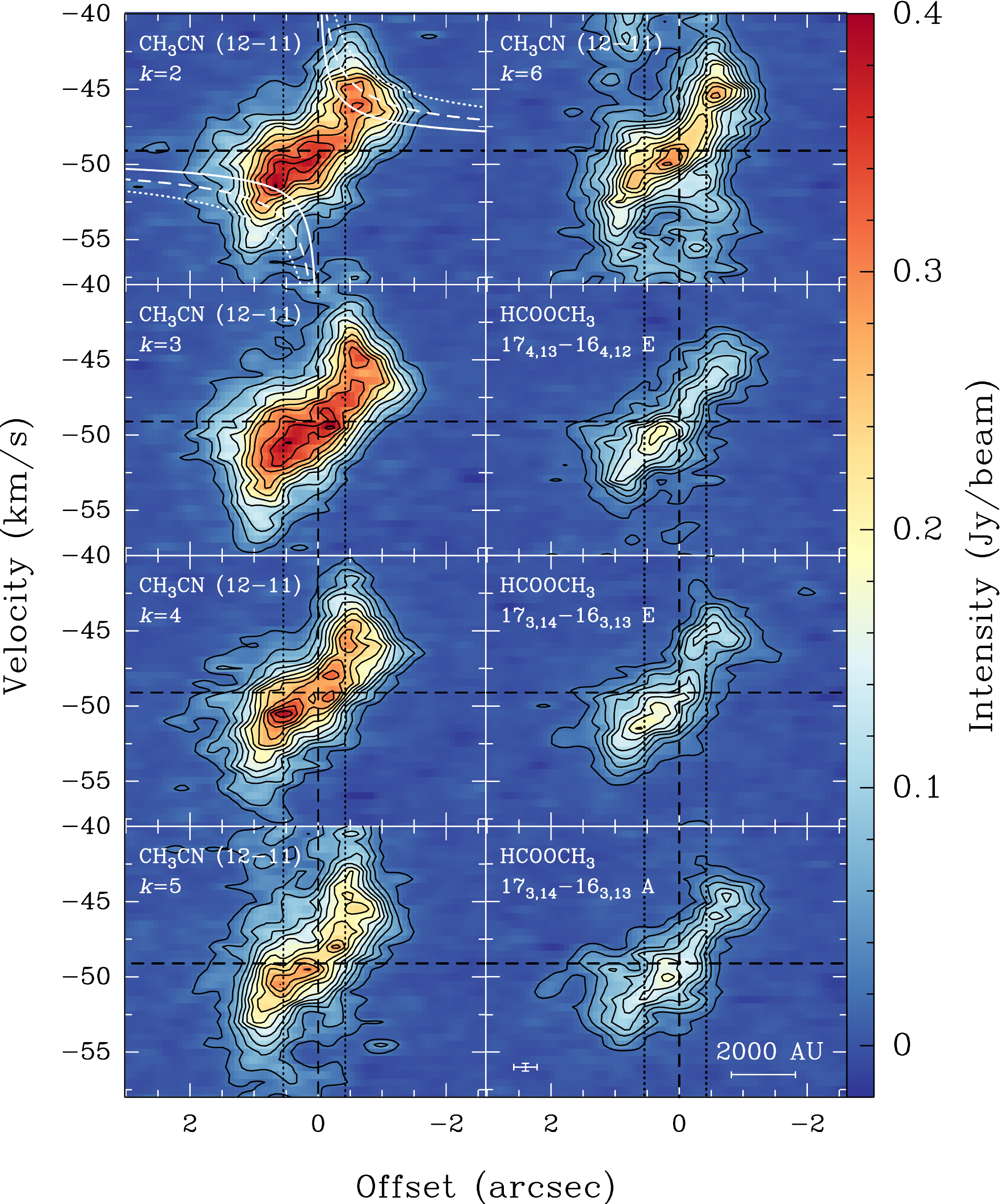}
      \caption{Position-velocity plots of \W\ for a cut in the direction of rotation as depicted by a dashed line in the bottom left panel of Fig.~\ref{f: w3_mom1_ABD} for various species and transitions in the ABD configuration. The vertical dashed lines correspond to the center of the cut. The vertical dotted lines correspond to the positions of continuum peaks corresponding to \W~E and \W~W. The horizontal dashed lines correspond to the LSR velocity of \W. The black contours start at $4 \sigma$ and increase in steps of $6 \sigma$. The white solid, dashed, and dotted lines in the top left panel correspond to the region within which emission is expected if the gas is in a disk in Keplerian rotation about a 10, 25, 50 \mo\ star, respectively. The white curves are not fits to the rotation curve, but are drawn to guide the eye. A scale-bar and a cross that corresponds to the spatial and spectral resolutions are shown in the bottom right panel.}
         \label{f: w3h2o_PV_ABD}
   \end{figure}

\subsubsection{Large-scale \label{ss: large_scale_kinematics}} 

The PV plots of \mc\ and \mf\ for \W\ are shown in Fig.~\ref{f: w3h2o_PV_ABD} for a cut in the direction of the velocity gradient going through the continuum peaks (dashed line in bottom left panel of Fig.~\ref{f: w3_mom1_ABD}) obtained from the NOEMA observations. White curves in the top left panel correspond to gas in Keplerian rotation with $V_\mathrm{rot} \propto R^{-1/2}$, about a 10, 25, and 50 \mo\ central object. These white curves are not fits to the PV diagram, but are merely drawn to guide the eye. It is clear that the gas is not in Keplerian rotation; however, higher velocity gas is observed closer to the center of \W\ which can be a signpost for differential rotation.

\citetads{1997ApJ...475..211O} created models for comparison to their interferometric data of the low-mass protostar L1527, using a thin disk with 2000~AU extent configured edge-on and present PV diagrams for cases with various degrees of infall and rotation. In the case of infall-only motions, their PV plots are axisymmetric with two peaks offset symmetrically in the velocity axis. With the addition of rotation, the peaks become blueshifted and redshifted away from the central positions such that in the case of a pure Keplerian rotation and in the absence of infall one would recover the classical butterfly-shaped rotation curve. Comparing our PV plots to the \citetads{1997ApJ...475..211O} scenarios, much of the emission that one would expect in the cases including infall motions would have to appear in the top-left and bottom-right quadrants of our plots, while we observe minimal contributions there. Therefore, we do not detect infall motions in our interferometric data, probably because the infalling envelope is too diffuse and filtered out. Furthermore, models by \citetads{2012ApJ...748...16T} for spherical rotating collapse and filamentary rotating collapse showed similar results to that of \citetads{1997ApJ...475..211O}, confirming that the absence of infall results in the lack of emission in those quadrants and not the projection or source morphology. Moreover, the PV plots of \mf, which is a less abundant species than \mc, are more representative of rigid-body-like rotation. 

\citetads{2006ApJ...639..975C} generated a binary model with radiative transfer post-processing of methyl cyanide for each source within \W, showing that the high-velocity deviations from solid body rotation in their PV plots could be a result of two spatially unresolved cores (on similar scales to our observations) with a small radial velocity difference. The detection of a smooth velocity gradient in all tracers across \W~E and W suggests that the observed angular velocity difference cannot be solely due to binary motion of the cores, but that there also exists a large-scale toroidal structure encompassing and rotating about the two resolved cores. Such circumbinary toroidal structures have indeed been observed previously at lower angular resolutions in other sources (\eg\ \citeads{2005A&A...435..901B}; \citeads{2007A&A...473..493B}).

\subsubsection{Small-scale \label{ss: small_scale_kinematics}} 

\begin{figure}
   \centering 
   \includegraphics[width=0.85\hsize]{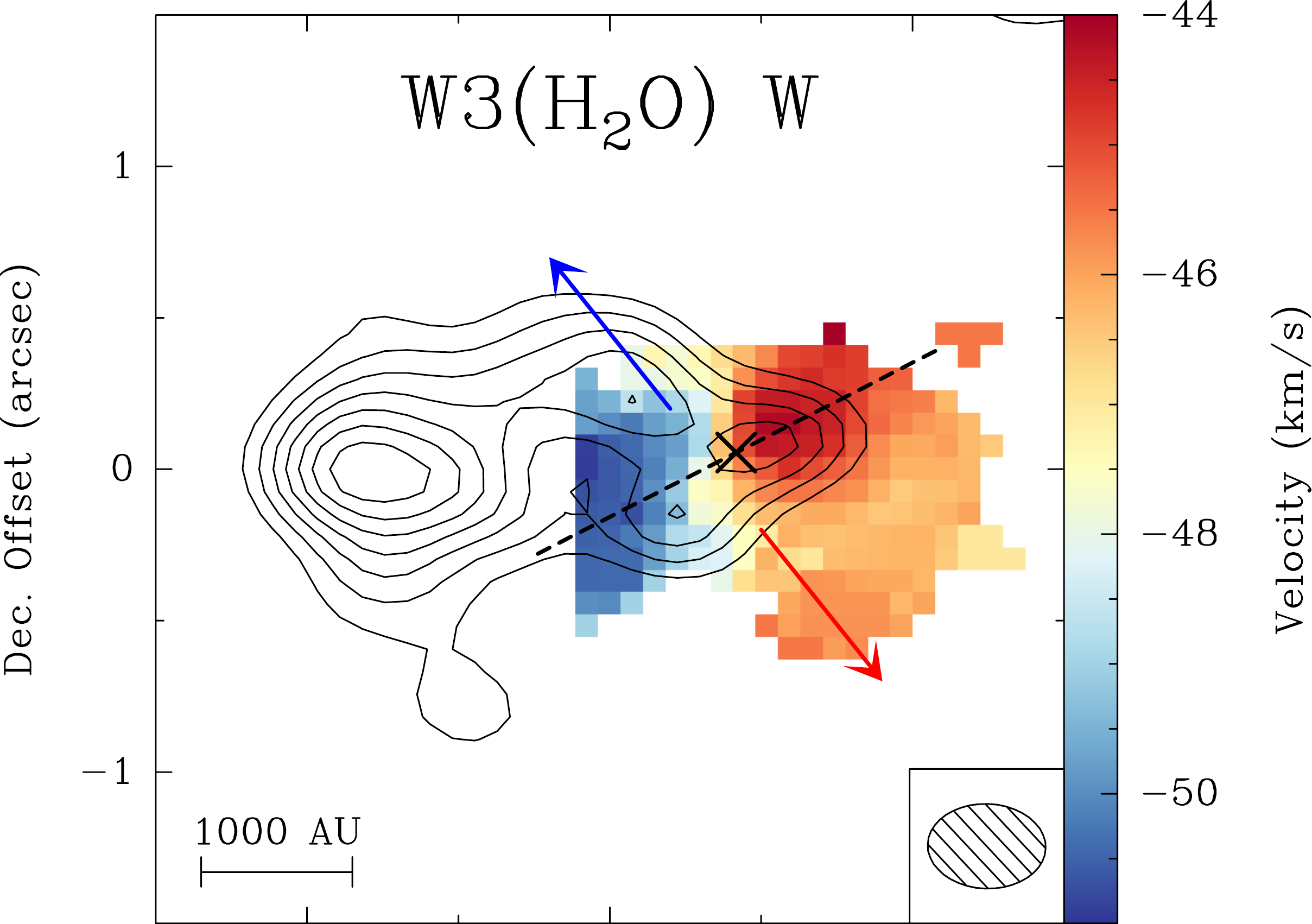} 
  
   \vspace{0.1cm}
   
   \includegraphics[width=0.85\hsize]{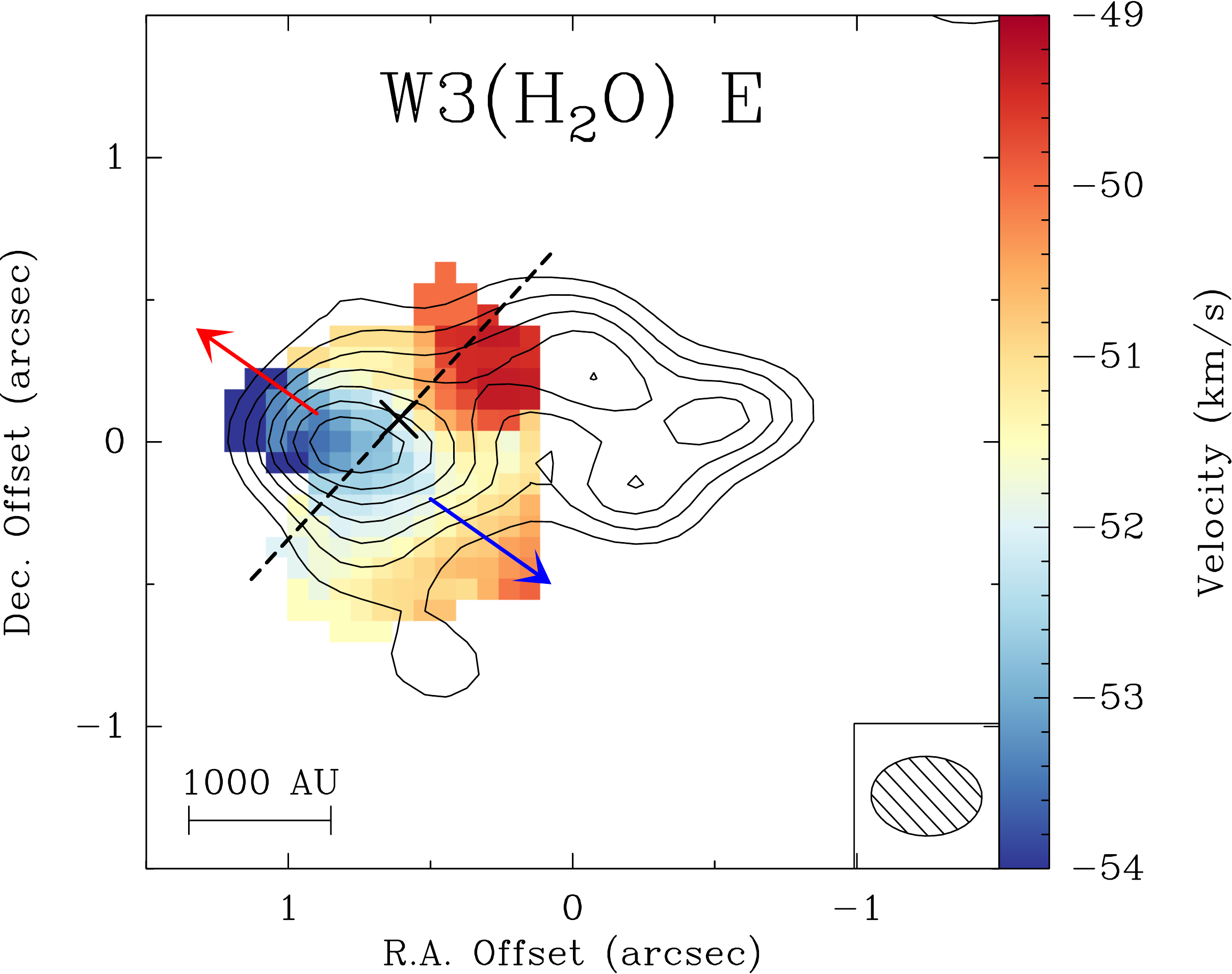}
       \caption{Intensity-weighted peak velocity (first moment) map of \mck{3} using only the A-array observations and masked out to show contributions from \W~W (\emph{top}) and \W~E (\emph{bottom}). The solid contours correspond to the 1.37~mm continuum in the A-array only observations and start at $6\sigma$ and increase in steps of $3\sigma$ ($1\sigma$ = 2.5~m\jpb). The dashed lines correspond to the cuts made for the PV plots (Fig.~\ref{f: w3h2o_PV_A_fragments}). The blue and red arrows show the directions of bipolar molecular outflows (Fig.~\ref{f: w3h2o_outflows}). A scale-bar and the synthesized beam ($0\farcs39\times0\farcs28$, PA=88\degr) are shown in the bottom. Note the different velocity ranges for the two cores.}
         \label{f: w3h2o_mom1_A_fragments}
   \end{figure}

Although we see a smooth velocity gradient in \mc\ across the entire \W\ structure, the existence of (at least) two cores within it as presented by our continuum results, and two collimated outflows requires further analysis. Imaging the data in the most extended configuration and therefore reaching our highest resolution, we can filter out the large-scale envelope to analyse the kinematics of gas around each of the cores. Figure~\ref{f: w3h2o_mom1_A_fragments} shows the first moment map of \mck{3} for the cores to the east and west using the A-array observations exclusively. The maps have been scaled and masked to highlight the velocity structure of each core. The mean line-of-sight velocity difference between the two main cores, \W~W and E, is a few \kms. Velocity differences of a few \kms\ are observed across each core, approximately perpendicular to the directions of the bipolar molecular outflows emanating from each core \citepads{2011ApJ...740L..19Z}. This indicates that the small-scale gradients across each individual core are most likely due to rotation. The two line-of-sight velocity gradients are on the order of $\sim$1000~\kms~$\mathrm{pc}^{-1}$, depending on the choice for the extent of the gradient, much faster than the rotational motion of 170~\kms~$\mathrm{pc}^{-1}$ for the whole of \W\ from the ABD data. Furthermore, the directions of the velocity gradients observed for each core are slightly inclined with respect to the overall east-west motion of the gas on larger scales (Fig.~\ref{f: w3_mom1_ABD}), suggesting that the general rotation seen around the two cores may be inherited from the large-scale rotation. Meanwhile, the direction of the blueshifted (redshifted) outflow emanating from \W~W is almost in the opposite direction of the blueshifted (redshifted) outflow ejected from \W~E. This implies that the inclination angles of the two rotating structures with respect to the plane of the sky are likely different. 

\begin{figure*}
   \centering
   \includegraphics[width=0.45\hsize]{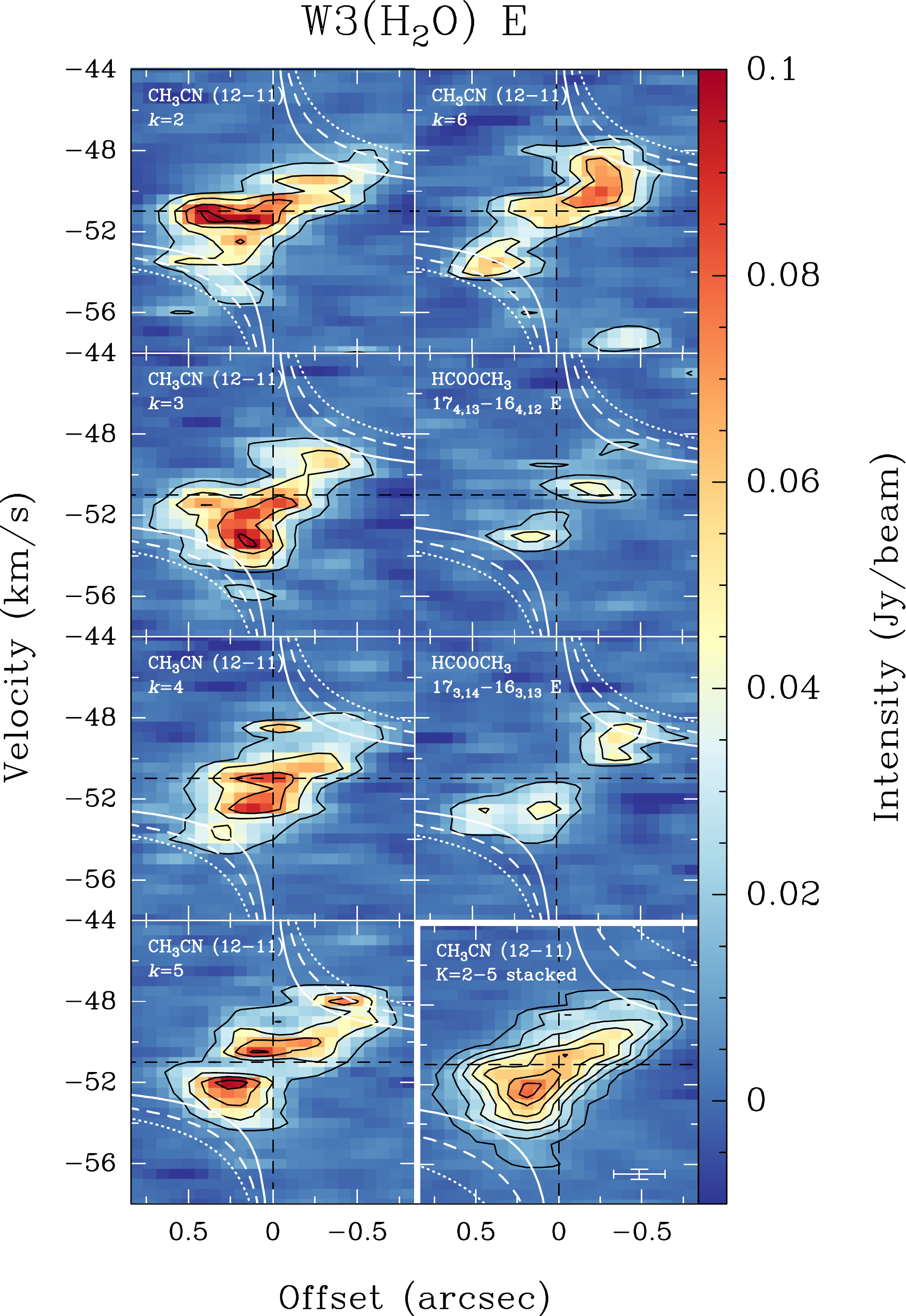} 
   \hspace{1cm}
   \includegraphics[width=0.45\hsize]{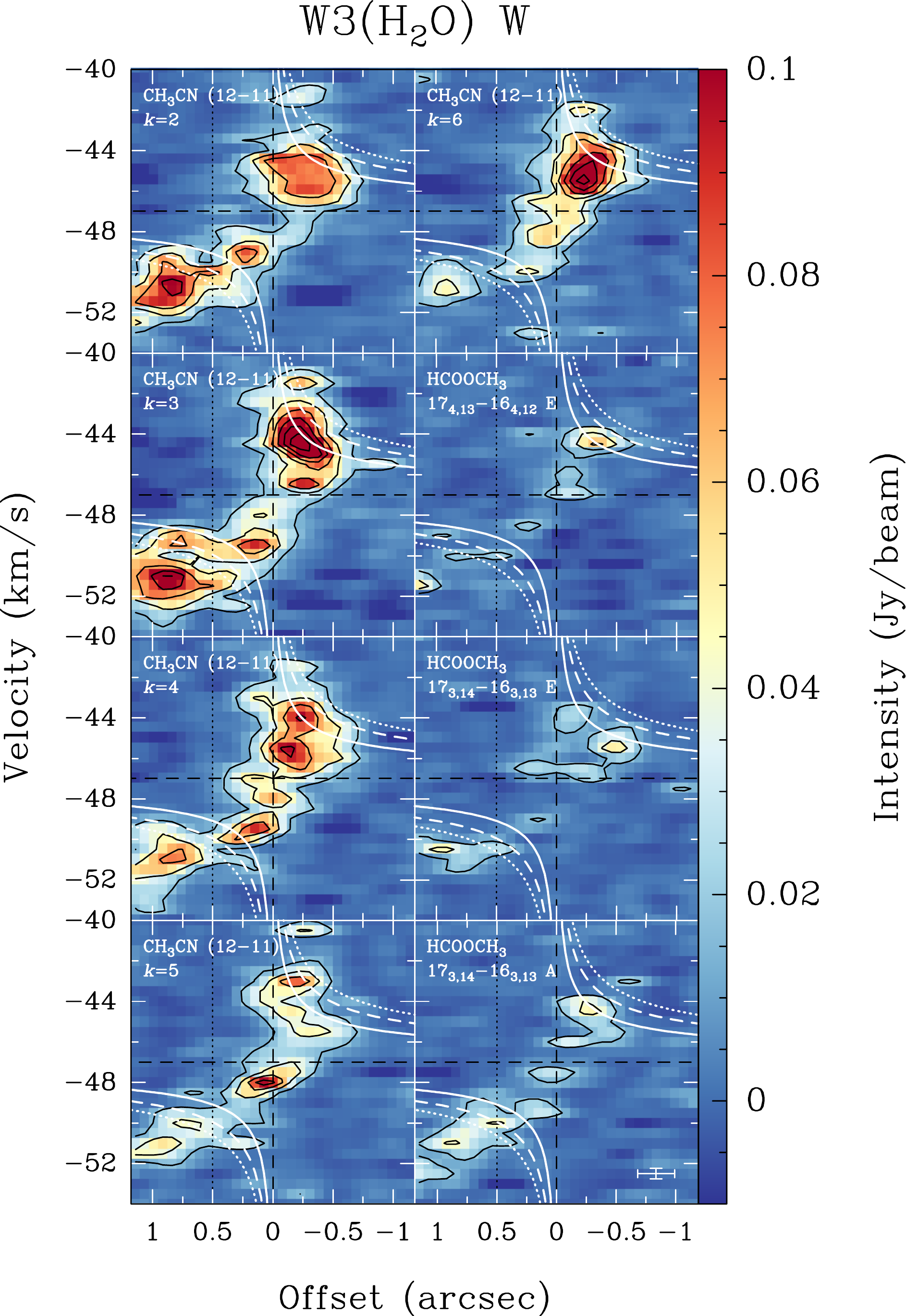}
   \caption{Position-velocity plots along a cut in the direction of rotation as depicted by dashed lines in Fig.~\ref{f: w3h2o_mom1_A_fragments} for fragment to the east (\emph{left}) and to the west (\emph{right}). The black contours start at $4 \sigma$ and increase in steps of $6 \sigma$. The white solid, dashed, and dotted lines correspond to the region within which emission is expected if the gas is in a disk in Keplerian rotation about a 5, 10, 15 \mo\ star, respectively. The crosses in the bottom right corners correspond to the spatial and spectral resolutions. Regions to the left of the dotted vertical line in the right figure contain contributions from \W~E.}
         \label{f: w3h2o_PV_A_fragments}
   \end{figure*}

Figure~\ref{f: w3h2o_PV_A_fragments} shows PV diagrams for \W~E (left) and \W~W (right) corresponding to cuts in the directions of rotation as depicted by dashed lines in Fig.~\ref{f: w3h2o_mom1_A_fragments}. Based on the PV plots, the Local Standard of Rest (LSR) velocities of \W~E and \W~W are estimated to be --51~\kms\ and --47~\kms, respectively. White curves correspond to gas in Keplerian rotation about a 5, 10, and 15 \mo\ centrally-dominated object. As on larger scales, these PV diagrams do not show the symmetric 4-quadrant shape expected if the velocity gradients were due to infall. 

The PV plot for \W~W contains contributions from \W~E due to the angle and extent of the cut, hence there exists added emission in quadrants toward positive offset, making it difficult to infer rotational signatures pertaining to the blueshifted emission of \W~W. The redshifted rotational signatures seen in the quadrants toward negative offsets however show signatures of increased gas velocities closer to the center of the core. Such a trend in the PV plot implies differential rotation of material, possibly within a disk-like object. 

PV plots for \W~E have a lower signal-to-noise ratio than \W~W as line emission is typically weaker for this fragment despite its 1.37~mm dust continuum peak being stronger. The linearity of the rotation curves in \mc\ does not reveal Keplerian signatures but is more consistent with rigid-body-like rotation. In order to increase the signal-to-noise ratio, we stacked the PV plots of \mckr{2}{5} and show this stacked PV plot in the bottom-right panel of Fig.~\ref{f: w3h2o_PV_A_fragments} for \W\ E. Stacking is a reasonable strategy as the variation in average linewidths of these transitions around this core is below our spectral resolution; therefore, assuming these lines to be probing the same surface is valid. In the stacked image, the 4$\sigma$ contour reveals a high-velocity feature close to the center of the core toward positive offsets. As this feature has an extent on the order of our spatial resolution, it is unclear whether the rotation observed for \W\ E is due to a disk-like object, an unresolved binary (or multiple) system, or a combination of the two.

\subsection{Temperature distribution}
As a symmetric top molecule, \mc\ is an excellent thermometer of hot molecular gas (\eg\ \citeads{1984ApJ...286..232L}; \citeads{1998ApJ...494..636Z}) since its relative populations in different $K$-levels are dominated by collisions. Our high spectral-resolution observations of \mc\ covers its $J=12-11$, $K=0-6$ transitions and some of their isotopologues which have upper energy levels in the range 70--325~K with 0.5~\kms\ spectral resolution (see Table~\ref{t:narrow_lines_info}). 

We made use of the eXtended \textsc{casa} Line Analysis Software Suite \citepads[\textsc{xclass\footnote{\url{https://xclass.astro.uni-koeln.de}}},][]{2017A&A...598A...7M} to model the observed spectra under the assumption of Local Thermodynamical Equilibrium (LTE) which is typically valid for \mc\ in such high-density environments (see Section~\ref{ss: mass_estimates}). In summary, \textsc{xclass} solves the radiative transfer equation in one dimension for a set of initial conditions (source size, column density, temperature, linewidth, and peak velocity) provided by the user, and through a $\chi^2$ minimisation routine changes the initial conditions within ranges that have been provided by the user to obtain the best fit to the observed spectra. The details of our \textsc{xclass} modelling are summarised in Appendix~\ref{a: xclass}.

In Fig.~\ref{f: XCLASS_output_maps}, we present our results of pixel-by-pixel \textsc{xclass} modelling of \mckr{4}{6}, including \mcisokr{0}{3}, in AB configuration which produces rotational temperature, column density, peak velocity, and linewidth maps for \mc. The column density map is doubly peaked, similar to the continuum emission, although the column density peaks are slightly offset to the northwest by a synthesized beam. This offset is consistent with the offset found between the continuum peaks and the integrated intensity maps of \mck{3} (see Fig.~\ref{f: w3_mom1_ABD}) and most high-density tracers (see Fig.~\ref{af: all_narrow_moment0} in Appendix~\ref{a: moment_maps_ABD}). The median \mc\ column density is $1.4\times10^{15}~\mathrm{cm^{-2}}$. The velocity gradient derived in this way is consistent with the large-scale east-west velocity gradient observed in the first moment map of \mc, confirming that our modelling strategy captures this accurately, regardless of its origin. The linewidths are also larger by a few \kms\ for \W~W, also seen in the second moment maps. The median rotational temperature of \W\ is $\sim$165~K and the temperature structure does not follow any particular pattern within either core. Two high-temperature features can be seen which may be associated with regions carved out by the molecular outflows allowing a deeper look into the cores, or regions that may have been additionally heated by the outflow. Nevertheless, the compactness of these features, which are on the same order as the size of our synthesized beam, along with decreased signal-to-noise at the edges of the map prevent us from making firm conclusions in this regard. 

   \begin{figure*}
   \centering
   \includegraphics[width=\hsize]{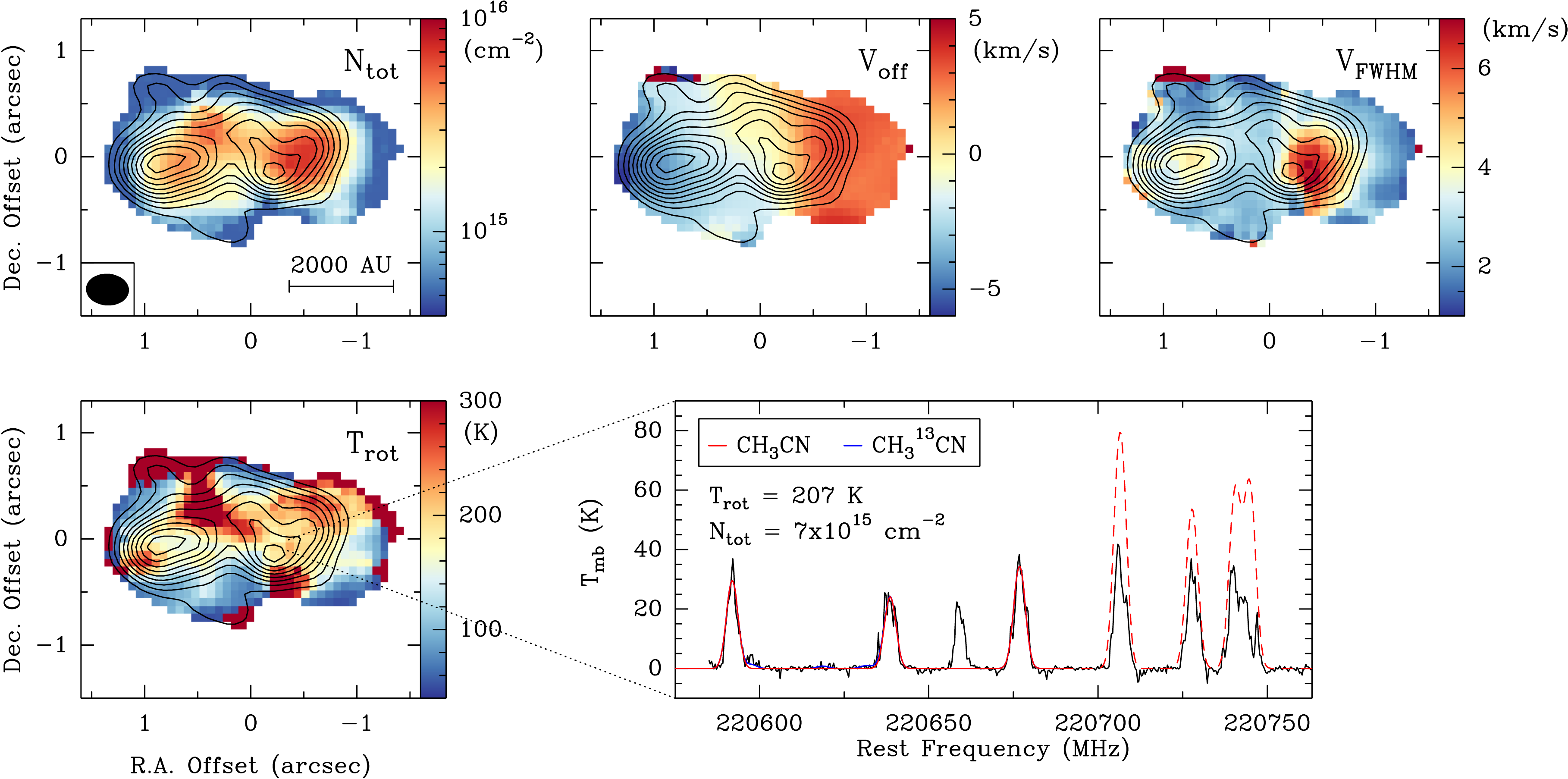}
      \caption{Column density (\emph{top left}), velocity offset (\emph{top middle}), full width at half maximum linewidth (\emph{top right}), and rotational temperature (\emph{bottom left}) maps obtained by fitting \mckr{4}{6}\ and \mcisokr{0}{3}\ lines simultaneously with \textsc{xclass}. The black contours correspond to the continuum image in the AB configuration, start at $6\sigma$ and increase in steps of $3\sigma$ (see Table~\ref{t: config_specs}). \emph{Bottom right:} observed spectrum of a given pixel drawn in black and overlaid with the resulting fit for \mckr{4}{6} in red and the marginally detected \mcisokr{0}{3} in blue. The dashed red line corresponds to the predicted fit for the \mc\ lines that were not used in the fitting process (see explanation in Appendix~\ref{a: xclass}). The bright line detected between $K=4$ and 5 components is identified as $\mathrm{C_2H_5CN}$. The corresponding fit parameters are provided in the panel. Regions outside of the most extended combination of 6$\sigma$ contours of integrated intensity of \mc\ lines are masked out.}
         \label{f: XCLASS_output_maps}
   \end{figure*}

\subsection{Mass estimates \label{ss: mass_estimates}} 

The combined bolometric luminosity of \W\ and \WOH, determined from fitting the SED from the near-IR to sub-mm, is $8.3\times10^{4}$\,\lo\ \citepads{2011A&A...525A.149M}. The contribution from OH can be estimated by first calculating the corresponding flux of ionising photons \citepads[see, e.g., Appendix B.2 of][for more details]{Sanchez-MongeThesis} using:

\begin{equation}
\left(\frac{Q_{0}}{\mathrm{photons\,s^{-1}}}\right)=8.852\times10^{40}\left(\frac{F_{\nu}}{\mathrm{Jy}}\right) \left (\frac{\nu}{\mathrm{GHz}}\right)^{0.1} \left (\frac{T_{\mathrm{e}}}{\mathrm{K}}\right)^{0.35} \left (\frac{d}{\mathrm{pc}}\right)^{2},
\end{equation}

\noindent where $F_{\nu}$ is the flux density of the free-free radio continuum emission at frequency $\nu$, $T_{\mathrm{e}}$ is the electron temperature and $d$ is the distance. The observed integrated radio flux at 15\,GHz for \WOH\ is 2.53\,Jy \citepads{1994ApJS...91..659K}, and assuming a typical electron temperature of 10$^{4}$\,K, this results in a value of $Q_{0}$ = 1.2$\times10^{48}$\,photons\,s$^{-1}$. Interpolating from Table 1 of \citetads{2011MNRAS.416..972D} and the relationships between spectral type and photospheric temperature from \citetads{2005A&A...436.1049M} and \citetads{1981ARA&A..19..295B} for O and B stars respectively, this ionising photon flux approximately corresponds to an O8.5 main-sequence star, with $M \approx 20$\,\mo\ and $L \approx 4.4 \times 10^{4}$\,\lo. This leaves a total bolometric luminosity of $3.9\times10^{4}$\,\lo\ which we assume to be evenly distributed between the two cores within \W, $1.95\times10^{4}$\,\lo\ each, which using the same tables would correspond to two 15\,\mo\ stars of spectral type B0. 

The above estimates are based on \citetads{2011MNRAS.416..972D} for zero-age main sequence (ZAMS) stars. High-mass protostars growing by accretion resemble ZAMS stars in terms of their luminosities and temperatures when core nuclear burning dominates other sources of luminosity such as accretion and envelope burning. When this occurs depends primarily on when the protostar gains sufficient mass but also on the accretion rate. Stellar structure calculations suggest this occurs at 5 to 10~\mo\ for accretion rates of $10^{-5}$ to $10^{-4}$~\mo\ $\mathrm{yr^{-1}}$, respectively (\citeads{2000A&A...359.1025N}; \citeads{2001A&A...373..190B}; \citeads{2006ApJ...637..850K}). The mass estimates further assume that all the emitted energy has been produced within the (proto)stars, ignoring contributions from episodic accretion to the luminosity. Therefore, the 15\,\mo\ estimates can be taken as upper limits, in agreement with calculations of \citetads{2006ApJ...639..975C} who find a minimum binary mass of 22~\mo\ for the protostars within \W. Furthermore, assuming the gas to be in gravito-centrifugal rotation around the two individual cores, our PV plots (see Fig.~\ref{f: w3h2o_PV_A_fragments}) suggest the range of $5-15$~\mo\ to be a reasonable estimate for the protostellar masses.

As dust is typically optically thin in the 1.3~mm wavelength regime and proven to be for this region in particular \citepads{2006ApJ...639..975C}, we use the prescription by \citetads{1983QJRAS..24..267H} to convert the flux density, $F_\nu$, of the continuum observations to a mass. In the form presented by \citetads{2009A&A...504..415S}, 
\begin{equation} \label{e: mass}
  M=\frac{d^2\,F_\nu\,R}{B_\nu (T_D)\,\kappa_{\nu}},
\end{equation}
where $R$ is the gas-to-dust mass ratio, $B_\nu (T_D)$ is the Planck function at a dust temperature of $T_D$, and $\kappa_\nu$ is the dust absorption coefficient. We adopt a gas-to-dust mass ratio of $R=150$ \citepads{2011piim.book.....D} and a value of $\kappa_\nu=0.9$~$\mathrm{cm^2\,g^{-1}}$ for the dust absorption coefficient from \citetads{1994A&A...291..943O}, corresponding to thin ice mantles after $10^5$ years of coagulation at a density of $10^6$~cm$^{-3}$. 

High-mass cores such as the ones we are studying typically have densities high enough to thermalize the methyl cyanide lines. To check this, using the spontaneous decay rate of \mck{4} obtained from the LAMDA database\footnote{\url{http://home.strw.leidenuniv.nl/~moldata/}}, $7.65\times10^{-4}\,\mathrm{s^{-1}}$, and the collision rate of $2.05\times10^{-10}\,\mathrm{cm^3\,s^{-1}}$ with H$_2$ at 140~K \citepads{1986ApJ...309..331G}, we calculate the simplified 2-level critical density of this line to be $n_\mathrm{crit} \approx 3.7\times10^6\,\mathrm{cm^{-3}}$. The effective density, once the radiation field is taken into account, is typically an order of magnitude lower (\citeads{1999ARA&A..37..311E}; \citeads{2015PASP..127..299S}). Following \citetads{2009A&A...504..415S}, the H$_2$ column density is calculated via
\begin{equation} \label{e: N_H2}
  N_\mathrm{H_2}=\frac{S_\nu\,R}{B_\nu(T_D)\,\theta_B\,\kappa_\nu\,\mu\,m_\mathrm{H}},
\end{equation}
where $S_\nu$ is the peak intensity, $\theta_B$ is the beam solid angle, $\mu$ is the mean molecular weight assumed to be equal to 2.8, and $m_\mathrm{H}$ is the mass of the hydrogen atom. At a temperature of 140~K and using our continuum intensity of a given position near the center, we can estimate the H$_2$ column density to be $4.5 \times 10^{24}$\,cm$^{-2}$. This can be converted to a volume density of $n_\mathrm{H_2}>7\times10^7\,\mathrm{cm}^{-3}$, assuming the extent of the third dimension to be at maximum the plane-of-sky size of the clump ($\sim$4000~AU). Therefore, since the density of molecular hydrogen is much higher than the critical density of the lines, the LTE assumption is valid and the rotational temperature map of \mc\ obtained from our modelling can be assumed to be tracing the gas kinetic temperature. 

Using Eq.~\ref{e: mass}, our continuum map with its unit converted to \jpp\ and the temperature map obtained from \textsc{xclass}, assuming dust and gas temperatures are coupled, we obtain a pixel-by-pixel mass density map (see Fig.~\ref{af: mass_map_AB} in Appendix~\ref{a: mass_map}). Summing over the pixels in our mass density map in the ABD observations, the total mass for \W\ is calculated to be $\sim$26.8~\mo, with 15.4~\mo\ contributed from the core to the east, and 11.4~\mo\ from the core to the west. Similarly, using the AB observations, we obtain a total mass of $\sim$11.4~\mo\ for \W, with a core mass of 6.7~\mo\ and 4.7~\mo\ from the cores to the east and west, respectively. The effect of spatial filtering of the interferometer is clear in these mass estimates as the exclusion of the D-array data removes more than half of the mass. 

Comparing our NOEMA observations to SCUBA-2 850~$\mu$m single-dish observations, about 25\% of the flux is filtered out by the interferometer in our ABD observations (assuming a $\nu^{-3.5}$ frequency-relation, \citeads{2018arXiv180501191B}), implying that our core mass estimates of 15.4 and 11.4~\mo\ are lower limits, but nevertheless reasonable. Masses estimated in this manner have contributions from the cores and the disk-like structures and not from any embedded (proto)stars. 

\subsection{Toomre stability}
For a differentially rotating disk, the shear force and gas pressure can provide added stability against gravitational collapse. This idea was originally introduced by \citetads{1960AnAp...23..979S} and further quantified by \citetads{1964ApJ...139.1217T} for a disk of stars, and has since been used in various applications ranging from planet formation to galaxy dynamics. We investigate the stability of the rotating structures in \W, assuming that they are disks in gravito-centrifugal equilibrium, against axisymmetric instabilities using the \tq\ parameter,
\begin{equation} \label{e: Toomre}
  Q=\frac{c_s\,\Omega}{\pi\,G\,\Sigma},
\end{equation}
where $c_s$ is the sound speed, and $\Omega$ is the epicyclic frequency of the disk which is equivalent to its angular velocity. The surface density of the disk, $\Sigma$, is calculated by multiplying the column density (Eq.~\ref{e: N_H2}) by the mean molecular weight and mass of the hydrogen atom ($\mu m_\mathrm{H}$) to convert the number column density to a mass surface density. A thin disk becomes unstable against axisymmetric gravitational instabilities if $Q < 1$.

Having obtained a temperature map representative of the kinetic temperature, we are able to calculate the \tq\ parameter pixel-by-pixel. In particular, the temperature is used in the calculation of $B_\nu(T_D)$ and the sound speed,
\begin{equation}
c_s=\sqrt{\frac{\gamma k_\mathrm{B} T}{\mu m_\mathrm{H}}},
\end{equation}
where $\gamma$ is the adiabatic index with a value of $5/3$.

We assume two 10~\mo\ (proto)stellar objects to be present, one at the position of each of the two continuum peaks (see AB image in Fig.~\ref{f: w3h2o_cont_allconfig}). The angular velocity maps are created by adding up mass within concentric circles starting at the position of each core and going outwards. In this way, we incorporate the mass of the rotating structures, calculated via Eq.~\ref{e: mass}, and the mass of the central objects. The angular velocity of a disk in gravito-centrifugal equilibrium at a radius $r$ is
\begin{equation} \label{e: omega}
\Omega(r)=\sqrt{\frac{G(M_\mathrm{\ast}+M_\mathrm{disk}(r))}{r^3}},
\end{equation}
where the mass of the central object, $M_\mathrm{\ast}$, is 10~\mo, and $M_\mathrm{disk}(r)$ is the gas mass enclosed within $r$. Given the radii involved, this means that in practice, most parts of the maps are dominated by $M_\mathrm{disk}$ rather than $M_\mathrm{\ast}$. 

   \begin{figure}
   \centering
   \includegraphics[width=\hsize]{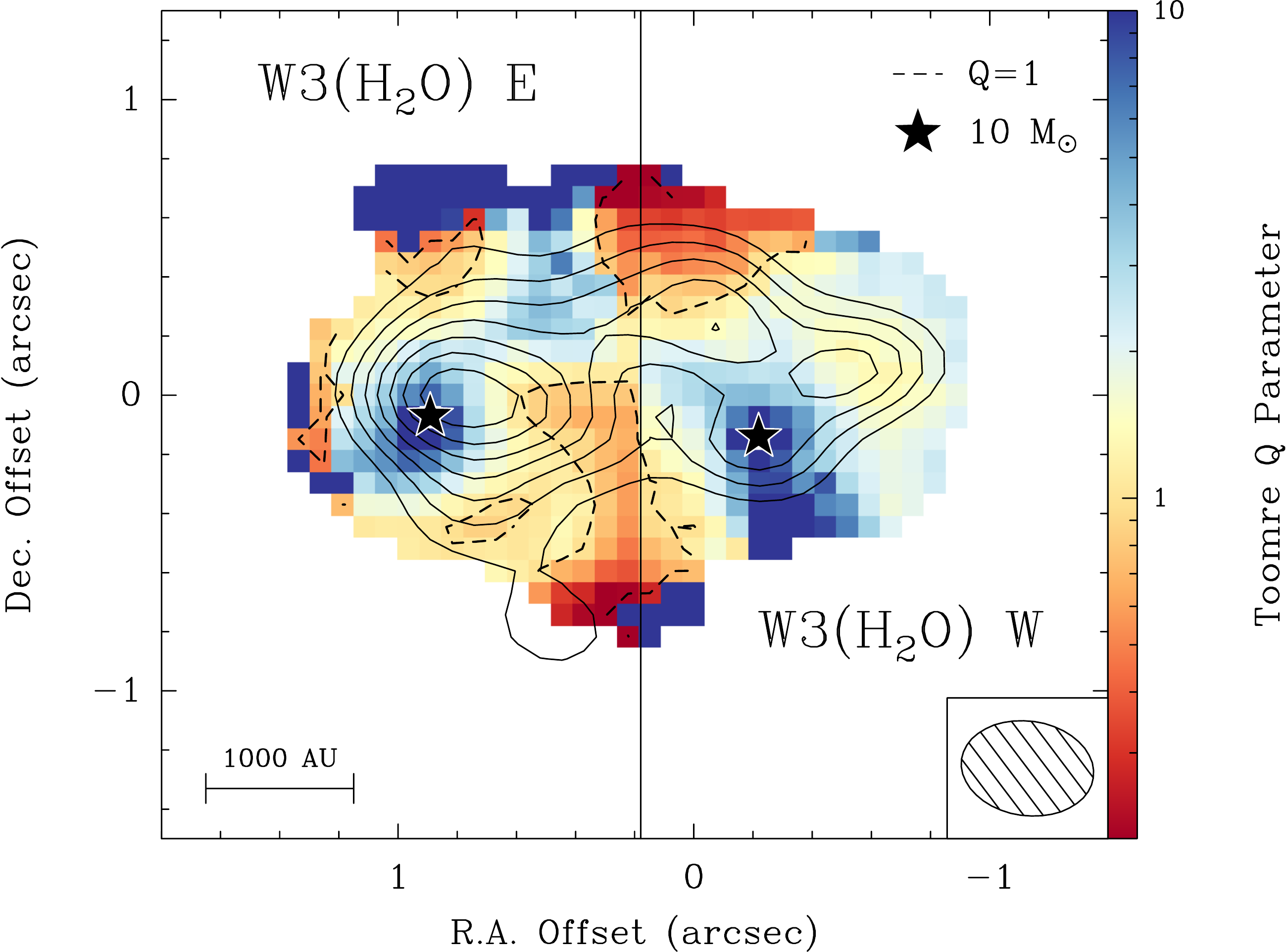}
      \caption{\tq\ map obtained by assuming two disk-like structures in gravito-centrifugal rotation about the positions of peak continuum emission as depicted by the two stars, each with a mass of 10~\mo. The \tq\ calculations and the positions of (proto)stars are based on the AB-array data (see Fig.~\ref{f: w3h2o_cont_allconfig}) with regions outside of the 6$\sigma$ mm continuum emission masked out. Solid contours correspond to our continuum data in the most extended (A-array) configuration, starting at 6$\sigma$ and increasing in steps of 3$\sigma$ ($1\sigma$ = 2.5~m\jpb). The solid vertical line corresponds to the stitching boundary. The dashed lines correspond to $Q=1$.}
         \label{f: ToomreQ_AB}
   \end{figure}

In Fig.~\ref{f: ToomreQ_AB} we present our \tq\ map of \W, which is created by stitching together the \tq\ maps of the two individual cores. The stitching boundary is shown by a solid vertical line and the positions of the two central objects corresponding to our continuum peaks in the AB image (see Fig.~\ref{f: w3h2o_cont_allconfig}) are depicted by stars. While the \tq\ calculations were performed on the AB-array data, we have drawn the continuum contours from the A-array only observations. The boundary where $Q=1$ is shown by a dashed line.

The most significant factor stabilizing the disk against Toomre instability is the high gas temperatures and the fast differential rotation of material closest to the (proto)star, therefore, we find the highest \tq\ values closest to the presumed locations of the (proto)stars depicted by stars in Fig.~\ref{f: ToomreQ_AB} as expected. We find low \tq\ values at the periphery of \W~E and W, suggesting that both rotating structures are only marginally stable against axisymmetric instabilities. Moreover, \W~W shows low \tq\ values at the positions of additional continuum peaks in the highest resolution A-array data (solid contours in Fig.~\ref{f: ToomreQ_AB}), suggesting further fragmentation at these positions may be possible. As the dimensions of the observed continuum peaks are smaller than the size of our synthesized beam, the validity of this hypothesis needs to be confirmed with higher resolution observations (at 345~GHz) to be taken with NOEMA.

A Toomre stability analysis done by \citetads{2016ApJ...823..125C} for the high-mass protostar IRAS 20126+4104 showed that the massive disk around that protostar is hot and stable to fragmentation, with \tq\ $>2.8$ at all radii. Hydrodynamic simulations of collapsing protostellar cores by \citetads{2016ApJ...823...28K} find massive accretion disks which evolve to become asymmetrically Toomre-unstable, creating large spiral arms with an increased rate of accretion of material onto the central object. They find that the self-gravity of these spiral arm segments is lower than the tidal forces from the star, causing the spiral arms in their simulations to remain stable against collapse. However, they find that the Toomre conditions combined with the cooling of the disk \citepads{2001ApJ...553..174G} would potentially yield the formation of a binary companion. Similarly, simulations by \citetads{2011ApJ...732...20K} see such spiral arms in a massive disk, and simulations by \citetads{2009Sci...323..754K} and more recently Meyer et al.~(\citeyearads{2017MNRAS.464L..90M}, \citeyearads{2018MNRAS.473.3615M}) see disk fragmentation on even smaller spatial scales, on the order of hundreds of AU. Therefore, disk fragmentation scenarios in the high-mass regime are valid from a theoretical stand-point. 

\paragraph{Uncertainties:} 

\begin{itemize}
\item
Among the assumptions that go into this Toomre analysis is the thin-disk approximation. According to \citetads{2001ApJ...553..174G}, the critical value of $Q$ for an isothermal disk of finite thickness should be 0.676 (as derived from \citeads{1965MNRAS.130..125G}). Even with such a lower critical value, the rotating structures around each core would still show signs of instability in their outskirts. To get an estimate on the importance of disk thickness, following \citetads{2001ApJ...553..174G}, the instability condition for a disk with scale height, $H\simeq c_s/\Omega$, can be written as
\begin{equation}
  M_\mathrm{disk}\gtrsim \frac{H}{r}M_\ast.
\end{equation}
Using our temperature and angular velocity maps, we create a map of the $H/r$ ratio. In this way, we calculate an average value of  $H/r \approx 0.3$. Assuming the mass of each (proto)star is roughly 10~\mo, we calculate that for the candidate disks to be unstable against gravitational collapse, the mass of the candidate disk must be $M_\mathrm{disk}\gtrsim 3$~\mo. Since the mass estimates for the cores obtained from the continuum maps using Eq.~\ref{e: mass} are 15 and 11~\mo, the masses are large enough to result in the rotating structures being Toomre-unstable. Therefore, the thickness of the candidate disks does not affect the general interpretation of our results. 
\item
The angular velocity maps used in the calculation of the \tq\ map have been created assuming a face-on geometry for the candidate disks. If the objects are inclined, we would be overestimating the surface density of the candidate disks and the line-of-sight angular velocities. Since these two parameters have the opposite effect on the \tq~value (see Eq.~\ref{e: Toomre}), our results are still reasonable. 
\item 
\W~E may harbour multiple (proto)stars within it (see Section~\ref{ss: kinematics}) such that the possibility of rotational contributions from multi-body rotation would require a different prescription for the epicyclic frequency than what we have assumed.
\item
The estimation of 10~\mo\ for the central (proto)stars have been deduced from our own and previous mass estimations (see Section~\ref{ss: mass_estimates}), as well as examining the emission in the PV plots with Keplerian curves that are just `grazing' the emission. The latter method has the caveat that the observed PV plot is the result of the convolution with the
instrumental beam and the line width. Therefore, the real emission is less extended in space and velocity than what we observe and as a result we may be overestimating the mass of the (proto)stars. Performing our Toomre calculations but with the assumption of 5~\mo\ (proto)stars, we obtain median \tq\ values of 1.3 as compared to a value of 1.7 in the 10~\mo\ calculations. Therefore, lower protostellar masses would lead to the decrease in the \tq\ value, and thus larger unstable regions, making the disk fragmentation scenario even more plausible. On the other hand, as the mass estimates using the bolometric luminosities suggest the central (proto)stellar masses to be in the 15~\mo\ regime, the median \tq\ value would be 2.0, with unstable regions still present in the outskirts. The \tq\ maps created assuming two 5 and 15~\mo\ (proto) stars at the position of the continuum peaks can be found in Appendix~\ref{a: toomre_maps}. The change of $\pm$5~\mo\ does not strongly affect the \tq\ maps because we add the mass of the candidate disks, estimated from the continuum emission to be 15 and 11~\mo, in rings (see Eq.~\ref{e: omega}) and they dominate over an addition or subtraction of 5~\mo, especially in the outermost regions. This, together with the root dependence on mass lead to uncertainties in $M_\mathrm{\ast}$ to only have a small effect on the \tq\ results.
\end{itemize}
We are currently conducting simulated observations from 3D hydrodynamic simulations to investigate the reliability of our methods in even more depth (Ahmadi et al. in prep.). Our preliminary results on the effect of inclination and (proto)stellar mass on the calculation of \tq\ maps using hydrodynamic simulations suggest that the \tq\ value is not strongly dependent on these parameters. A forthcoming paper is dedicated to this topic. 

\subsection{Cooling}

For a marginally stable disk ($Q \sim 1$), the locally unstable regions compress as they start to collapse, providing compressional heating in these regions. This increase in the local thermal pressure can counteract the local collapse and put the disk back into a state of equilibrium such that the region will not fragment unless heat can be dissipated on a short timescale. Through numerical simulations, \citetads{2001ApJ...553..174G} and \citetads{2003ApJ...597..131J} have introduced a cooling parameter, $\beta$, to study the effect of cooling for marginally stable thin disks in gravito-centrifugal equilibrium. In their prescription,
\begin{equation}
  \beta = \Omega \,t_\mathrm{cooling},
\end{equation}
where the cooling time, $t_\mathrm{cooling}$, is assumed to be constant and a function of the surface density and internal energy of the disk. If the local collapse needs more than a few orbits, due to a long cooling timescale, a locally collapsing region within the disk can be ripped apart by shear. Therefore, there exists a critical $\beta$ value below which a locally collapsing region would be rapidly cooling, and if it is sufficiently self-gravitating, it would be prone to fragmentation. Conversely, values above this critical $\beta$ value would put a marginally stable disk into a self-regulating scenario such that heating acts as a stabilizing force and directly counteracts the cooling rate. 

The Gammie cooling criterion is relevant only if the disk is marginally stable, with regions that are locally unstable against axisymmetric gravitational instabilities. A Toomre-stable disk will not fragment regardless of how quickly it may be cooling.

This critical value is determined through numerical simulations and varies depending on the heating and cooling recipes and convergence issues. For an overview of these issues, see \citetads{2016ARA&A..54..271K}, and more recently \citetads{2017ApJ...848...40B}. Regardless of the specific numerical simulation issues, it is generally assumed that the critical $\beta$ value ranges between $1-5$.

Following the prescription of \citetads{2016ApJ...823...28K},
\begin{equation}
  t_\mathrm{cooling} = \frac{E_z}{F_\mathrm{grey}},
\end{equation}
where $E_z$ is the column internal energy integrated in the $z$-direction (along the axis of rotation), and $F_\mathrm{grey}$ is the radiative flux away from the disk adopting a greybody approximation. The internal energy of an ideal gas is calculated via
\begin{equation}
  E_{int}= c_v \, \rho \, T,
\end{equation}
where $\rho$ is the volume density and $c_v$  is the specific heat capacity defined as
\begin{equation}
  c_v=\frac{N_A k_B}{M_u(\gamma-1)},
\end{equation}
  with $N_A$ as Avogadro's number and $M_u$ the molar mass. By replacing the volume density with the surface density, the column internal energy is then
\begin{equation}
  E_z=\frac{N_A  k_B \Sigma T  }{3M_u(\gamma-1)}.
\end{equation}

\begin{figure}
   \centering
   \includegraphics[width=\hsize]{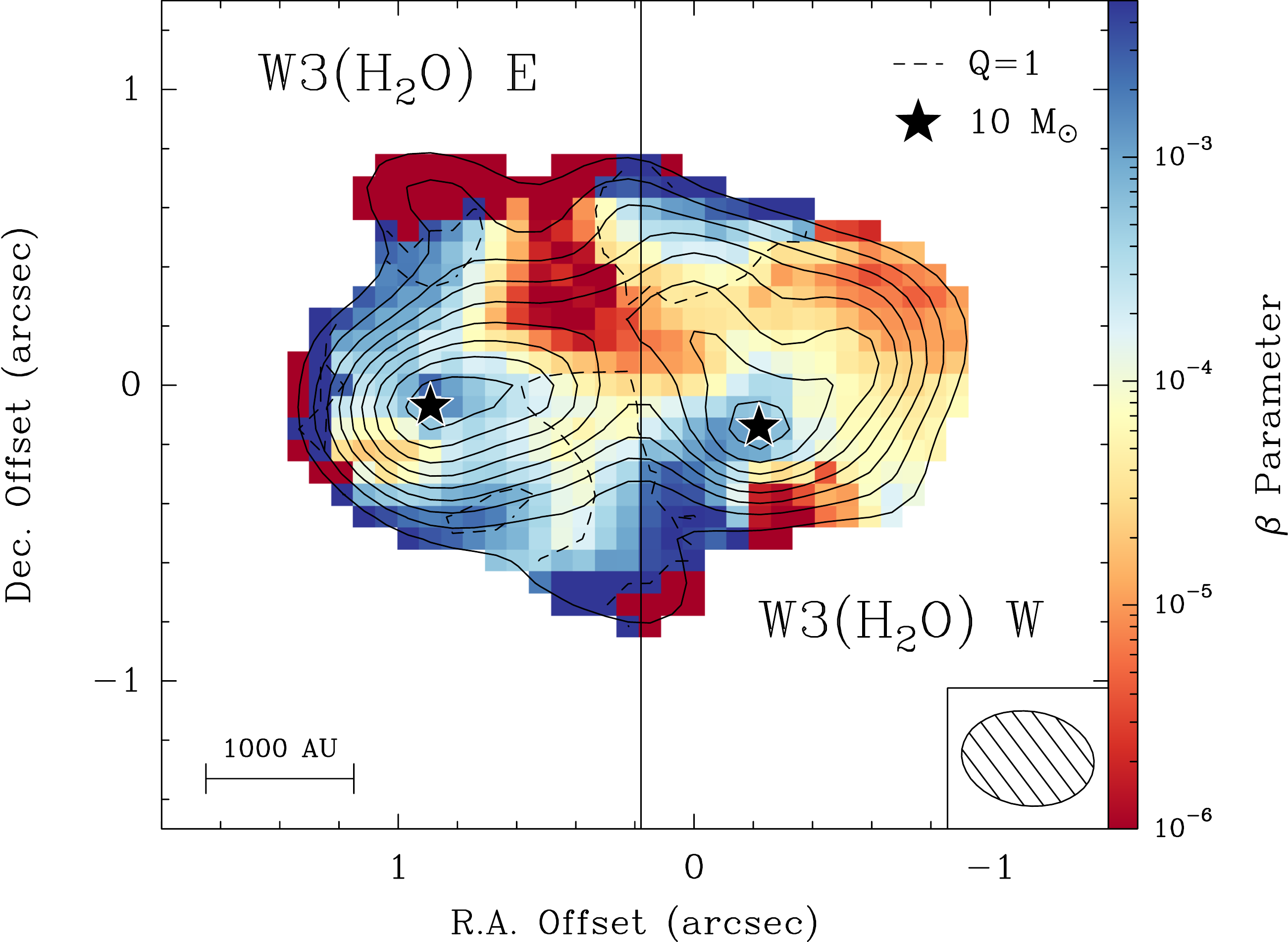}
      \caption{Map of $\beta$ cooling parameter obtained by assuming two disk-like structures in gravito-centrifugal rotation about the positions of peak continuum emission as depicted by stars. The solid contours correspond to our continuum data in the most extended configuration, starting at 6$\sigma$ and increasing in steps of 3$\sigma$ ($1\sigma$ = 2.5~m\jpb). The solid vertical line corresponds to the stitching boundary. The dashed line corresponds to Toomre $Q=1$. Regions outside of the 6$\sigma$ mm continuum emission contour in the AB configuration are masked out.}
         \label{f: beta_AB}
   \end{figure}

Knowing the cooling time, which is on the order of months, and the rotation rate at each pixel, we construct a local $\beta$ map. Fig.~\ref{f: beta_AB} shows the $\beta$ map for the two disks presumed in gravito-centrifugal equilibrium about the positions of peak continuum emission as depicted by stars. The median $\beta$ value is $1.8\times10^{-4}$, and the maximum value found in the outskirts is on the order of $10^{-2}$, which is much lower than the critical values of 1--5. This finding is consistent with that of \citetads{2016ApJ...823...28K} and \citetads{2018MNRAS.473.3615M} who also find $\beta \ll 1$ in their simulation of collapsing protostellar cores with a mass of 100~\mo\ at later times. The rotating bodies within \W\ are therefore able to cool rapidly and any local collapse induced by the gravitational instabilities in the disks will lead to fragmentation. Rapid cooling in this context means that the cooling timescale is comparable to the local dynamical timescale.

\section{Conclusions \& Summary \label{s: conclusion}} 
Our IRAM large program, CORE, aims to study fragmentation and the kinematics of a sample of 20 high-mass star forming regions. We have performed a case-study for one of the sources in our sample, the prototypical hot core \W. In this paper we present details of the spectral line setup for our NOEMA observations in the 1.37~mm band which covers transitions of important tracers of dense gas and disks (\eg\ \mc, \mf, CH$_3$OH), inner envelopes (\eg\ $\mathrm{H_2CO}$), and outflows (\eg\ $\mathrm{^{13}CO}$, SO). We cover the range of 217\,078.6~$-$~220\,859.5~MHz with a spectral resolution of 2.7~\kms, and eight narrower bands centered on specific transitions to provide higher spectral resolution of 0.5~\kms. 

With the aim of studying the fragmentation and kinematic properties of \W, the following is a summary of our findings:

\begin{itemize}
  \item At an angular resolution of $\sim$0$\farcs35$ ($\sim$700~AU at 2~kpc), \W\ fragments into two cores, which we refer to as \W~E and \W~W, separated by $\sim$2300~AU, as seen in both line and thermal dust continuum emission.
  \item Based on the integrated dust continuum emission, \W\ has a total mass of $\sim$26.8~\mo, with 15.4~\mo\ contributed from \W\ E, and 11.4~\mo\ from \W\ W. 
  \item On large scales, there exists a clear velocity gradient in the east-west direction across \W, spanning $\sim$6000~AU, attributed to a combination of circumbinary and circumstellar motions. On smaller scales, velocity gradients of a few \kms\ are observed across each of the cores, perpendicular to the directions of two bipolar molecular outflows, one emanating from each core, as traced by $^{12}$CO and $^{13}$CO. The direction of motion of the gas around the individual cores deviates little from the overall larger-scale rotation of \W, suggesting that these motions, which we interpret as rotation, seen around the cores may be inherited from the large-scale rotation. 
  \item The kinematics of the rotating structure about \W\ W shows differential rotation of material about a (proto)star as deduced from the redshifted part of its PV plot, suggesting that the rotating structure may be a disk-like object. The radio source with a rising spectrum at this position can be attributed to a circumstellar jet or wind ionised by the embedded (proto)star at this position.
  \item The PV plots of the rotating structure about \W\ E is inconclusive on whether the observed rotation is due to a disk-like rotating object, an unresolved binary (or multiple) system, or a combination of both.
  \item We fit the emission of \mckr{4}{6} and \mcisokr{0}{3} with \textsc{xclass} and produce temperature, column density, peak velocity, and velocity dispersion maps. On average, the entire structure is hot ($\sim$165~K) with no particular temperature structure. The column density map of \mc\ is doubly peaked, similar to the continuum and line emission maps, with a median column density of $\sim$1.4$\times10^{15}~\mathrm{cm^{-2}}$. Close to the center of the cores, the H$_2$ column density is estimated to be $\sim 5 \times 10^{24}~\mathrm{cm^{-2}}$.
  \item We investigate the axis-symmetric stability of the two rotating structures using the Toomre criterion. Our \tq\ map shows low values in the outskirts of both rotating structures, suggesting that they are unstable to fragmentation. Some regions with low \tq\ values in the vicinity of \W\ W coincide with unresolved thermal dust continuum peaks (in our highest resolution observations), hinting at the possibility of further fragmentation in this core. 
 \item The Toomre-unstable regions within \W~E and \W~W are able to cool rapidly and any local collapse induced by the gravitational instabilities will lead to further fragmentation. 
\end{itemize}

In this work, we showcased our in-depth analysis for the kinematics and stability of the rotating structures within \W. We showed that high-mass cores can be prone to fragmentation induced by gravitational instabilities at $\sim$1000~AU scales, and core fragmentation at larger scales. Therefore, different modes of fragmentation contribute to the final stellar mass distribution of a given region. The question still remains, how universal are these findings for the high-mass star formation process? To this end, we aim to benchmark our methods using hydrodynamic simulations and extend our analysis to the full CORE sample in future papers. 

\begin{acknowledgements}
The authors would like to thank the anonymous referee whose comments helped the clarity of this paper. We also thank Thomas M\"oller for his technical support with the \textsc{xclass} analysis, and Kaitlin Kratter for fruitful discussions regarding the \tq\ stability analysis. AA, HB, JCM, and FB acknowledge support from the European Research Council under the European Community's Horizon 2020 framework program (2014-2020) via the ERC Consolidator Grant `From Cloud to Star Formation (CSF)' (project number 648505). RK acknowledges financial support via the Emmy Noether Research Group on Accretion Flows and Feedback in Realistic Models of Massive Star Formation funded by the German Research Foundation (DFG) under grant no. KU 2849/3-1. TCs acknowledges support from the \emph{Deut\-sche For\-schungs\-ge\-mein\-schaft, DFG\/} via the SPP (priority programme) 1573 `Physics of the ISM'. AP acknowledges financial support from UNAM and CONACyT, M\'exico.
\end{acknowledgements}



\appendix

\section{Moment maps \label{a: moment_maps_ABD}} 
\begin{figure*}[!h]
\centering
\includegraphics[width=0.9\hsize]{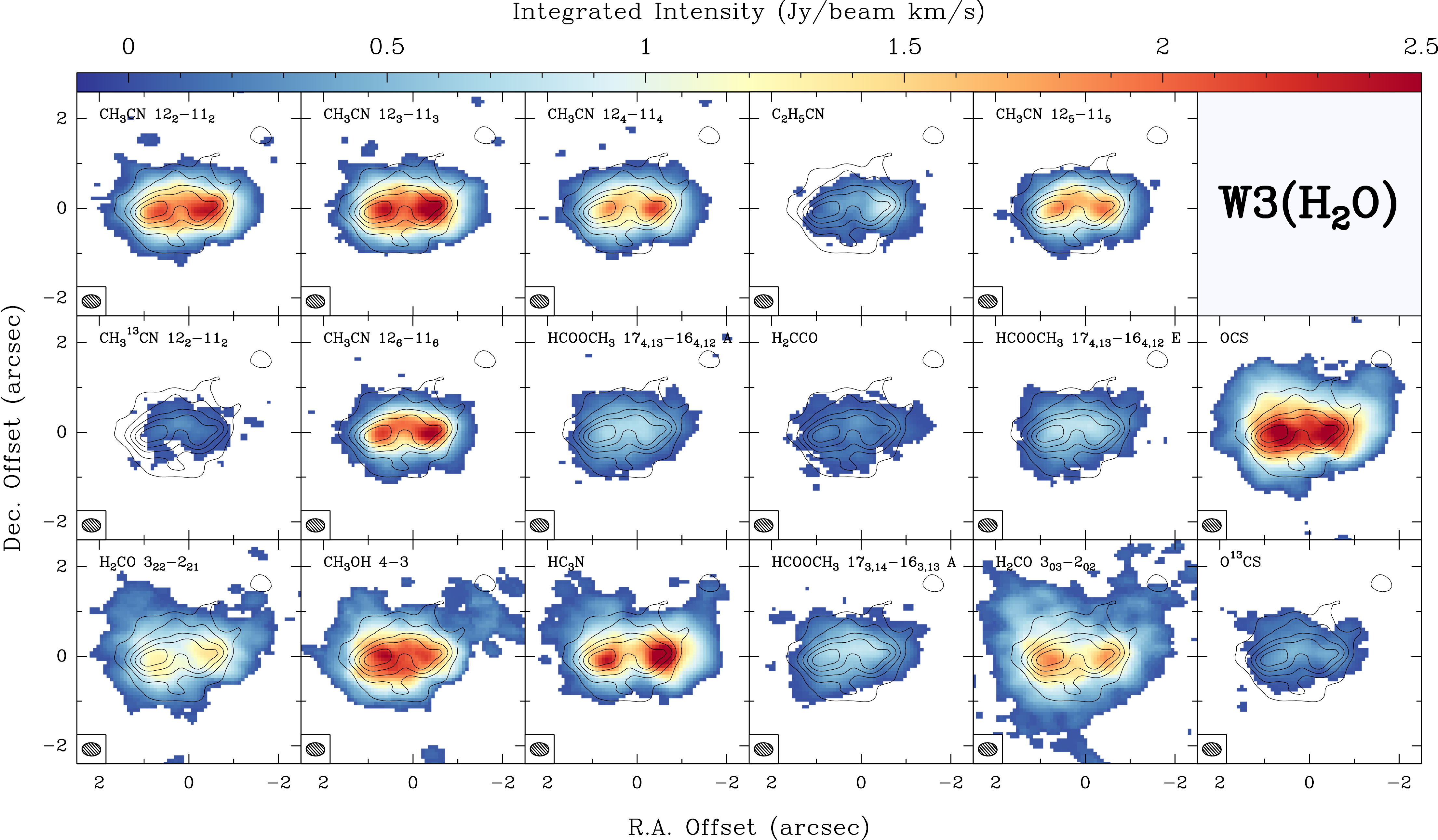}

\vspace{1cm}

\includegraphics[width=0.9\hsize]{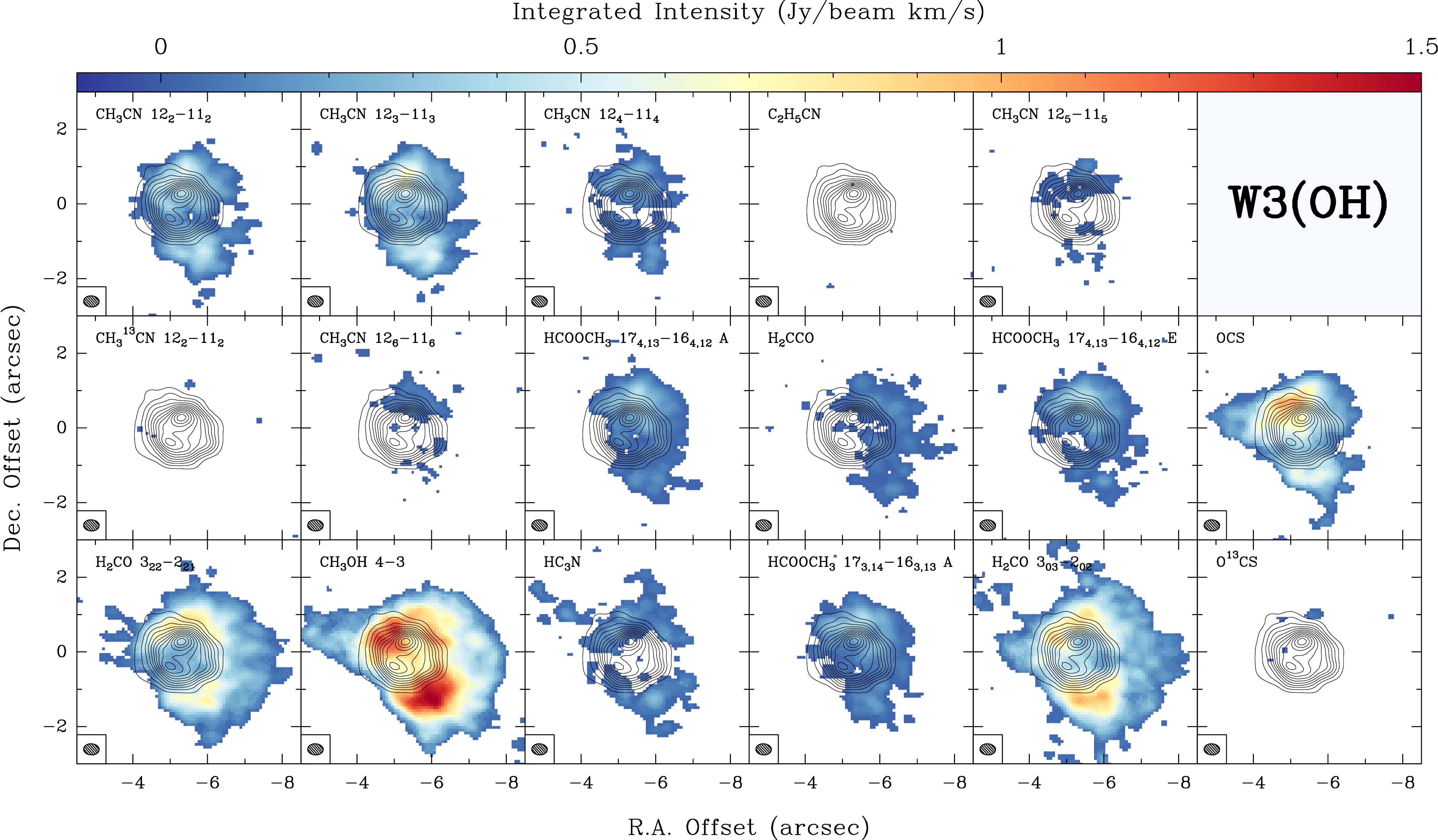}
\caption{Integrated intensity (zeroth moment) maps of most important lines covered in the narrow-band receiver for the observations in the ABD configuration for \W\ (\emph{top}) and \WOH\ (\emph{bottom}). The solid contours correspond to the dust continuum and start at and increase by $6\sigma$ ($1\sigma$ = 3.2~m\jpb). The size of the synthesized beam is shown in the bottom left of each panel. The map of \mck{5} may not be accurate because it is blended with other lines.}
\label{af: all_narrow_moment0}
\end{figure*}

\begin{figure*}[!ht]
\centering
\includegraphics[width=0.87\hsize]{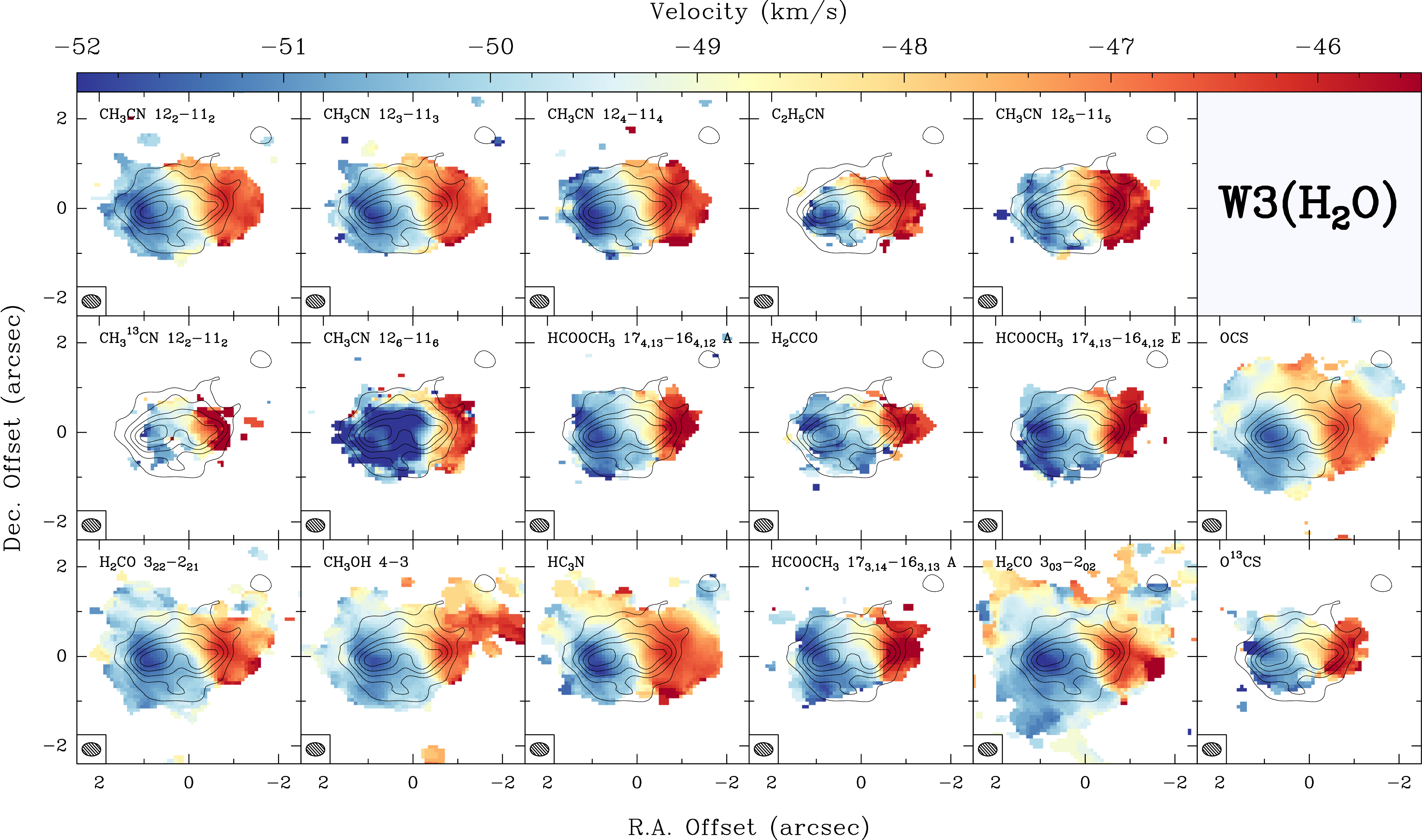}

\vspace{1cm}

\includegraphics[width=0.9\hsize]{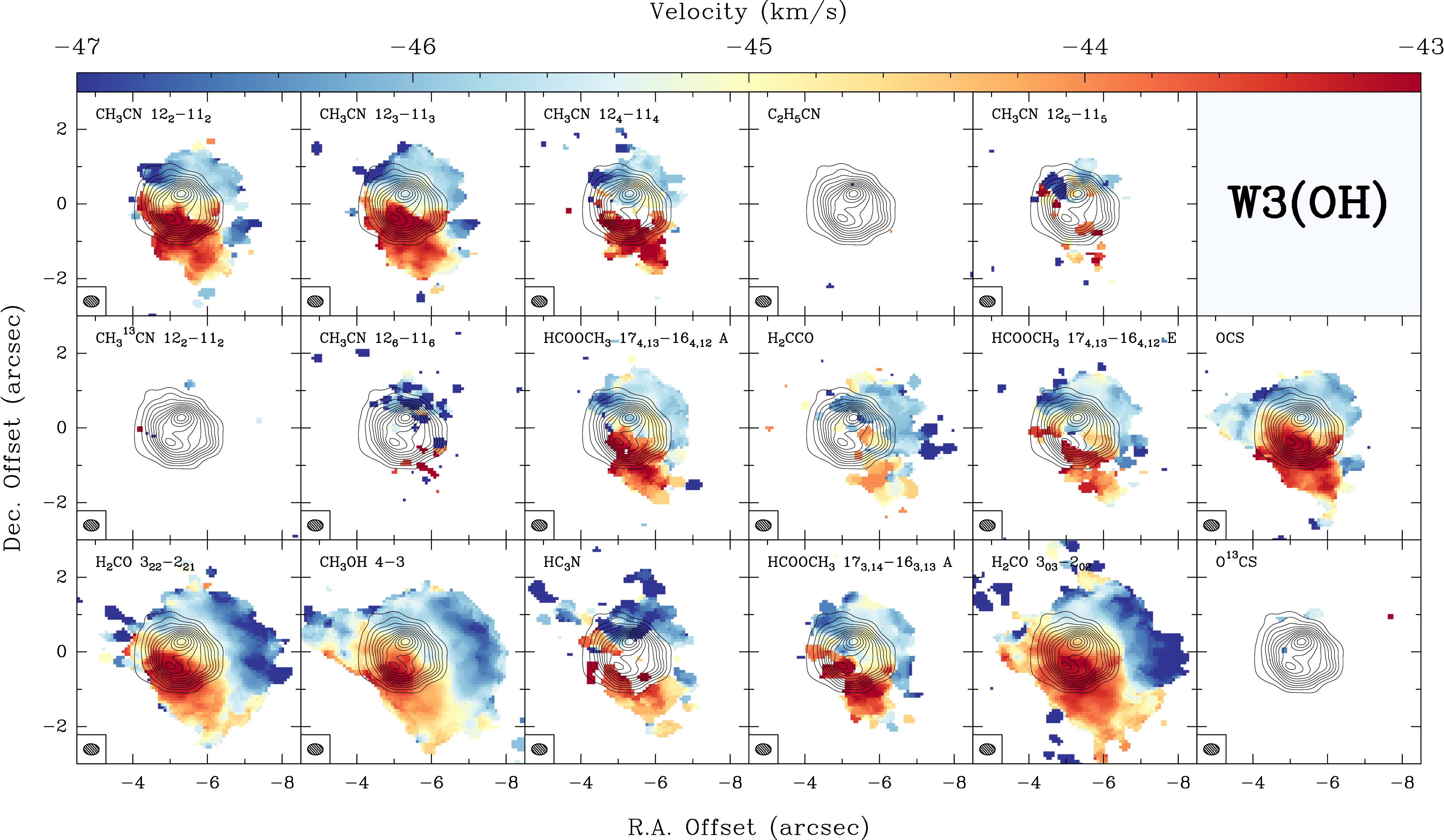}
\caption{Intensity-weighted peak velocity (first moment) maps of most important lines covered in the narrow-band receiver for the observations in the ABD configuration for \W\ (\emph{top}) and \WOH\ (\emph{bottom}). The solid contours correspond to the dust continuum and start at and increase by $6\sigma$ ($1\sigma$ = 3.2~m\jpb). The size of the synthesized beam is shown in the bottom left of each panel. The map of \mck{5} may not be accurate because it is blended with other lines.}
\label{af: all_narrow_moment1}
\end{figure*}

\begin{figure*}[!ht]
\centering
\includegraphics[width=0.87\hsize]{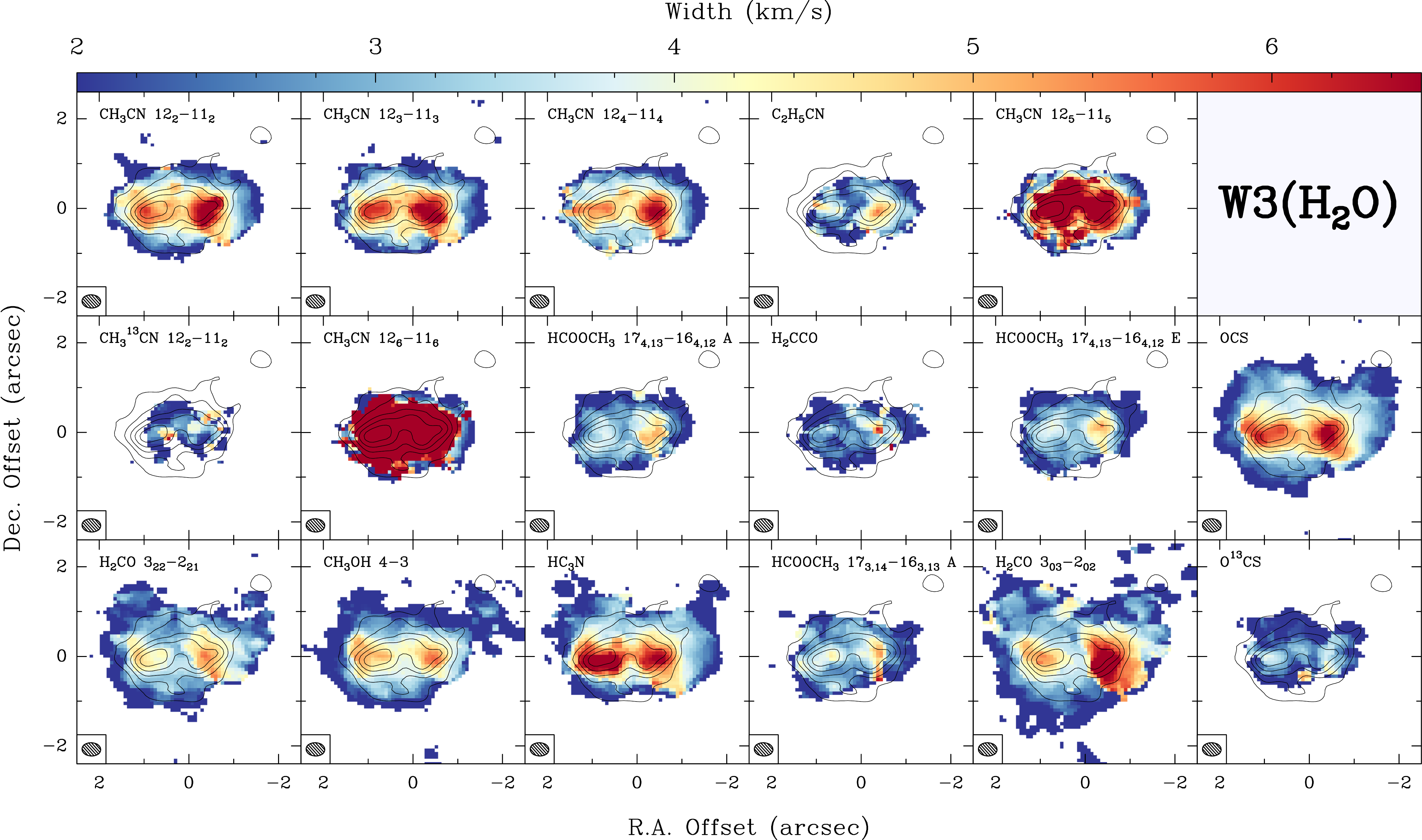}

\vspace{1cm}

\includegraphics[width=0.87\hsize]{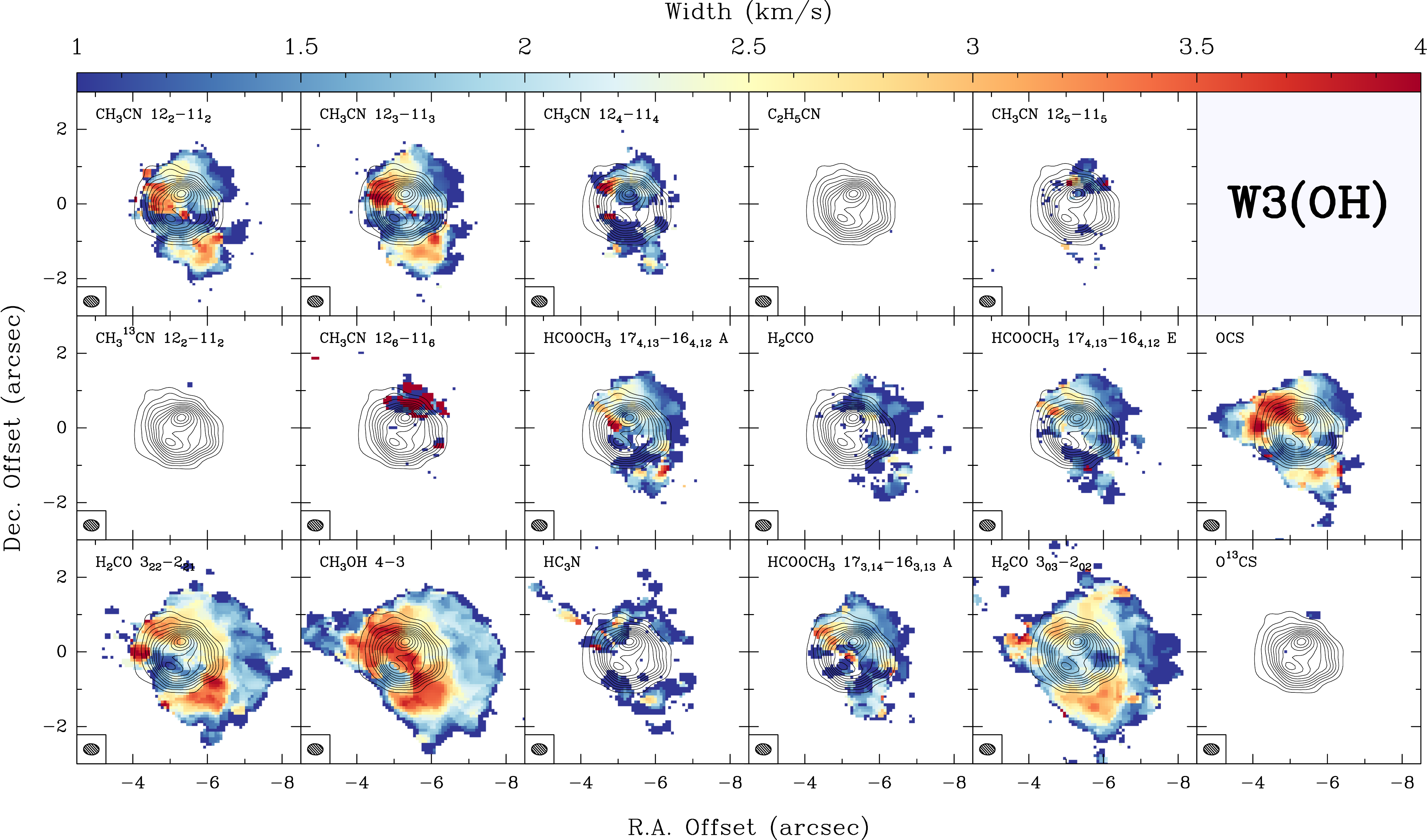}
\caption{Root mean square velocity dispersion (second moment) maps of most important lines covered in the narrow-band receiver for the observations in the ABD configuration for \W\ (\emph{top}) and \WOH\ (\emph{bottom}). The solid contours correspond to the dust continuum and start at and increase by $6\sigma$ ($1\sigma$ = 3.2~m\jpb). The size of the synthesized beam is shown in the bottom left of each panel. The map of \mck{5} may not be accurate because it is blended with other lines.}
\label{af: all_narrow_moment2}
\end{figure*}

In this section we present zeroth (Fig.~\ref{af: all_narrow_moment0}), first (Fig.~\ref{af: all_narrow_moment1}), and second (Fig.~\ref{af: all_narrow_moment2}) moment maps of various detected lines in the narrow-band receiver for \W\ (top panels) and \WOH\ (bottom panels) in the combined A-, B-, and D-array observations. All moment maps have been created inside regions where the signal-to-noise is greater than 5$\sigma$.

\clearpage 

\section{Details of \textsc{xclass} fitting \label{a: xclass}} 

\begin{figure}[!ht]
   \centering
   \includegraphics[width=\hsize]{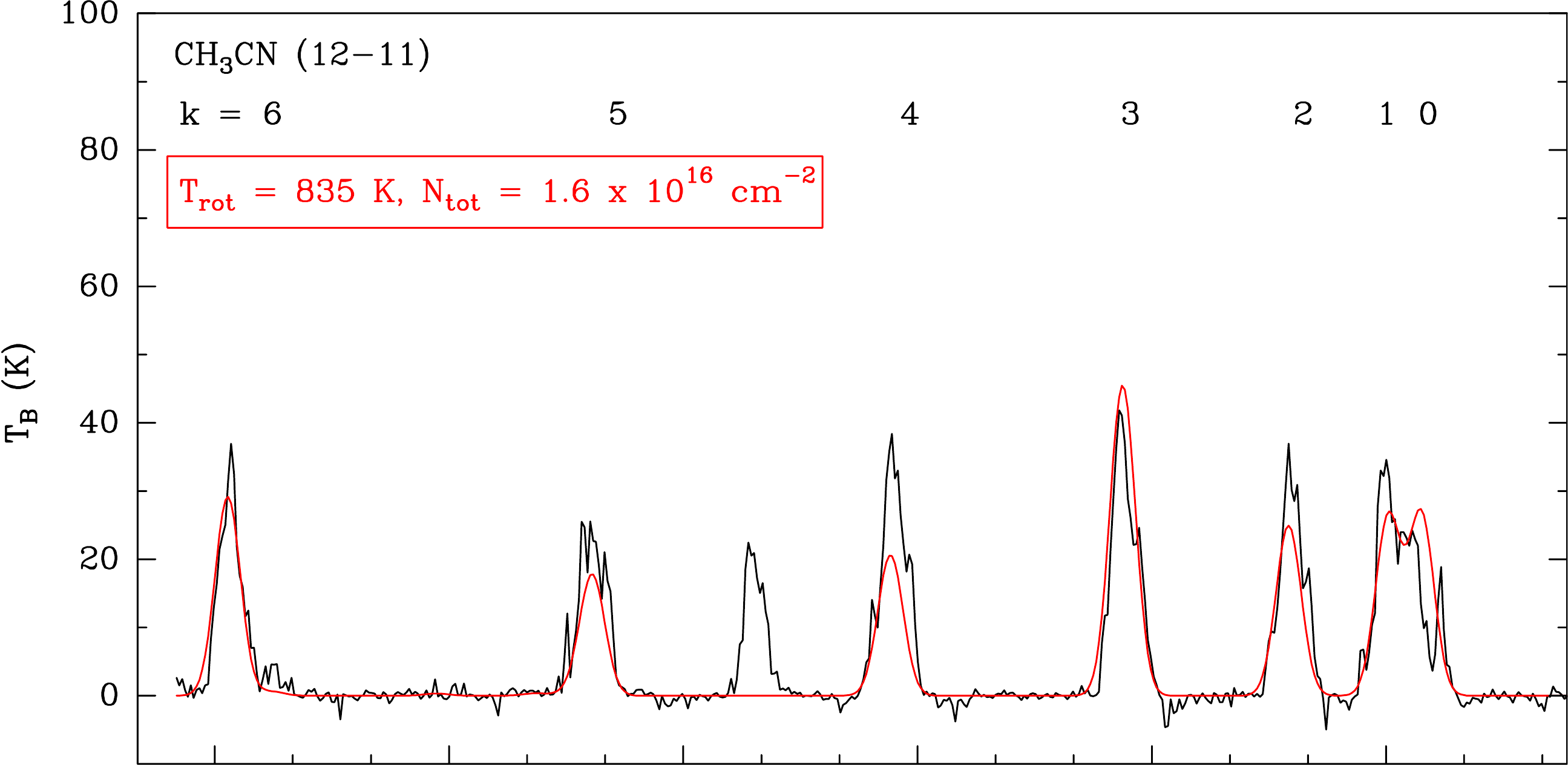}
   \includegraphics[width=\hsize]{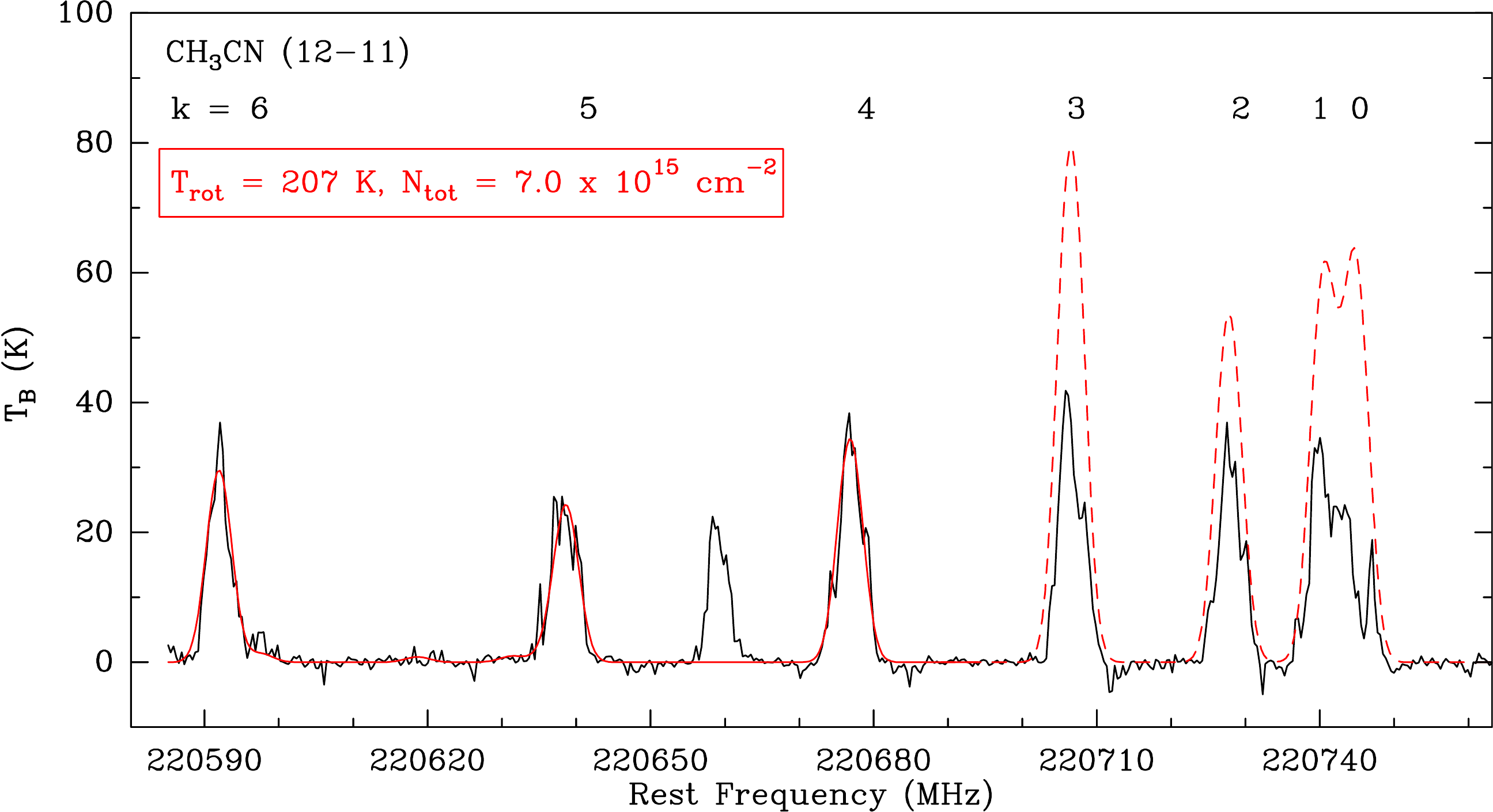}
      \caption{\emph{Top}: Spectrum of a given pixel in black along with the \textsc{xclass} fit for \mckr{0}{6} and \mcisokr{0}{3} lines in solid red. The corresponding fit parameters are provided in the panel. \emph{Bottom}: Spectrum of the same pixel as above in black with the \textsc{xclass} fit for \mckr{4}{6} and \mcisokr{0}{3} in solid red. The dashed red line corresponds to the predicted spectrum for the \mc\ lines that were not used in the fitting process. This shows that the exclusion of the low-$K$ lines in the fitting process allows \textsc{xclass} to provide a better fit for the optically-thinner high-$K$ lines. The bright line detected between $K=4$ and 5 components is $\mathrm{C_2H_5CN}$.}
         \label{af: ch3cn_XCLASS_AB_specfit}
   \end{figure}

\textsc{xclass} explores the parameter space (source size, column density, temperature, linewidth, and peak velocity) using as many algorithms as the user demands, and stops when the maximum number of iterations has been reached. In our analysis, we employed a combination of the Genetic and Levenberg-Marquardt algorithms\footnote{See the \textsc{xclass} manual for algorithm descriptions.}, and an isothermal model such that one temperature is used for reproducing the observed population of lines for a given species in a given spectrum. We assumed the size of the source to be larger than the beam and therefore did not fit the beam filling factor parameter. Our initial fits to the spectra of \mc\ included the full $K=0-6$ ladder, along with its \mcisokr{0}{3}\ isotopologue, prescribing the $^{12}$C/$^{13}$C ratio to be 76 which is consistent with the findings of \citetads{1982A&A...109..344H} and the calculations of \citetads{2015ApJ...803...39Q} for \W. The top panel of Fig.~\ref{af: ch3cn_XCLASS_AB_specfit} shows an example of the best-fit spectrum for one pixel in the AB observations, which yielded a high rotational temperature of 835~K. The lower intensities of the low-$K$ lines of \mc\ compared to the transitions higher on the $K$-ladder indicates that the low-$K$ transitions are optically thick. Furthermore, the fits to the optically thinner high-$K$ lines are not satisfactory. Repeating the same procedure but only fitting the \mckr{4}{6} lines along with the \mcisokr{0}{3}\ isotopologues results in a better fit to these lines, for a lower rotational temperature of 207~K (see bottom panel of Fig.~\ref{af: ch3cn_XCLASS_AB_specfit}). These findings are in agreement with line fitting analysis of \citetads{2015A&A...581A..71F} in Orion KL in which they find that fitting all \mc\ lines, assuming they are optically thin and in LTE, yields higher temperatures than when fitting them with an optical depth correction (see their Fig.~8). The exclusion of the optically thick and low energy lower-$K$ lines along with the inclusion of the $^{13}$C isotopologues, although barely detected, allows the software to avoid prioritizing the fitting of these optically thick lines and therefore derive the temperature more accurately. This finding is also related to the existence of temperature and density gradients which our one-component model cannot properly reproduce. As lower-$K$ lines are more easily excited than the higher-$K$ lines, they can probe the temperature in the envelope as well as the disk surface, while the higher-$K$ lines are better at tracing the disk and may not be as excited in the envelope. In fact, the reason why the brightness temperature of the low-$K$ lines is lower than the high-$K$ lines may be due to self-absorption of the photons from the warmer inner region by molecules in the cooler envelope material between the disk and the observer. Another explanation can be that the optically thick low-$K$ transitions are more extended and therefore may be partially resolved out. 

\section{Mass density map \label{a: mass_map}} 
\begin{figure}[!ht]
   \centering
   \includegraphics[width=\hsize]{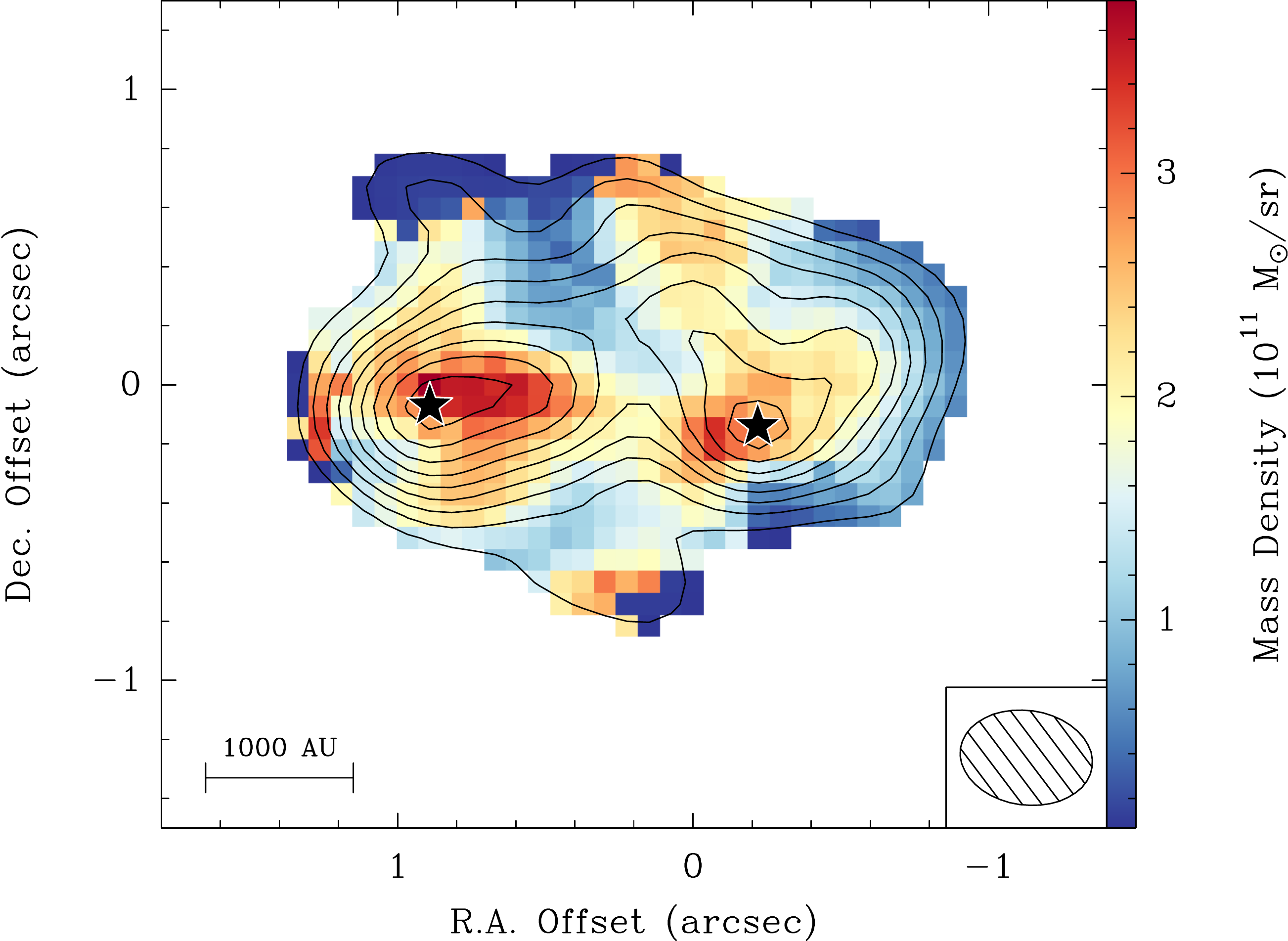}
      \caption{Mass density map obtained using the temperature and continuum maps in Eq.~\ref{e: mass}. The solid contours correspond to our continuum observations in the AB configuration, starting at 6$\sigma$ and increasing in steps of 3$\sigma$ ($1\sigma$ = 2.5~m\jpb). Each of the peak continuum positions, as depicted by stars, is expected to host at least one 10~\mo\ (proto)star or more lower mass sources.}
         \label{af: mass_map_AB}
   \end{figure}

In this section we present the mass density map for \W\ in Fig.~\ref{af: mass_map_AB}, which has been constructed using Eq.~\ref{e: mass} with the continuum map converted to \jpp\ units and the temperature map obtained from modelling \mckr{4}{6} lines with \textsc{xclass}. This map is used in the calculation of the angular velocity in the construction of the \tq\ map.

\section{\tq\ maps \label{a: toomre_maps}} 
\begin{figure}[!ht]
   \centering
   \includegraphics[width=\hsize]{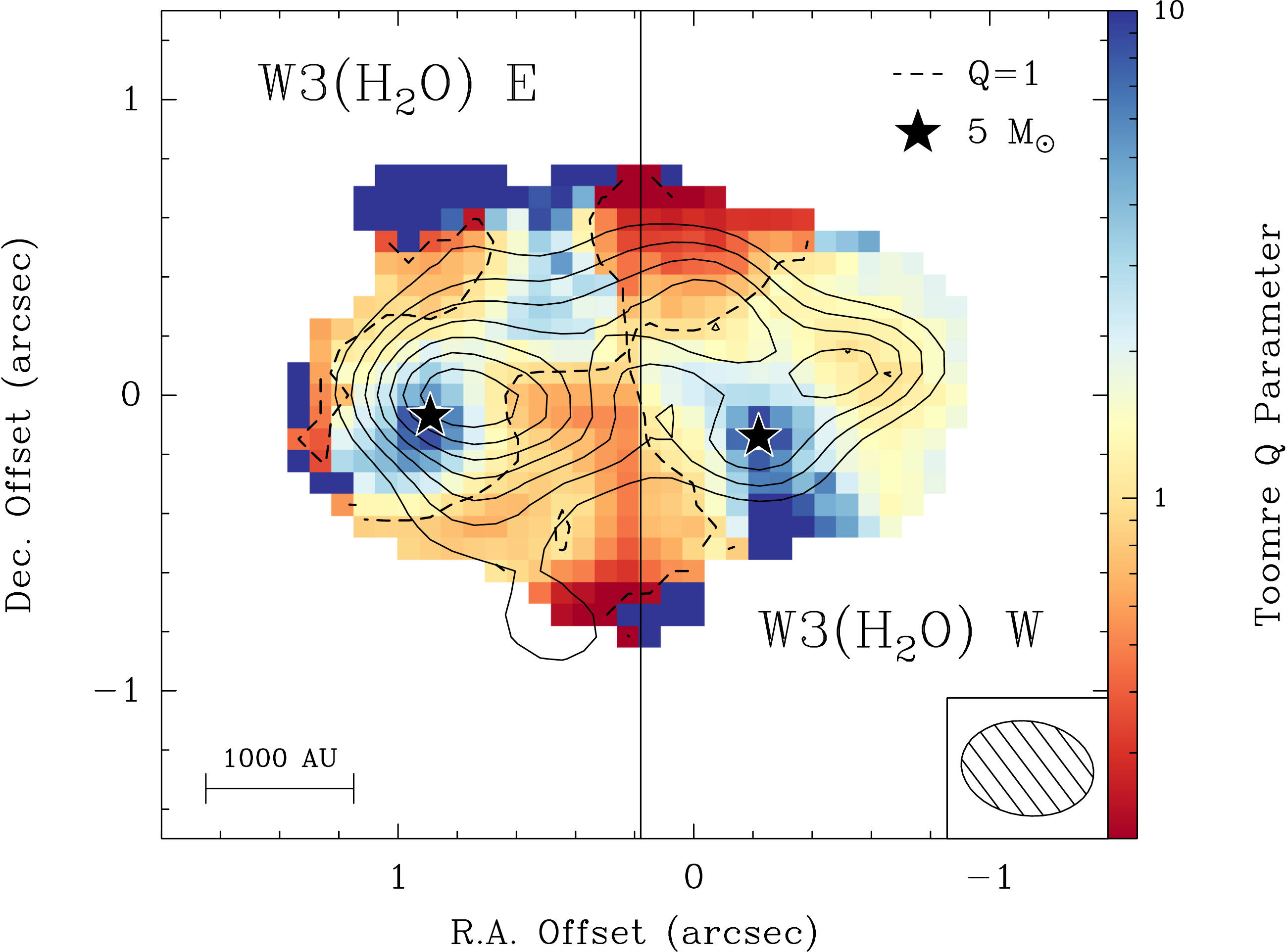}
   \includegraphics[width=\hsize]{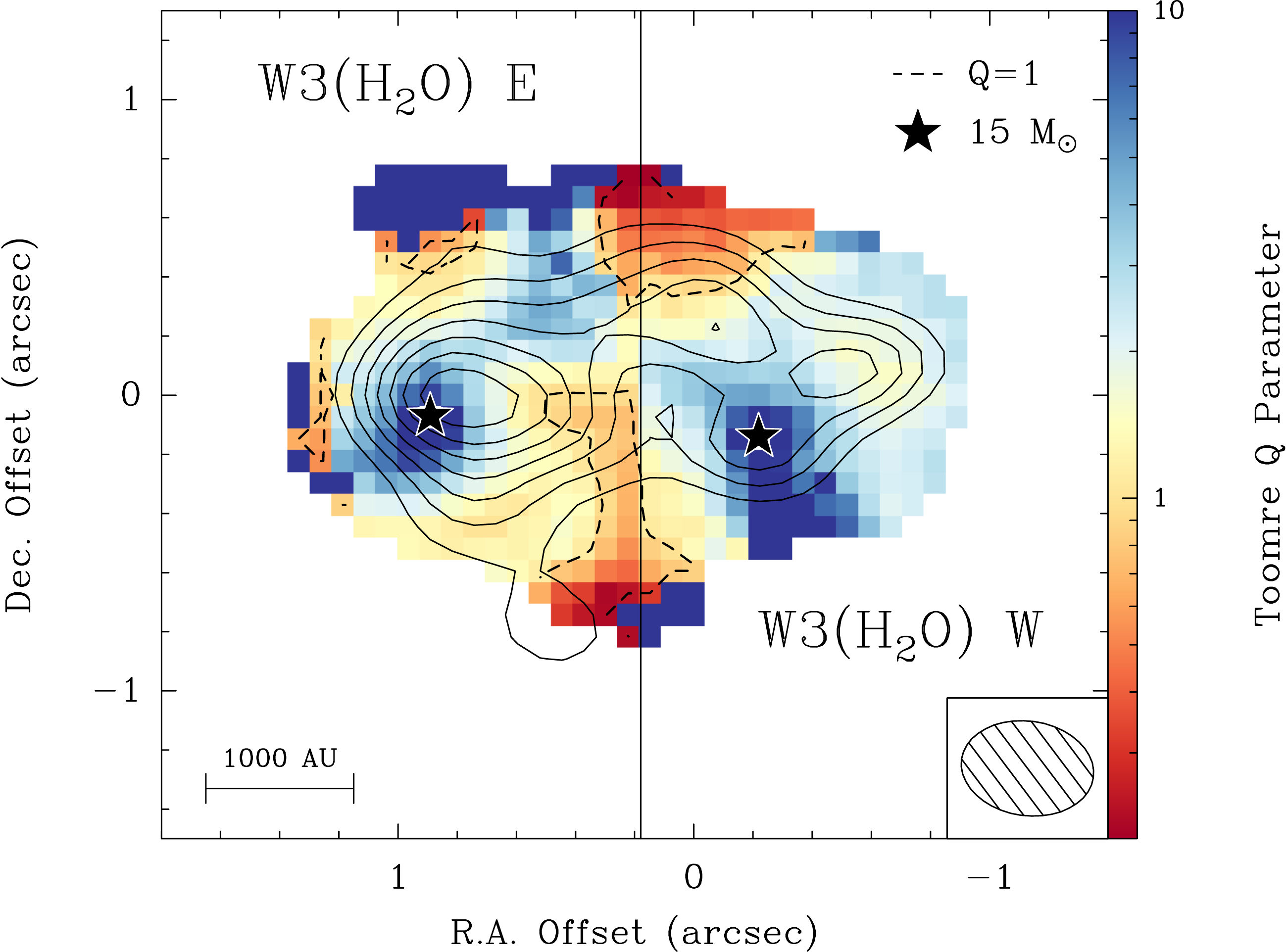}
      \caption{\tq\ map obtained by assuming two disks in gravito-centrifugal rotation about two 5~\mo\ (proto)stars (\emph{top}) and two 15~\mo\ (proto)stars (\emph{bottom}) at the positions of peak continuum emission as depicted by the two stars. The solid contours correspond to our continuum data in the most extended configuration, starting at 6$\sigma$ and increasing in steps of 3$\sigma$ ($1\sigma$ = 2.5~m\jpb). The solid vertical line corresponds to the stitching boundary. The dashed line corresponds to $Q=1$. Regions outside of the 6$\sigma$ mm continuum emission contour in the AB configuration are masked out.}
         \label{af: ToomreQ_diff_masses}
   \end{figure}

In Fig.~\ref{af: ToomreQ_diff_masses}, we present Toomre Q maps created assuming either two 5~\mo\ (proto)stars at the positions of the two continuum peaks (top panel), or two 15~\mo\ (proto)stars (bottom panel).

\begin{thebibliography}{}
\bibitem[{{Baehr} {et~al.}(2017){Baehr}, {Klahr}, \&
  {Kratter}}]{2017ApJ...848...40B}
{Baehr}, H., {Klahr}, H., \& {Kratter}, K.~M. 2017, \apj, 848, 40

\bibitem[{{Behrend} \& {Maeder}(2001)}]{2001A&A...373..190B}
{Behrend}, R. \& {Maeder}, A. 2001, \aap, 373, 190

\bibitem[{{Beltr{\'a}n} {et~al.}(2005){Beltr{\'a}n}, {Cesaroni}, {Neri},
  {Codella}, {Furuya}, {Testi}, \& {Olmi}}]{2005A&A...435..901B}
{Beltr{\'a}n}, M.~T., {Cesaroni}, R., {Neri}, R., {et~al.} 2005, \aap, 435, 901

\bibitem[{{Beltr{\'a}n} \& {de Wit}(2016)}]{2016A&ARv..24....6B}
{Beltr{\'a}n}, M.~T. \& {de Wit}, W.~J. 2016, \aapr, 24, 6

\bibitem[{{Beltr{\'a}n} {et~al.}(2014){Beltr{\'a}n}, {S{\'a}nchez-Monge},
  {Cesaroni}, {Kumar}, {Galli}, {Walmsley}, {Etoka}, {Furuya}, {Moscadelli},
  {Stanke}, {van der Tak}, {Vig}, {Wang}, {Zinnecker}, {Elia}, \&
  {Schisano}}]{2014A&A...571A..52B}
{Beltr{\'a}n}, M.~T., {S{\'a}nchez-Monge}, {\'A}., {Cesaroni}, R., {et~al.}
  2014, \aap, 571, A52

\bibitem[{{Beuther} {et~al.}(2018){Beuther}, {Mottram}, {Ahmadi}, {Bosco},
  {Linz}, {Henning}, {Klaassen}, {Winters}, {Maud}, {Kuiper}, {Semenov},
  {Gieser}, {Peters}, {Urquhart}, {Pudritz}, {Ragan}, {Feng}, {Keto},
  {Leurini}, {Cesaroni}, {Beltran}, {Palau}, {Sanchez-Monge}, {Galvan-Madrid},
  {Zhang}, {Schilke}, {Wyrowski}, {Johnston}, {Longmore}, {Lumsden}, {Hoare},
  {Menten}, \& {Csengeri}}]{2018arXiv180501191B}
{Beuther}, H., {Mottram}, J.~C., {Ahmadi}, A., {et~al.} 2018, ArXiv e-prints,
  arXiv:1805.01191

\bibitem[{{Beuther} {et~al.}(2002){Beuther}, {Schilke}, {Gueth}, {McCaughrean},
  {Andersen}, {Sridharan}, \& {Menten}}]{2002A&A...387..931B}
{Beuther}, H., {Schilke}, P., {Gueth}, F., {et~al.} 2002, \aap, 387, 931

\bibitem[{{Beuther} {et~al.}(2007){Beuther}, {Zhang}, {Hunter}, {Sridharan}, \&
  {Bergin}}]{2007A&A...473..493B}
{Beuther}, H., {Zhang}, Q., {Hunter}, T.~R., {Sridharan}, T.~K., \& {Bergin},
  E.~A. 2007, \aap, 473, 493

\bibitem[{{Beuther} {et~al.}(2005){Beuther}, {Zhang}, {Sridharan}, \&
  {Chen}}]{2005ApJ...628..800B}
{Beuther}, H., {Zhang}, Q., {Sridharan}, T.~K., \& {Chen}, Y. 2005, \apj, 628,
  800

\bibitem[{{Boehm-Vitense}(1981)}]{1981ARA&A..19..295B}
{Boehm-Vitense}, E. 1981, \araa, 19, 295

\bibitem[{{Bontemps} {et~al.}(2010){Bontemps}, {Motte}, {Csengeri}, \&
  {Schneider}}]{2010A&A...524A..18B}
{Bontemps}, S., {Motte}, F., {Csengeri}, T., \& {Schneider}, N. 2010, \aap,
  524, A18

\bibitem[{{Carrasco-Gonz{\'a}lez} {et~al.}(2012){Carrasco-Gonz{\'a}lez},
  {Galv{\'a}n-Madrid}, {Anglada}, {Osorio}, {D'Alessio}, {Hofner},
  {Rodr{\'{\i}}guez}, {Linz}, \& {Araya}}]{2012ApJ...752L..29C}
{Carrasco-Gonz{\'a}lez}, C., {Galv{\'a}n-Madrid}, R., {Anglada}, G., {et~al.}
  2012, \apjl, 752, L29

\bibitem[{{Cesaroni} {et~al.}(2007){Cesaroni}, {Galli}, {Lodato}, {Walmsley},
  \& {Zhang}}]{2007prpl.conf..197C}
{Cesaroni}, R., {Galli}, D., {Lodato}, G., {Walmsley}, C.~M., \& {Zhang}, Q.
  2007, in Protostars and Planets V, 197--212

\bibitem[{{Cesaroni} {et~al.}(2017){Cesaroni}, {S{\'a}nchez-Monge},
  {Beltr{\'a}n}, {Johnston}, {Maud}, {Moscadelli}, {Mottram}, {Ahmadi},
  {Allen}, {Beuther}, {Csengeri}, {Etoka}, {Fuller}, {Galli},
  {Galv{\'a}n-Madrid}, {Goddi}, {Henning}, {Hoare}, {Klaassen}, {Kuiper},
  {Kumar}, {Lumsden}, {Peters}, {Rivilla}, {Schilke}, {Testi}, {van der Tak},
  {Vig}, {Walmsley}, \& {Zinnecker}}]{2017A&A...602A..59C}
{Cesaroni}, R., {S{\'a}nchez-Monge}, {\'A}., {Beltr{\'a}n}, M.~T., {et~al.}
  2017, \aap, 602, A59

\bibitem[{{Chen} {et~al.}(2006){Chen}, {Welch}, {Wilner}, \&
  {Sutton}}]{2006ApJ...639..975C}
{Chen}, H.-R., {Welch}, W.~J., {Wilner}, D.~J., \& {Sutton}, E.~C. 2006, \apj,
  639, 975

\bibitem[{{Chen} {et~al.}(2016){Chen}, {Keto}, {Zhang}, {Sridharan}, {Liu}, \&
  {Su}}]{2016ApJ...823..125C}
{Chen}, H.-R.~V., {Keto}, E., {Zhang}, Q., {et~al.} 2016, \apj, 823, 125

\bibitem[{{Clark}(1980)}]{1980A&A....89..377C}
{Clark}, B.~G. 1980, \aap, 89, 377

\bibitem[{{Davies} {et~al.}(2011){Davies}, {Hoare}, {Lumsden}, {Hosokawa},
  {Oudmaijer}, {Urquhart}, {Mottram}, \& {Stead}}]{2011MNRAS.416..972D}
{Davies}, B., {Hoare}, M.~G., {Lumsden}, S.~L., {et~al.} 2011, \mnras, 416, 972

\bibitem[{{Draine}(2011)}]{2011piim.book.....D}
{Draine}, B.~T. 2011, {Physics of the Interstellar and Intergalactic Medium}
  (Princeton University Press)

\bibitem[{{Dreher} \& {Welch}(1981)}]{1981ApJ...245..857D}
{Dreher}, J.~W. \& {Welch}, W.~J. 1981, \apj, 245, 857

\bibitem[{{Evans}(1999)}]{1999ARA&A..37..311E}
{Evans}, II, N.~J. 1999, \araa, 37, 311

\bibitem[{{Fallscheer} {et~al.}(2009){Fallscheer}, {Beuther}, {Zhang}, {Keto},
  \& {Sridharan}}]{2009A&A...504..127F}
{Fallscheer}, C., {Beuther}, H., {Zhang}, Q., {Keto}, E., \& {Sridharan}, T.~K.
  2009, \aap, 504, 127

\bibitem[{{Feng} {et~al.}(2015){Feng}, {Beuther}, {Henning}, {Semenov},
  {Palau}, \& {Mills}}]{2015A&A...581A..71F}
{Feng}, S., {Beuther}, H., {Henning}, T., {et~al.} 2015, \aap, 581, A71

\bibitem[{{Frank} {et~al.}(2014){Frank}, {Ray}, {Cabrit}, {Hartigan}, {Arce},
  {Bacciotti}, {Bally}, {Benisty}, {Eisl{\"o}ffel}, {G{\"u}del}, {Lebedev},
  {Nisini}, \& {Raga}}]{2014prpl.conf..451F}
{Frank}, A., {Ray}, T.~P., {Cabrit}, S., {et~al.} 2014, in Protostars and
  Planets VI, ed. H.~{Beuther}, R.~{Klessen}, C.~{Dullemond}, \& T.~{Henning}
  (Univ. of Arizona Press, Tucson), 451--474

\bibitem[{{Gammie}(2001)}]{2001ApJ...553..174G}
{Gammie}, C.~F. 2001, \apj, 553, 174

\bibitem[{{Goldreich} \& {Lynden-Bell}(1965)}]{1965MNRAS.130..125G}
{Goldreich}, P. \& {Lynden-Bell}, D. 1965, \mnras, 130, 125

\bibitem[{{Green}(1986)}]{1986ApJ...309..331G}
{Green}, S. 1986, \apj, 309, 331

\bibitem[{{Hachisuka} {et~al.}(2006){Hachisuka}, {Brunthaler}, {Menten},
  {Reid}, {Imai}, {Hagiwara}, {Miyoshi}, {Horiuchi}, \&
  {Sasao}}]{2006ApJ...645..337H}
{Hachisuka}, K., {Brunthaler}, A., {Menten}, K.~M., {et~al.} 2006, \apj, 645,
  337

\bibitem[{{Henkel} {et~al.}(1982){Henkel}, {Wilson}, \&
  {Bieging}}]{1982A&A...109..344H}
{Henkel}, C., {Wilson}, T.~L., \& {Bieging}, J. 1982, \aap, 109, 344

\bibitem[{{Hildebrand}(1983)}]{1983QJRAS..24..267H}
{Hildebrand}, R.~H. 1983, \qjras, 24, 267

\bibitem[{{Ilee} {et~al.}(2016){Ilee}, {Cyganowski}, {Nazari}, {Hunter},
  {Brogan}, {Forgan}, \& {Zhang}}]{2016MNRAS.462.4386I}
{Ilee}, J.~D., {Cyganowski}, C.~J., {Nazari}, P., {et~al.} 2016, \mnras, 462,
  4386

\bibitem[{{Johnson} \& {Gammie}(2003)}]{2003ApJ...597..131J}
{Johnson}, B.~M. \& {Gammie}, C.~F. 2003, \apj, 597, 131

\bibitem[{{Johnston} {et~al.}(2015){Johnston}, {Robitaille}, {Beuther}, {Linz},
  {Boley}, {Kuiper}, {Keto}, {Hoare}, \& {van Boekel}}]{2015ApJ...813L..19J}
{Johnston}, K.~G., {Robitaille}, T.~P., {Beuther}, H., {et~al.} 2015, \apjl,
  813, L19

\bibitem[{{Kahn}(1974)}]{1974A&A....37..149K}
{Kahn}, F.~D. 1974, \aap, 37, 149

\bibitem[{{Keto} \& {Wood}(2006)}]{2006ApJ...637..850K}
{Keto}, E. \& {Wood}, K. 2006, \apj, 637, 850

\bibitem[{{Keto} {et~al.}(1995){Keto}, {Welch}, {Reid}, \&
  {Ho}}]{1995ApJ...444..765K}
{Keto}, E.~R., {Welch}, W.~J., {Reid}, M.~J., \& {Ho}, P.~T.~P. 1995, \apj,
  444, 765

\bibitem[{{Klassen} {et~al.}(2016){Klassen}, {Pudritz}, {Kuiper}, {Peters}, \&
  {Banerjee}}]{2016ApJ...823...28K}
{Klassen}, M., {Pudritz}, R.~E., {Kuiper}, R., {Peters}, T., \& {Banerjee}, R.
  2016, \apj, 823, 28

\bibitem[{{Kratter} \& {Lodato}(2016)}]{2016ARA&A..54..271K}
{Kratter}, K. \& {Lodato}, G. 2016, \araa, 54, 271

\bibitem[{{Krumholz} {et~al.}(2009){Krumholz}, {Klein}, {McKee}, {Offner}, \&
  {Cunningham}}]{2009Sci...323..754K}
{Krumholz}, M.~R., {Klein}, R.~I., {McKee}, C.~F., {Offner}, S.~S.~R., \&
  {Cunningham}, A.~J. 2009, Science, 323, 754

\bibitem[{{Kuiper} {et~al.}(2010){Kuiper}, {Klahr}, {Beuther}, \&
  {Henning}}]{2010ApJ...722.1556K}
{Kuiper}, R., {Klahr}, H., {Beuther}, H., \& {Henning}, T. 2010, \apj, 722,
  1556

\bibitem[{{Kuiper} {et~al.}(2011){Kuiper}, {Klahr}, {Beuther}, \&
  {Henning}}]{2011ApJ...732...20K}
{Kuiper}, R., {Klahr}, H., {Beuther}, H., \& {Henning}, T. 2011, \apj, 732, 20

\bibitem[{{Kuiper} \& {Yorke}(2013)}]{2013ApJ...772...61K}
{Kuiper}, R. \& {Yorke}, H.~W. 2013, \apj, 772, 61

\bibitem[{{Kurtz} {et~al.}(1994){Kurtz}, {Churchwell}, \&
  {Wood}}]{1994ApJS...91..659K}
{Kurtz}, S., {Churchwell}, E., \& {Wood}, D.~O.~S. 1994, \apjs, 91, 659

\bibitem[{{Leurini} {et~al.}(2011){Leurini}, {Codella}, {Zapata},
  {Beltr{\'a}n}, {Schilke}, \& {Cesaroni}}]{2011A&A...530A..12L}
{Leurini}, S., {Codella}, C., {Zapata}, L., {et~al.} 2011, \aap, 530, A12

\bibitem[{{Loren} \& {Mundy}(1984)}]{1984ApJ...286..232L}
{Loren}, R.~B. \& {Mundy}, L.~G. 1984, \apj, 286, 232

\bibitem[{{Martins} {et~al.}(2005){Martins}, {Schaerer}, \&
  {Hillier}}]{2005A&A...436.1049M}
{Martins}, F., {Schaerer}, D., \& {Hillier}, D.~J. 2005, \aap, 436, 1049

\bibitem[{{Matsumoto} \& {Hanawa}(2003)}]{2003ApJ...595..913M}
{Matsumoto}, T. \& {Hanawa}, T. 2003, \apj, 595, 913

\bibitem[{{Maud} {et~al.}(2015){Maud}, {Moore}, {Lumsden}, {Mottram},
  {Urquhart}, \& {Hoare}}]{2015MNRAS.453..645M}
{Maud}, L.~T., {Moore}, T.~J.~T., {Lumsden}, S.~L., {et~al.} 2015, \mnras, 453,
  645

\bibitem[{{Meyer} {et~al.}(2018){Meyer}, {Kuiper}, {Kley}, {Johnston}, \&
  {Vorobyov}}]{2018MNRAS.473.3615M}
{Meyer}, D.~M.-A., {Kuiper}, R., {Kley}, W., {Johnston}, K.~G., \& {Vorobyov},
  E. 2018, \mnras, 473, 3615

\bibitem[{{Meyer} {et~al.}(2017){Meyer}, {Vorobyov}, {Kuiper}, \&
  {Kley}}]{2017MNRAS.464L..90M}
{Meyer}, D.~M.-A., {Vorobyov}, E.~I., {Kuiper}, R., \& {Kley}, W. 2017, \mnras,
  464, L90

\bibitem[{{M{\"o}ller} {et~al.}(2017){M{\"o}ller}, {Endres}, \&
  {Schilke}}]{2017A&A...598A...7M}
{M{\"o}ller}, T., {Endres}, C., \& {Schilke}, P. 2017, \aap, 598, A7

\bibitem[{{Motte} {et~al.}(2017){Motte}, {Bontemps}, \&
  {Louvet}}]{2017arXiv170600118M}
{Motte}, F., {Bontemps}, S., \& {Louvet}, F. 2017, ArXiv e-prints
  [\eprint[arXiv]{1706.00118}]

\bibitem[{{Mottram} {et~al.}(2011){Mottram}, {Hoare}, {Urquhart}, {Lumsden},
  {Oudmaijer}, {Robitaille}, {Moore}, {Davies}, \&
  {Stead}}]{2011A&A...525A.149M}
{Mottram}, J.~C., {Hoare}, M.~G., {Urquhart}, J.~S., {et~al.} 2011, \aap, 525,
  A149

\bibitem[{{M{\"u}ller} {et~al.}(2005){M{\"u}ller}, {Schl{\"o}der}, {Stutzki},
  \& {Winnewisser}}]{2005JMoSt.742..215M}
{M{\"u}ller}, H.~S.~P., {Schl{\"o}der}, F., {Stutzki}, J., \& {Winnewisser}, G.
  2005, Journal of Molecular Structure, 742, 215

\bibitem[{{M{\"u}ller} {et~al.}(2001){M{\"u}ller}, {Thorwirth}, {Roth}, \&
  {Winnewisser}}]{2001A&A...370L..49M}
{M{\"u}ller}, H.~S.~P., {Thorwirth}, S., {Roth}, D.~A., \& {Winnewisser}, G.
  2001, \aap, 370, L49

\bibitem[{{Norberg} \& {Maeder}(2000)}]{2000A&A...359.1025N}
{Norberg}, P. \& {Maeder}, A. 2000, \aap, 359, 1025

\bibitem[{{Ohashi} {et~al.}(1997){Ohashi}, {Hayashi}, {Ho}, \&
  {Momose}}]{1997ApJ...475..211O}
{Ohashi}, N., {Hayashi}, M., {Ho}, P.~T.~P., \& {Momose}, M. 1997, \apj, 475,
  211

\bibitem[{{Ossenkopf} \& {Henning}(1994)}]{1994A&A...291..943O}
{Ossenkopf}, V. \& {Henning}, T. 1994, \aap, 291, 943

\bibitem[{{Palau} {et~al.}(2015){Palau}, {Ballesteros-Paredes},
  {V{\'a}zquez-Semadeni}, {S{\'a}nchez-Monge}, {Estalella}, {Fall}, {Zapata},
  {Camacho}, {G{\'o}mez}, {Naranjo-Romero}, {Busquet}, \&
  {Fontani}}]{2015MNRAS.453.3785P}
{Palau}, A., {Ballesteros-Paredes}, J., {V{\'a}zquez-Semadeni}, E., {et~al.}
  2015, \mnras, 453, 3785

\bibitem[{{Palau} {et~al.}(2013){Palau}, {Fuente}, {Girart}, {Estalella}, {Ho},
  {S{\'a}nchez-Monge}, {Fontani}, {Busquet}, {Commer{\c c}on}, {Hennebelle},
  {Boissier}, {Zhang}, {Cesaroni}, \& {Zapata}}]{2013ApJ...762..120P}
{Palau}, A., {Fuente}, A., {Girart}, J.~M., {et~al.} 2013, \apj, 762, 120

\bibitem[{{Peters} {et~al.}(2010){Peters}, {Banerjee}, {Klessen}, {Mac Low},
  {Galv{\'a}n-Madrid}, \& {Keto}}]{2010ApJ...711.1017P}
{Peters}, T., {Banerjee}, R., {Klessen}, R.~S., {et~al.} 2010, \apj, 711, 1017

\bibitem[{{Pickett} {et~al.}(1998){Pickett}, {Poynter}, {Cohen}, {Delitsky},
  {Pearson}, \& {M{\"u}ller}}]{1998JQSRT..60..883P}
{Pickett}, H.~M., {Poynter}, R.~L., {Cohen}, E.~A., {et~al.} 1998, \jqsrt, 60,
  883

\bibitem[{{Pudritz} {et~al.}(2007){Pudritz}, {Ouyed}, {Fendt}, \&
  {Brandenburg}}]{2007prpl.conf..277P}
{Pudritz}, R.~E., {Ouyed}, R., {Fendt}, C., \& {Brandenburg}, A. 2007, in
  Protostars and Planets V, ed. B.~{Reipurth}, D.~{Jewitt}, \& K.~{Keil} (Univ.
  of Arizona Press, Tucson), 277--294

\bibitem[{{Qin} {et~al.}(2015){Qin}, {Schilke}, {Wu}, {Wu}, {Liu}, {Liu}, \&
  {S{\'a}nchez-Monge}}]{2015ApJ...803...39Q}
{Qin}, S.-L., {Schilke}, P., {Wu}, J., {et~al.} 2015, \apj, 803, 39

\bibitem[{{Reid} {et~al.}(1995){Reid}, {Argon}, {Masson}, {Menten}, \&
  {Moran}}]{1995ApJ...443..238R}
{Reid}, M.~J., {Argon}, A.~L., {Masson}, C.~R., {Menten}, K.~M., \& {Moran},
  J.~M. 1995, \apj, 443, 238

\bibitem[{{Safronov}(1960)}]{1960AnAp...23..979S}
{Safronov}, V.~S. 1960, Annales d'Astrophysique, 23, 979

\bibitem[{{S{\'a}nchez-Monge}(2011)}]{Sanchez-MongeThesis}
{S{\'a}nchez-Monge}, {\'A}. 2011, PhD thesis, Universitat de Barcelona, Spain

\bibitem[{{S{\'a}nchez-Monge} {et~al.}(2013){S{\'a}nchez-Monge}, {Cesaroni},
  {Beltr{\'a}n}, {Kumar}, {Stanke}, {Zinnecker}, {Etoka}, {Galli}, {Hummel},
  {Moscadelli}, {Preibisch}, {Ratzka}, {van der Tak}, {Vig}, {Walmsley}, \&
  {Wang}}]{2013A&A...552L..10S}
{S{\'a}nchez-Monge}, {\'A}., {Cesaroni}, R., {Beltr{\'a}n}, M.~T., {et~al.}
  2013, \aap, 552, L10

\bibitem[{{Schuller} {et~al.}(2009){Schuller}, {Menten}, {Contreras},
  {Wyrowski}, {Schilke}, {Bronfman}, {Henning}, {Walmsley}, {Beuther},
  {Bontemps}, {Cesaroni}, {Deharveng}, {Garay}, {Herpin}, {Lefloch}, {Linz},
  {Mardones}, {Minier}, {Molinari}, {Motte}, {Nyman}, {Reveret}, {Risacher},
  {Russeil}, {Schneider}, {Testi}, {Troost}, {Vasyunina}, {Wienen}, {Zavagno},
  {Kovacs}, {Kreysa}, {Siringo}, \& {Wei{\ss}}}]{2009A&A...504..415S}
{Schuller}, F., {Menten}, K.~M., {Contreras}, Y., {et~al.} 2009, \aap, 504, 415

\bibitem[{{Shchekinov} \& {Sobolev}(2004)}]{2004A&A...418.1045S}
{Shchekinov}, Y.~A. \& {Sobolev}, A.~M. 2004, \aap, 418, 1045

\bibitem[{{Shirley}(2015)}]{2015PASP..127..299S}
{Shirley}, Y.~L. 2015, \pasp, 127, 299

\bibitem[{{Steer} {et~al.}(1984){Steer}, {Dewdney}, \&
  {Ito}}]{1984A&A...137..159S}
{Steer}, D.~G., {Dewdney}, P.~E., \& {Ito}, M.~R. 1984, \aap, 137, 159

\bibitem[{{Tobin} {et~al.}(2012){Tobin}, {Hartmann}, {Bergin}, {Chiang},
  {Looney}, {Chandler}, {Maret}, \& {Heitsch}}]{2012ApJ...748...16T}
{Tobin}, J.~J., {Hartmann}, L., {Bergin}, E., {et~al.} 2012, \apj, 748, 16

\bibitem[{{Toomre}(1964)}]{1964ApJ...139.1217T}
{Toomre}, A. 1964, \apj, 139, 1217

\bibitem[{{Turner} \& {Welch}(1984)}]{1984ApJ...287L..81T}
{Turner}, J.~L. \& {Welch}, W.~J. 1984, \apjl, 287, L81

\bibitem[{{Williams} {et~al.}(2000){Williams}, {Blitz}, \&
  {McKee}}]{2000prpl.conf...97W}
{Williams}, J.~P., {Blitz}, L., \& {McKee}, C.~F. 2000, in Protostars and
  Planets IV, ed. V.~{Mannings}, A.~{Boss}, \& S.~{Russell} (Univ. of Arizona
  Press, Tucson), 97

\bibitem[{{Wilner} {et~al.}(1999){Wilner}, {Reid}, \&
  {Menten}}]{1999ApJ...513..775W}
{Wilner}, D.~J., {Reid}, M.~J., \& {Menten}, K.~M. 1999, \apj, 513, 775

\bibitem[{{Wolfire} \& {Cassinelli}(1987)}]{1987ApJ...319..850W}
{Wolfire}, M.~G. \& {Cassinelli}, J.~P. 1987, \apj, 319, 850

\bibitem[{{Wyrowski} {et~al.}(1999){Wyrowski}, {Schilke}, {Walmsley}, \&
  {Menten}}]{1999ApJ...514L..43W}
{Wyrowski}, F., {Schilke}, P., {Walmsley}, C.~M., \& {Menten}, K.~M. 1999,
  \apjl, 514, L43

\bibitem[{{Xu} {et~al.}(2006){Xu}, {Reid}, {Zheng}, \&
  {Menten}}]{2006Sci...311...54X}
{Xu}, Y., {Reid}, M.~J., {Zheng}, X.~W., \& {Menten}, K.~M. 2006, Science, 311,
  54

\bibitem[{{Yorke} \& {Sonnhalter}(2002)}]{2002ApJ...569..846Y}
{Yorke}, H.~W. \& {Sonnhalter}, C. 2002, \apj, 569, 846

\bibitem[{{Zapata} {et~al.}(2011){Zapata}, {Rodr{\'{\i}}guez-Garza},
  {Rodr{\'{\i}}guez}, {Girart}, \& {Chen}}]{2011ApJ...740L..19Z}
{Zapata}, L.~A., {Rodr{\'{\i}}guez-Garza}, C., {Rodr{\'{\i}}guez}, L.~F.,
  {Girart}, J.~M., \& {Chen}, H.-R. 2011, \apjl, 740, L19

\bibitem[{{Zhang} {et~al.}(1998){Zhang}, {Ho}, \&
  {Ohashi}}]{1998ApJ...494..636Z}
{Zhang}, Q., {Ho}, P.~T.~P., \& {Ohashi}, N. 1998, \apj, 494, 636

\end{thebibliography}
\end{document}